\documentclass[preprint,superscriptaddress,amsmath,amssymb,floatfix, longbibliography, nofootinbib]{revtex4-1}
\usepackage{graphicx}
\usepackage{float}
\usepackage{latexsym,amsmath,amssymb,bm,euscript,multirow,makecell}
\usepackage{color,esvect}
\usepackage{wasysym}
\usepackage{epstopdf}
    \usepackage[colorlinks=true,linkcolor=blue,citecolor=blue,pdfencoding=auto, psdextra,urlcolor=blue]{hyperref}
\usepackage{type1cm}
\usepackage{appendix} 

\usepackage{booktabs,multirow}

\definecolor{amber}{rgb}{1.0, 0.49, 0.0}

\begin{document}
\title{Non-Abelian currents bootstrap}

\author{Yin-Chen He}
\affiliation{Perimeter Institute for Theoretical Physics, Waterloo, Ontario N2L 2Y5, Canada}
\author{Junchen Rong}
\affiliation{Institut des Hautes \'Etudes Scientifiques, 91440 Bures-sur-Yvette, France
}
\author{Ning Su}
\affiliation{Department of Physics, University of Pisa and INFN, \\Largo Pontecorvo 3, I-56127 Pisa, Italy
}
\author{Alessandro Vichi}
\affiliation{Department of Physics, University of Pisa and INFN, \\Largo Pontecorvo 3, I-56127 Pisa, Italy
}

\begin{abstract}
We initiate the study of correlation functions of non-Abelian spin-1 conserved current in three dimensional conformal field theories using numerical conformal bootstrap. We discuss the general framework and apply it to the particular cases of $SU(N)$ and $O(N)$ global symmetry. In both cases we obtain general bounds on operator dimensions. In the large-$N$ limit our bounds show features in correspondence of the expected position of fermionic QED$_3$ in three dimensions, as well as other interesting theories.
By imposing gaps inspired by the spectrum of QED$_3$ at large-$N$, we manage to restrict the plane of certain operator dimensions to a small island, where QED$_3$ must live.
\end{abstract}
\maketitle

\tableofcontents

\section{Introduction}

The conformal bootstrap \cite{Rattazzi:2008pe,Rychkov:2009ij} has proven itself a successful tool which allowed us to precisely determine critical exponents of many conformal field theories (CFTs) \cite{Kos:2016ysd,Atanasov:2022bpi,Chester:2019ifh,Chester:2020iyt}, such as the O(N) vector model and Gross-Neveu Yukawa CFTs \cite{Erramilli:2022kgp} (see also \cite{Poland:2018epd} for a review).
Despite the outstanding results obtained, CFTs arising as infrared (IR) fixed points of gauge theories have still eluded any bootstrap attempt to narrow them \cite{Chester:2016wrc,Li:2018lyb,He:2020azu,Li:2020bnb,He:2021xvg,Albayrak:2021xtd,He:2021sto,Li:2021emd,Reehorst:2020phk,Manenti:2021elk}. 

Quantum electrodynamics in 2+1 dimensions, besides being an interesting theoretical model by itself, is also an interesting model because of its connection to important condensed matter systems. 
In particular, the IR fixed point of $N=4$ flavors of Dirac fermions coupled to U(1) gauge theory were proposed to describe the so-called U(1) Dirac spin liquid (DSL) phase \cite{affleck1988large,wen1996theory,hastings2000dirac,hermele2005algebraic,hermele2008properties,song2020spinon,song2019unifying}. 
Similar to the famous Kosterlitz–Thouless phase \cite{kosterlitz1973ordering} in 1+1 dimensions, the U(1) Dirac spin liquid phase is a conformal phase of matter. 
Various theoretical and even real life materials were proposed which could potentially realize the U(1) DSL phase in their phase diagram \cite{song2020spinon,song2019unifying,ran2007projected,iqbal2013gapless,iqbal2016spin,he2017signatures,hu2019dirac}.   

The validity of these theoretical predictions and correspondingly the realization of the conformal U(1) DSL phase in nature depend on two necessary conditions: 1) the $N=4$ QED$_3$ theory indeed flow to an IR conformal fixed point under renormalization, 2) certain operators that are not protected by the lattice symmetry must be irrelevant. 

For QED$_3$ theory with $N$ flavors of Dirac fermions coupled to U(1) gauge theory, it is believed that there exists a critical $N^{critical}$, above which the theory flows to a conformal fixed point. Below $N^{critical}$, on the other hand, the conformal phase is replaced by a gapped phase analogous to the chiral symmetry breaking phase of four dimension Quantum chromodynamics (QCD). Remember that for (4-dimensional) QCD with SU(3) gauge symmetry coupled to $N$ quarks, there also exists a critical $N^{critical}$, above which the theory is believed to flow to the famous Banks-Zaks fixed point \cite{Belavin:1974gu,caswell1974asymptotic,banks1982phase}. 
Even though there have been many attempts in determining the $N^{critical}$ for 2+1 dimensional QED$_3$ using various methods \cite{dagotto1990chiral,Kaveh2005,Grover2012Chiral,Karthik2015,Karthik2016,Giombi2015,Lorenzo2016}, just like the 3+1D QCD, the number has not been concretely determined, mainly due to the strong coupling nature of the theory. 

Among the many models proposed to realize the U(1) Dirac spin liquid, one promising type is the spin-1/2 quantum magnet on various non-bipartite lattices (e.g. kagome and triangular lattice).
On these lattices, the lattice model usually preserves a UV symmetry group which is strictly smaller than the symmetry group of the IR conformal QED$_3$ theory. 
From the renormalization group flow point of view, this means that if an operator is not protected by symmetry, it will be generated by the Renormalization Group (RG) flow. 
Since the conformal fixed point lives in the IR of the RG flow, if the operator is an irrelevant operator of the CFT, its coupling constant will grow along the RG flow and eventually will drive the system away from the conformal fixed point.
In other words, if the conformal phase of QED$_3$ can be realized as a conformal phase of these lattice Hamiltonians, all operators allowed by the UV symmetry must be irrelevant. 

In order to check both of the necessary conditions concretely, it is very important to study the corresponding CFT non-perturbatively, this makes the conformal bootstrap method a very promising tool to attack this problem.
However, even though there have been various attempts to precisely pin-point of $N=4$ QED$_3$ among the vast region consistent with the bootstrap constraints \cite{He:2021sto,Albayrak:2021xtd}, its precise location has not been identified yet. 
The second necessary condition was used in \cite{He:2021sto} to get lower bounds on the order parameter critical exponents. 
The lower bounds seem to be consistent with various Monte Carlo studies. 

One possible approach to bootstrap QED$_3$ is to first locate the theory in a certain perturbative limit, where perturbative results provide a concrete guide in the choice of assumptions on the operator dimensions. 
Indeed, bootstrap islands corresponding to the large-$N$ limits of scalar QED$_3$ theories has been discovered in \cite{He:2021xvg}. 
After locating QED$_3$ in the perturbative limits, one may hope to follow the bootstrap islands to the non-perturbative regime by continuously changing the fermion/boson flavor number $N$.
Recently, various advanced numerical techniques \cite{reehorst2021navigator} have made such a calculation possible \cite{SlavaUnpublished2}.

It is not our goal in this paper to completely solve the problem of bootstrapping QED$_3$. 
We will rather try to attack the first part of the problem, finding QED$_3$ in a bootstrap setup in the large-$N$ limit.
One of the main features of CFTs corresponding to gauge theories is the absence of low dimensional scalar operators, since the fundamental fields appearing in the UV Lagrangian are not gauge invariant. 
Unfortunately conformal bootstrap starts to lose strength when the scaling dimension is far from the unitarity bound, leading to bounds that are not very constraining. 
On the other hand, a common feature of gauge theories is the presence of conserved currents associated with non-abelian global symmetries. 
One might hope that considering such universal operators inside correlation functions might lead to stronger constraints.

In the context of supersymmetric CFTs the inclusion of conserved operators is greatly simplified since most conserved currents belong to supersymmetric multiplets also containing scalar operators. Since supersymmetry relates correlation functions of operators in the same multiplet, one can get non trivial constraints by restricting to scalars only. In the general case instead one needs to deal with correlation functions of spinning operators. 
Previous works have considered conserved spin-1 currents associated to a global abelian $U(1)$ symmetry, both alone \cite{Dymarsky:2017xzb} and mixed with a charged scalar \cite{Reehorst:2019pzi}, as well as the stress tensor \cite{Dymarsky:2017yzx}.
In this work we initiate a systematic study of correlation functions of non-abelian conserved currents. 
We begin by considering the non-abelian currents alone and study the bounds on the CFT data contributing to this particular four point (4pt) function in the case of $SU(N)$ and $O(N)$ global symmetry. 
We plan to consider mixed correlation functions in future works. 
We will comment on this in the conclusions.

One of the main difficulties of the present analysis is that any CFT with a given global symmetry represents a solution of our crossing equations, it being free, bosonic, fermionic, based on gauge theory or not. Thus, in order to focus on a specific theory it is important to get a rough intuition of its expected spectrum and how it differs from other CFTs. After reviewing the general setup in Sec.~\ref{sec:setup}, we will concentrate on the case of $SU(N)$ and $O(N)$ global symmetries in Sec.~\ref{sec:SUN} and Sec.~\ref{sec:ON}. At the beginning of each section we discuss possible assumptions that can distinguish among gauge theories in the large-$N$ limit and less interesting free theories (more details are also given in Appendix~\ref{sec:spectrum}). Next, we present various bounds that apply to the space of all CFTs with a given global symmetry. Finally, in order to focus on a specific theory, we will impose ad-hoc assumptions on the CFT spectrum and obtain more stringent bounds. Appendix~\ref{sec:abeliancase} contains a detailed review of the bootstrap constraints for four identical currents, while Appendix~\ref{sec:nonabeliancase} generalizes the analysis to the non abelian case and Appendix~\ref{sec:cb} formulates the boostrap problem.

\section{Setup} 
\label{sec:setup}

In this work we consider the constraints arising from imposing CFT axioms on the 4pt functions of four conserved currents associated to a global symmetry $\mathcal G$. 

Since the current $J^a_\mu$ satisfies Ward identities, its normalization is fixed by this condition. Hence the two point function normalization acquire a physical meaning, as for the stress energy tensor:
\begin{align}
    \langle J_\mu^a(x)J_\nu^b(0)\rangle  =C_J  \frac{\eta_{\mu\nu}-2\frac{x_\mu x_\nu}{x^2}}{x^{2(d-1)}} \delta^{ab}\,.
\end{align}
The coefficient $C_J$ is called current central charge.

Whenever we take the diverge of $J_\mu^a$, up to contact terms\footnote{Since we always work at non coincident points, we will discard all contact terms, unless explicitly stated.}, one has the conservation condition
\begin{align}
    \langle \partial^\mu J_\mu^a(x) \mathcal O_1(x_1)\ldots \mathcal O_n(x_n)\rangle = 0.
\end{align}
for any set of operators $\mathcal O_i(x_i)$. As we review in the appendices, when restricted to 3pt function, the above conditions translates to an algebraic relation among OPE coefficients involving $J$, while when imposed on 4pt function, it produces a set of differential equations linking some of the functions parametrizing the correlator. 

In Appendix~\ref{sec:abeliancase} we first review the bootstrap setup for abelian currents and we generalize it to the non-abelian case in Appendix~\ref{sec:nonabeliancase}. In the main text we will maintain a schematic notation. Since currents transform in the adjoint representation of $\mathcal G$ we have the schematic OPE\footnote{We suppress dependence on coordinates that take care of right scaling dimensions. We also omit descendants.}
\begin{align}
\label{eq:schematic-OPE}
 J_\mu^a \times J_\nu^b  \sim \sum_{r \in Adj\otimes Adj}  \widetilde{T}_{r}^{ab,A}\sum_{\mathcal O} \sum_{i=1}^{n_3} \lambda_{JJO}^{(i)} \widetilde{Q}_i^{\mu\nu,\alpha_1\ldots\alpha_\ell}\mathcal O^{A}_{\alpha_1\ldots\alpha_\ell}\,.
\end{align}
In the above expression we denoted $\widetilde{T}_{r}^{ab,A}$ a tensor in flavor space that converts adjoint indices $a,b$ to representation $r$ indices, collectively denoted $A$. Similarly $\widetilde{Q}_i^{\mu\nu,\alpha_1\ldots\alpha_\ell}$ translates the vector indices $\mu,\nu$ to spin-$\ell$ representation. 
Notice that for a given operator there are multiple OPE coefficients $\lambda_{JJO}^{(i)}$: this is different from the OPE of two scalar operators.

Similarly, we write the 4pt correlation function as
\begin{align}
\label{eq:schematic-4pt}
   & \langle J_\mu^a(x_1)J_\nu^b(x_2)J_\rho^c(x_3)J_\sigma^d(x_4) \rangle = 
    \sum_{r} \sum_{s=1}^{n_4} \mathcal{K}_4(x_i) T_{r}^{abcd} \mathcal{Q}_{\mu\nu\rho\sigma}^{(s)}(x_i) g_s^{(r)}(u,v) \,,\\
   & \mathcal{K}_4(x_i) = x_{12}^{-4}x_{34}^{-4}, \quad u=\frac{x_{12}^2x_{34}^2}{x_{13}^2x_{24}^2}, \quad v=\frac{x_{14}^2x_{23}^2}{x_{13}^2x_{24}^2} , \quad x_{ij}^2 = (x_i^\mu-x_j^\mu)(x_i^\mu-x_j^\mu)\,,
   \end{align}
where the sum over $r$ represents a finite sum over irreducible representations of $\mathcal G$ while $s$ labels the independent tensor structures $\mathcal{Q}_{\mu\nu\rho\sigma}^{(s)}(x_i)$ one can construct using four points. The number of independent four point structures is $n_4=41$ when $J$'s are non-conserved currents, the number will be greatly reduced if $J$ is conserved. 
Finally $T_{r}^{abcd}$ are invariant tensors of the flavor symmetry group. The indices $a,b,c,d$ live in the adjoint representation of the flavor symmetry group, the superscript $r$ denotes the irreducible representations that appears in the (reducible) tensor product representation $Adj\otimes Adj$ \eqref{tensorreps}. $T_{r}^{abcd}$ are projectors that project the $Adj\otimes Adj$ \eqref{tensorreps} into invariant sub-spaces that correspond to the irreducible representations. 
The conformal invariant functions $g_s^{(r)}(u,v)$ also admit a conformal block decomposition in terms of operators in a given flavor symmetry irreducible representation, as in \eqref{eq:schematic-OPE}. The coefficients in such an expansion are related to the OPE coefficients in \eqref{eq:schematic-OPE}.  
The details will be discussed in Appendix \ref{sec:abeliancase} and \ref{sec:nonabeliancase}.

\subsection{Special operators in \texorpdfstring{$J^a_\mu \times J^b_\nu $}{}}

As reviewed in Appendix~\ref{sec3pt-:abeliancase}, in the case of abelian currents, there are no spin-1 operators appearing in the OPE $J\times J$, neither parity even, nor parity odd. One important novelty in the present setup is instead the possibility to access these operators. In particular, $J^a_\mu$ appears in its own OPE. Following \cite{Li:2015itl} we can parametrize the 3pt function $ \langle J_\mu^a(x_1)J_\nu^b(x_2)J_\rho^c(x_3) \rangle $ using the embedding formalism \cite{Costa:2011mg} (see appendix \ref{sec:abeliancase} for all definitions)
\begin{align}\label{eq:JJJ}
    \langle J^a(P_1,Z_1)J^b(P_2,Z_2)J^c(P_3,Z_3) \rangle = f^{abc} \frac{\lambda_1 V_1 V_2 V_3+\lambda_2 (V_1 H_{23}  + H_{13}V_2   +H_{12} V_3)}{(-2P_1\cdot P_2)^\frac32(-2P_1\cdot P_3)^\frac32(-2P_2\cdot P_3)^\frac32}\,,
\end{align}
where $P_i,Z_i$ are coordinates and polarizations in the embedding space.
Using Ward identities one can relate the coefficients $\lambda_1,\lambda_2$ to the current central charge $C_J$ and a second parameter:
\begin{align}
    \lambda_1 = \frac{3C_J}{S_3} - 5\lambda_{JJJ}\,, \qquad \lambda_2 = -\lambda_{JJJ} \,.
\end{align}
Here $S_d$ is the volume of the $(d-1)$ dimensional sphere $S_d=\frac{2\pi^{d/2}}{\Gamma(d/2)}$. 
An important point we should stress is that the 3pt function of three conserved currents is parametrized by 2 independent coefficients while the 3pt function of two currents and a generic non-conserved parity even spin-1 operator by only one.

Another important operator is the stress tensor, which appears in the parity even singlet sector:
\begin{align}\label{eq:JJT}
        \langle J^a(P_1,Z_1)J^b(P_2,Z_2)T(P_3,Z_3) \rangle = \delta^{ab} \frac{\lambda_1 V_1 V_2 V_3^2+\lambda_2 H_{12} V_3^2+\lambda_3 (V_1 H_{23}  + H_{13}V_2) + \lambda_4 H_{13}H_{23}  )}{(-2P_1\cdot P_2)^\frac12(-2P_1\cdot P_3)^\frac12(-2P_2\cdot P_3)^\frac12}\,.
\end{align}
A convenient parametrization of the coefficients $\lambda_i$ is \cite{Li:2015itl,Dymarsky:2017xzb}:
\begin{align}
\label{eq:gamma}
 \lambda_{1}&= -\left(7+108 \gamma \right)C\,,  &\lambda_{2}&= (5+36 \gamma )C\,,  &C=&\frac3{32\pi}C_J\,,  \nonumber  \\
 \lambda_{3}&=-6(1+4 \gamma ) C \,,  &\lambda_{4}&=2  \left(1-12 \gamma \right)C\,. &
\end{align}

We recall that the parameter $\gamma$ is subject to the conformal collider bounds $|\gamma|\leq 1/2$, as shown in \cite{HofmanConformalCollider2008}.

\section{\texorpdfstring{$SU(N)$}{} currents }
\label{sec:SUN}
After the general introduction of the previous sections, let us now concentrate on two cases of interest: in this section we analyze theories with a global $SU(N)$ global symmetry, while the next section will focus on  $O(N)$ symmetry. In a similar fashion as for the case of abelian currents \cite{Dymarsky:2017xzb}, it is possible to obtain bounds on the smallest dimension operator with a given spin, parity and transforming in a given irreducible representation of the global symmetry group. In the main text we only present plots relevant to the discussion.

\subsection{OPE and targets}

The tensor product of two $SU(N)$ adjoint representations $Adj_{i}$, i.e. rank-2 traceless tensors with one index in the fundamental and one in the anti-fundamental, is given by

\begin{equation}\label{tensorreps}
Adj_1 \times Adj_2 = S^+ +Adj^+ + S\bar{S}^+ + A\bar{A}^+ +Adj^- + S\bar{A}^- + A\bar{S}^-\,.
\end{equation}
Here $S$ and $Adj$ refer to $SU(N)$ singlet and adjoint representations. Instead $A\bar A$, $S\bar S$, $ S\bar A$, $S\bar A$ are rank-4 tensors with two upper and two lower indices. The naming convention refers to whether the upper and lower indices are symmetric (S) or antisymmetric (A). In addition the label $\pm$ refers to the symmetry properties under exchange of the irreps $Adj_i$. In the case of OPE of scalars, this would translate into the parity of spins exchanged in a given representation. In the case of spinning operators the connection is less straightforward and is outlined in Appendix \ref{sec:nonabeliancase}.

Below we will use the notation (Space-time Parity, Group representation, Spin) to refer to the OPE channels. For example, $(-, S, \ell=2)$ means parity odd, $SU(N)$ singlet, and $\ell=2$ channel.

Since our approach is very general, our bounds will contain all CFTs with an $SU(N)$ global symmetry, including free theories of fermions, free theories of scalars and generalized free fields (GFFs) of currents. The latter are defined as a conserved current of dimension $d-1$ and standard two point function, while all other correlation functions are defined though Wick contractions. All these theories only contain operators with integer dimension.

In order to focus on more interesting interacting theories one has to carefully understand the expected spectrum of the target theory and hopefully identify a series of assumptions that eliminates unwanted solutions. Let us focus on the difference among QED$_3$, GNY and QCD$_3$ at large-$N$. 

\begin{figure}[b]
    \centering
    \includegraphics[width=0.49\textwidth]{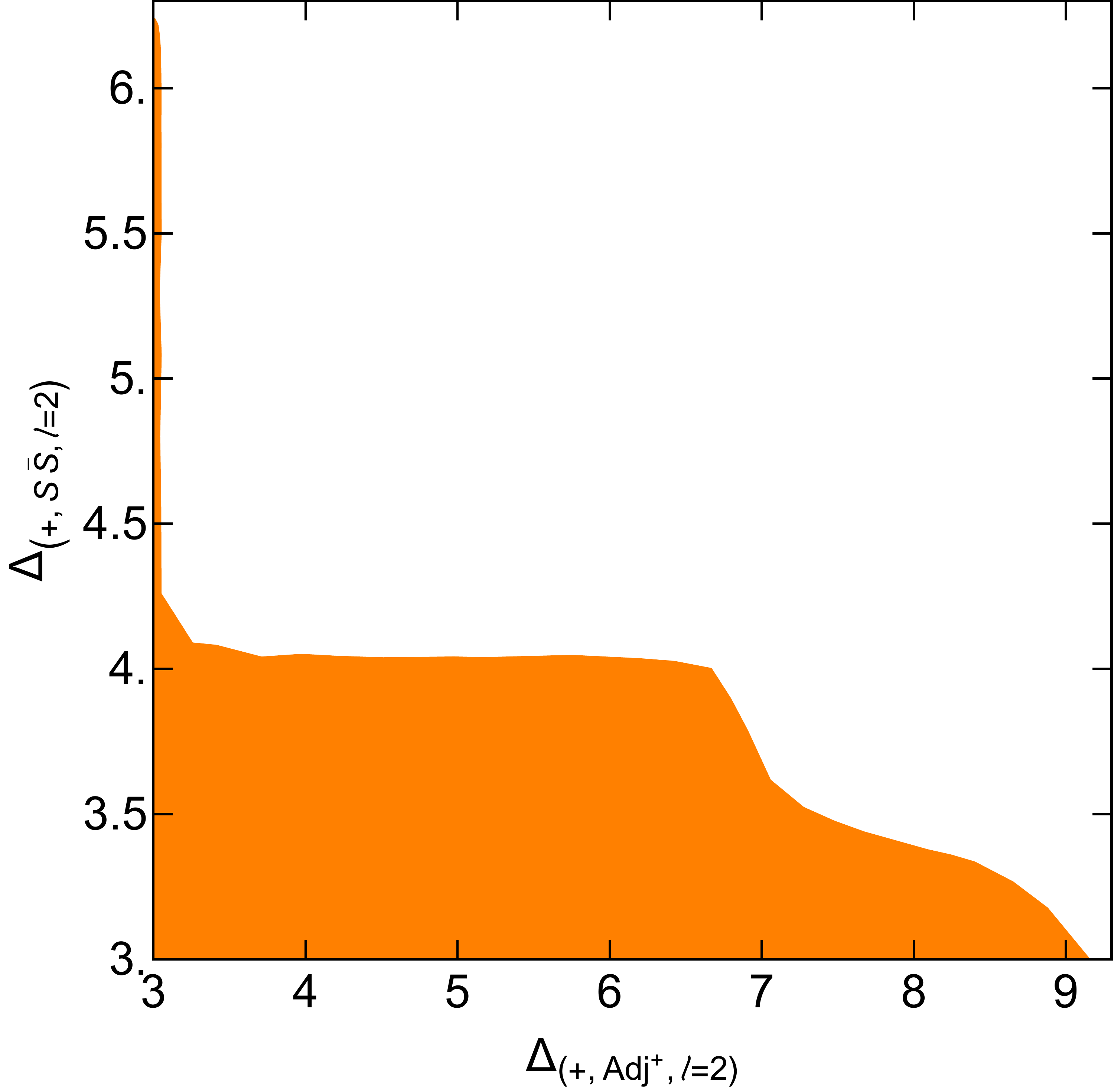}
    \includegraphics[width=0.49\textwidth]{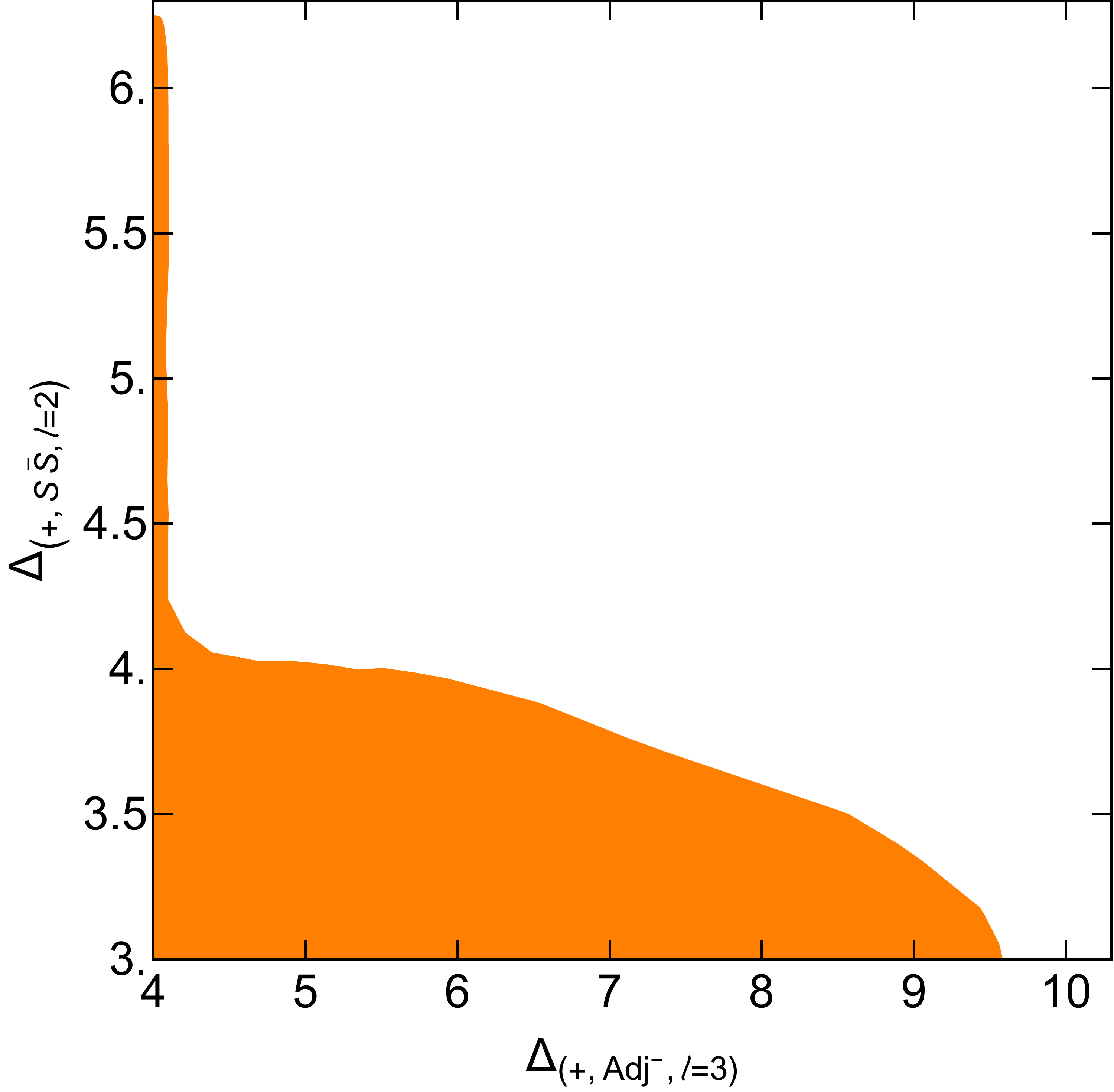}
    \caption{The lowest operator in the $(+, S\bar S$, $\ell=2)$ sector v.s. the sector $(+, Adj^+, \ell=2)$ and $(+, Adj^-, \ell=3)$: $SU(100)$ CFT. $\Lambda=19$ and no gap is imposed.}
    \label{fig:SSbl2-nogap}
\end{figure}

The key feature distinguishing QED$_3$ from QCD$_3$ is in the leading operators in the $(+, S\bar S, \ell=2)$ sector, where QED$_3$ and QCD$_3$ take values around 6 and 4 respectively. This difference is due to the fact QED$_3$ has no color index. In the Lagrangian picture, in order to construct such an operator in QED$_3$, some derivatives need to be inserted, which increases the scaling dimension. Such a mechanism to distinguish different color groups is explained in \cite{Reehorst:2020phk,Manenti:2021elk} and in \cite{,He:2021xvg}, where the operator is called the decoupling operator. Demanding a large gap closer to 6 in the sector $(+, S\bar S, \ell=2)$, we are left with QED$_3$, free fermion and GNY. Finally, a sharp difference between GNY and QED$_3$ is that GNY has a singlet scalar of dimension 2, while QED$_3$ has no relevant singlet scalar at large-$N$. In the next section we make use of these expectations to strengthen our bounds and zoom on QED$_3$ at large-$N$.

\begin{figure}[h]
    \centering
    \includegraphics[width=0.49\textwidth]{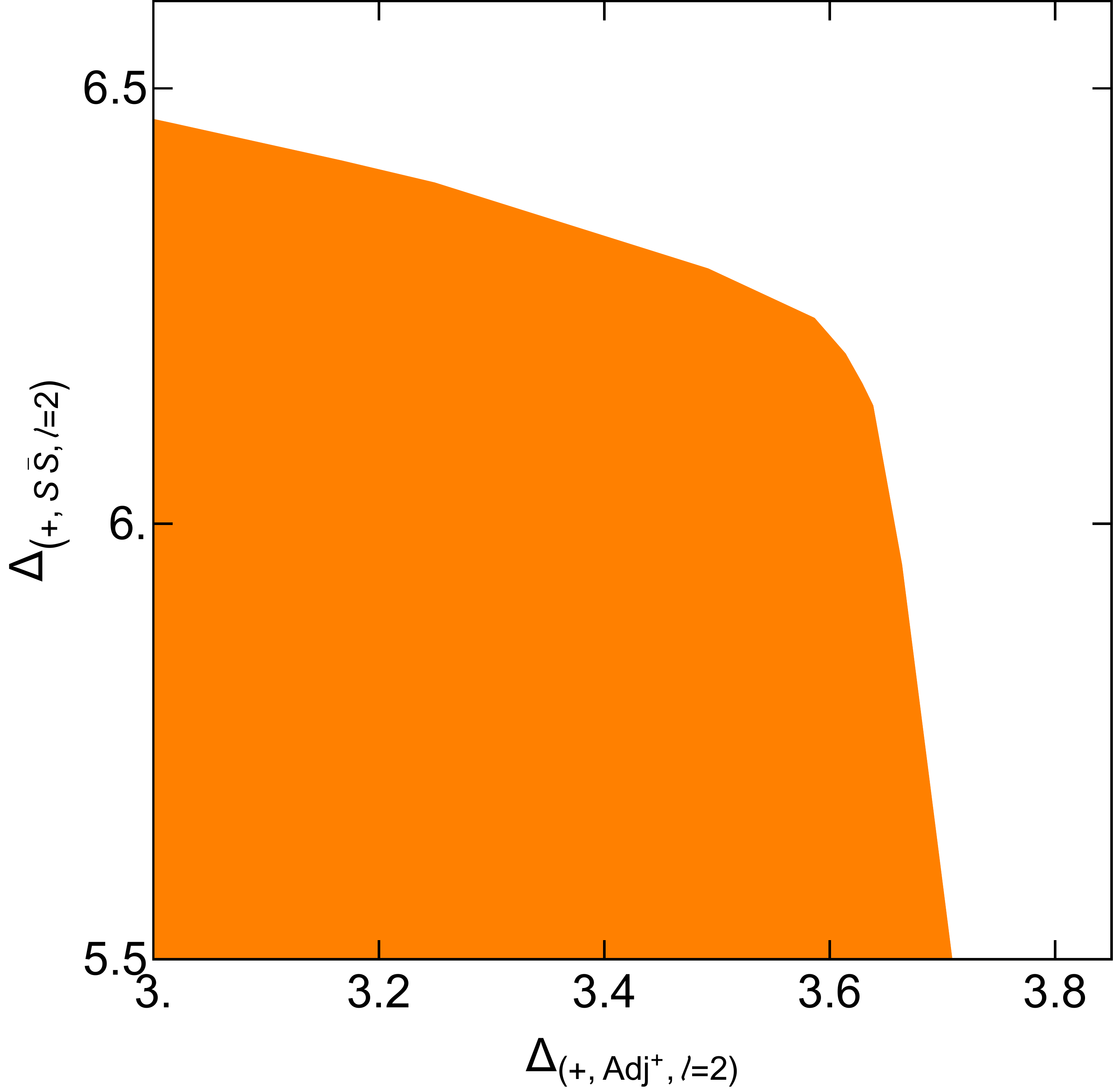}
    \includegraphics[width=0.49\textwidth]{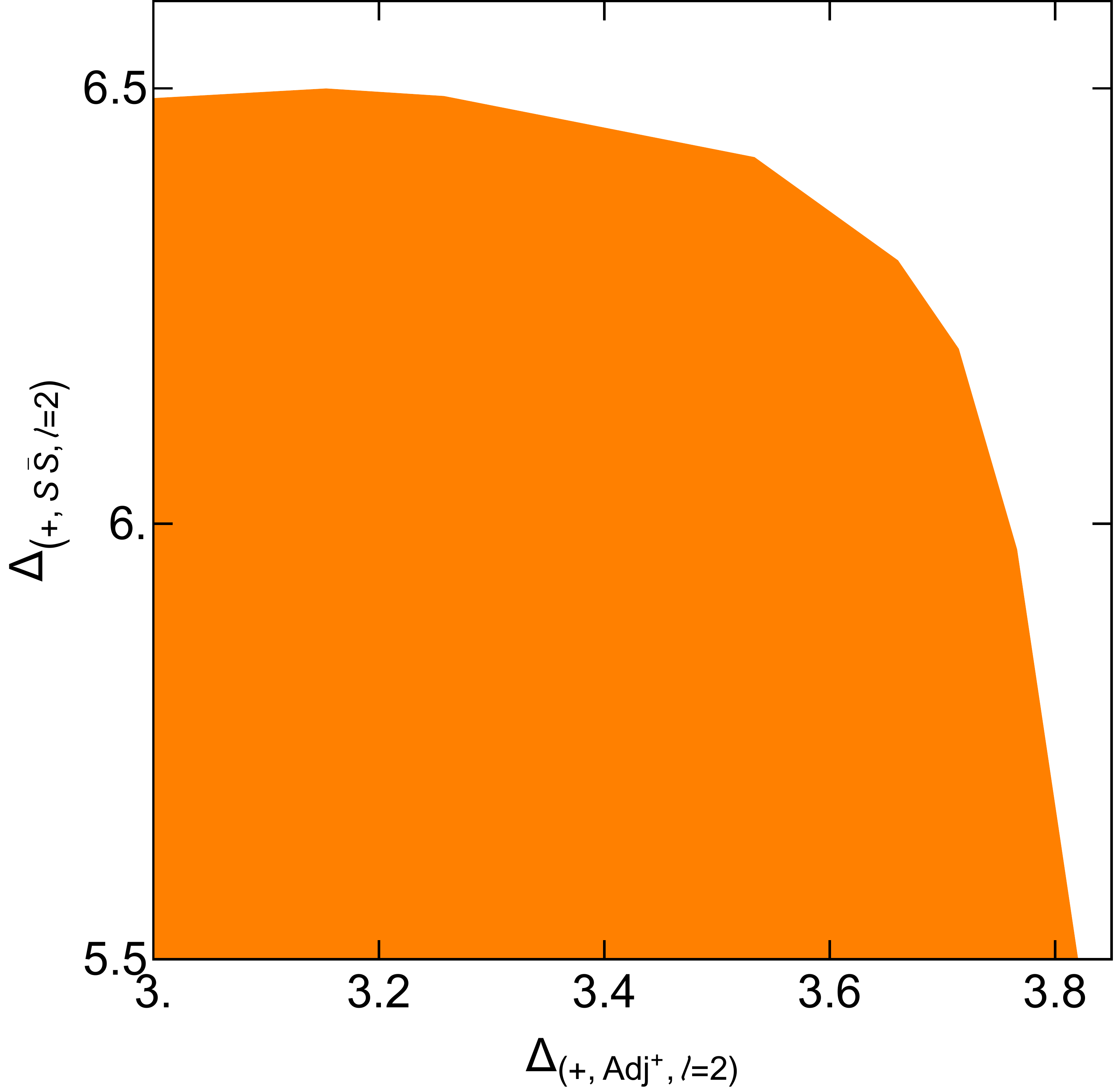}
    \caption{The lowest operator in the $(+, S\bar S$, $\ell=2)$ sector v.s. the sector $(+, Adj^+, \ell=2)$: $SU(4)$ and $SU(3)$ CFT. $\Lambda=19$ and  small gaps in several sectors are imposed: $(+, S, \ell=0)\ge 3$, $(+, Adj^+, \ell=0)\ge 2$, $(+, A\bar{A}, \ell=0)\ge 2$, $(+,S\bar{S},\ell=0)\ge 3$. These gaps are topological, namely the numerical bounds will not change if the gaps are either relaxed or tightened up.  }
    \label{fig:SSbl2_smallN}
\end{figure}

\subsection{Numerical results}

Let us begin with a general bound on spin $\ell=2,3$ operators in the adjoint and $S\bar{S}$ representation at large-$N$, say $N=100$. As shown in Fig.~\ref{fig:SSbl2-nogap} these bounds contain several features, the most peculiar being a prominent spike in the dimensions of the smallest operator in the $(+, S\bar{S}, \ell=2)$ sector. As observed in the previous subsection, a large gap in the $S\bar{S}$ sector is expected in the free fermion theory and in QED$_3$. These bounds imply that once we demand a large gap for the $S\bar S$ spin 2 operator, a slightly broken higher spin current has to be present in the spectrum. When the gap for the $S\bar S$ spin 2 operator is relaxed to around 4, we observe a kink, where higher spin currents are strongly broken. This corresponds to the GFF solution.

\begin{figure}[h]
    \centering
    \includegraphics[width=0.49\textwidth]{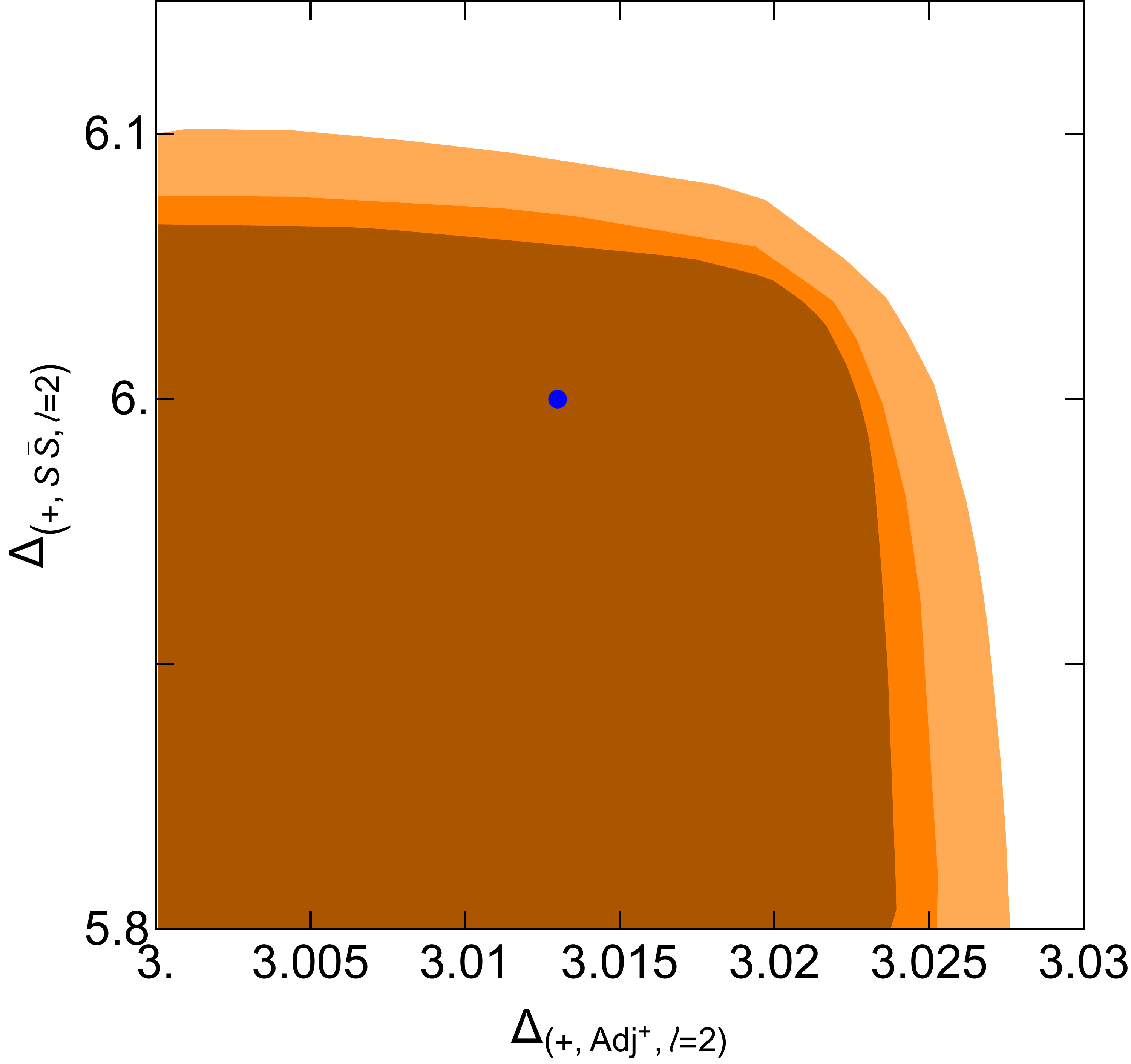}
    \includegraphics[width=0.49\textwidth]{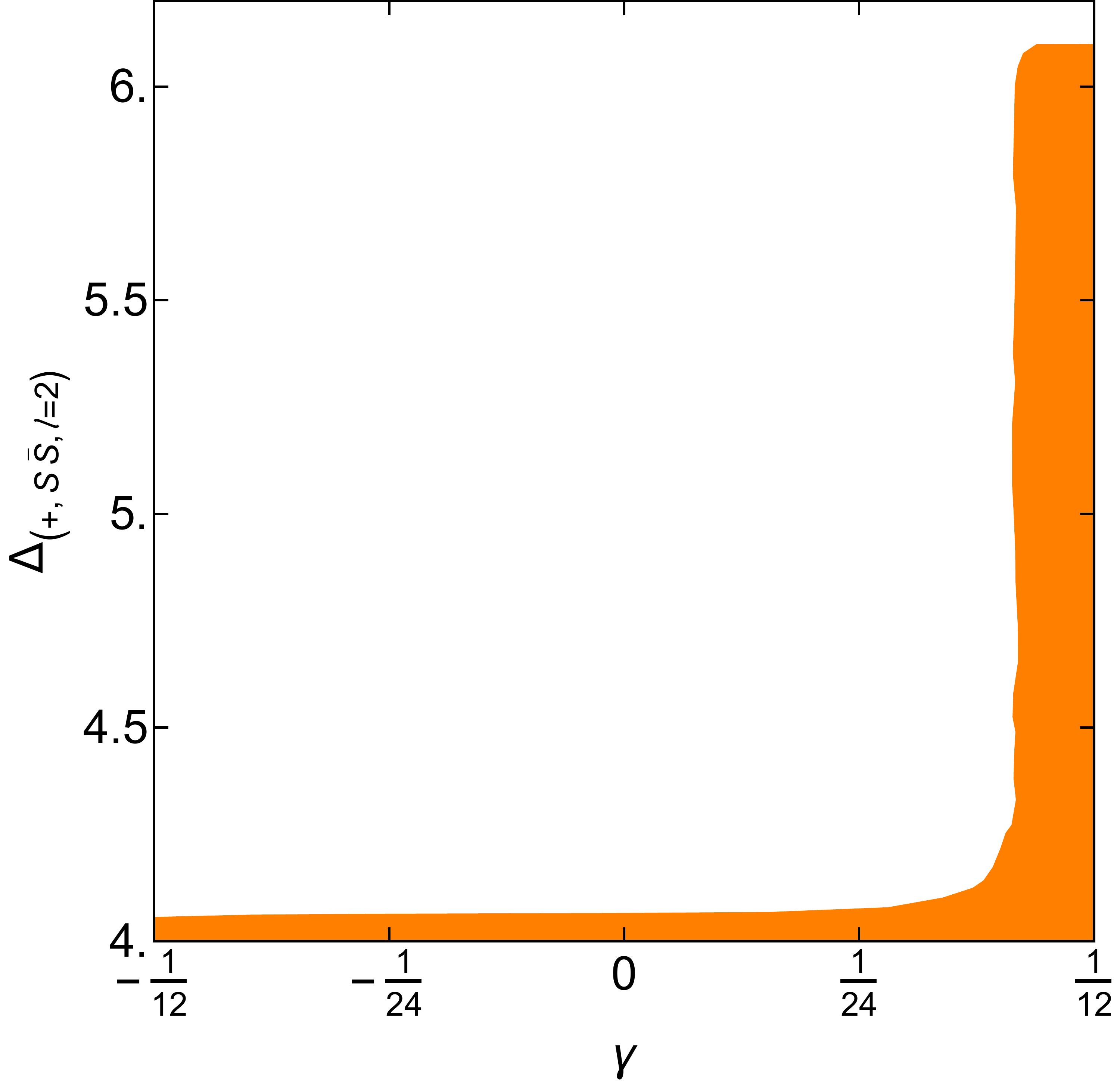}\label{fig:SSbl2_vs_gamma}
    \caption{\textbf{Left panel}: The lowest operator in the $(+, S\bar S$, $\ell=2)$ sector v.s. the sector $(+, Adj^+, \ell=2)$: $SU(100)$ CFT. $\Lambda=19,23,27$ and mild gaps in several sectors are imposed:  $(+, S, \ell=0)\ge 3$, $(+, Adj, \ell=0)\ge 2$, $(+, A\bar{A}, \ell=0)\ge 2$, $(+, S\bar{S}, \ell=0)\ge 3$. \textbf{Right panel}: $\gamma$ versus the lowest operator in the $(+, S\bar S$, $\ell=2)$ sector: $SU(100)$ CFT. $\Lambda=19$ and mild gaps in several sectors are imposed: $(+, S, \ell=0)\ge 3$, $(+, S, \ell=2)_{2nd}\ge 3.5$, $(+, Adj, \ell=0)\ge 2$, $(+, A\bar{A}, \ell=0)\ge 2$, $(+, S\bar{S}, \ell=0)\ge 3$.
    These gaps are topological, namely the numerical bounds will not change if the gaps are either relaxed or tightened up. }
    \label{fig:SSbl2}
\end{figure}

In Fig.~\ref{fig:SSbl2} we compare the position of the kink with the prediction for the anomalous dimension of the slightly broken higher spin current of \cite{Zhou2022HigherSpin} in presence of additional assumptions and we observe a good agreement. The sharp spikes evolve to smoother kinks for smaller values of $N$, as shown in Fig.~\ref{fig:SSbl2_smallN}, but in this case the large-$N$ prediction cannot be trusted.

The fact that a large gap in the $S\bar S$ sector corresponds to a theory resembling the free fermion is also suggested by the bound in Fig.~\ref{fig:SSbl2_vs_gamma}. There we bound the dimension of the lowest operator in the $(+, S\bar S$, $\ell=2)$ sector v.s. the parameter $\gamma$ defined in \eqref{eq:gamma}. We recall that the ANEC imposes the constraints $|\gamma|\leq 1/12$, where the two extremes are saturated by free theories.  For free fermion, it is known $\gamma=1/12$ \cite{Osborn:1993cr}. We observed a sharp pike around $\gamma=1/12$. It’s likely that QED$_3$ sits on the top of the pike with $\gamma<1/12$. In this bound we also add a gap of 3 in the scalar singlet, in principle to avoid a contamination from the GNY model.\footnote{However, at $\Lambda=19$ we are not sure whether this gap is fully effective. If we relax this gap to around 2, the bound does not change much.}

\begin{figure}[h]
    \centering
    \includegraphics[width=0.49\textwidth]{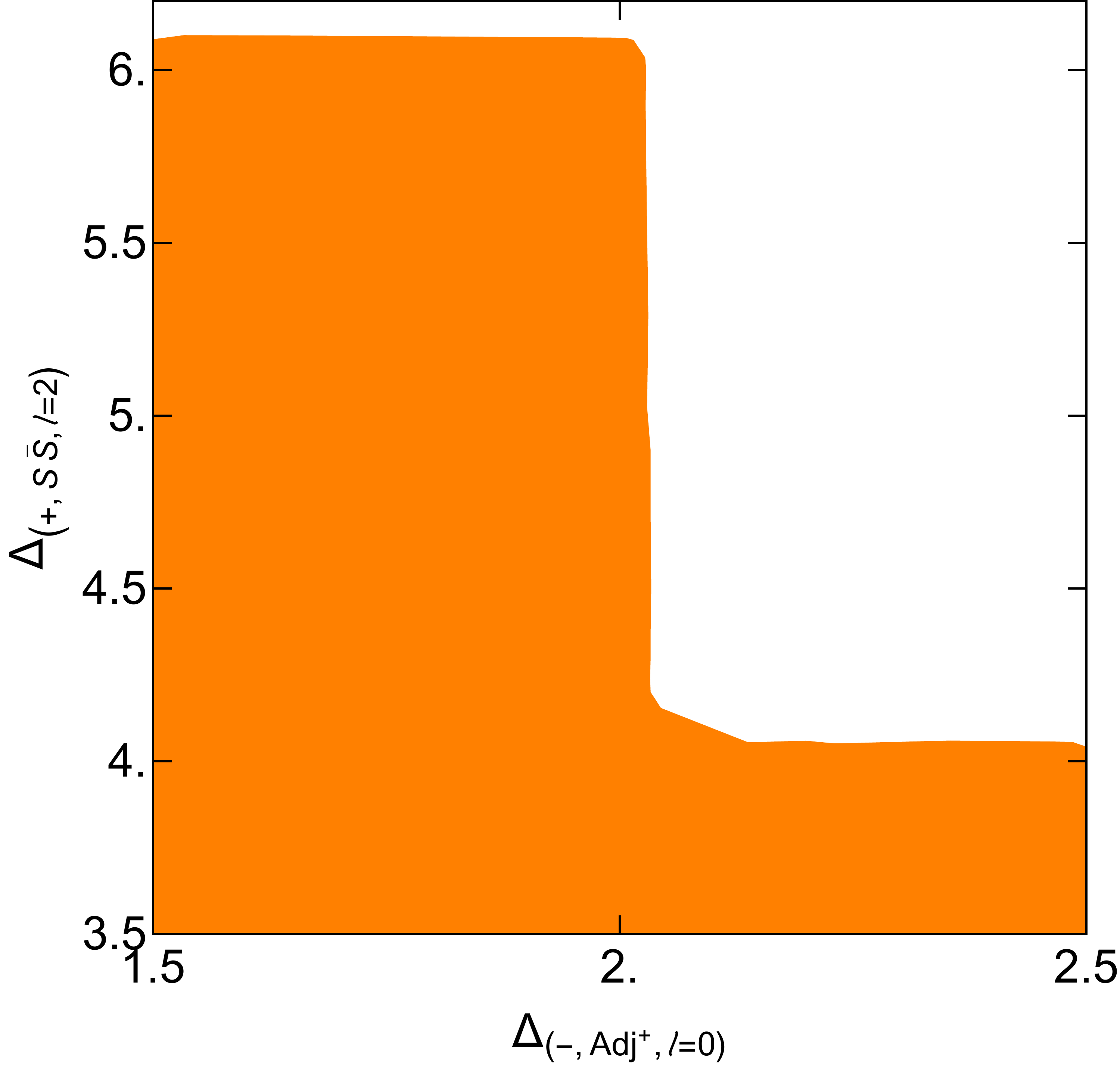}
    \caption{The lowest operator in the $(+, S\bar S$, $\ell=2)$ sector v.s. the sector $(-, Adj^+, \ell=0)$: $SU(100)$ CFT. $\Lambda=19$ and mild gaps in several sectors are imposed:  $(+, S, \ell=0)\ge 3$, $(+, Adj, \ell=0)\ge 2$, $(+, A\bar{A}, \ell=0)\ge 2$, $(+, S\bar{S}, \ell=0)\ge 3$.  .}
    \label{fig:adjP}
\end{figure}

We conclude this subsection by observing that the present setup, if combined with the correlation functions of scalars in the adjoint representation (namely the parity odd fermion bilinear $\bar\psi_i \psi_j$) has the potential to create an island with minimal assumptions. Indeed, the request $\Delta_{(+,S\bar S,\ell=2)}\sim 6$ in the single correlator bootstrap of adjoint scalars forces $\Delta_{\bar\psi\psi} \gtrsim 2$ (assuming the bounds has the usual GFF behavior); on the other hand our setup requires $\Delta_{\bar\psi\psi}\equiv \Delta_{(-,Adj,\ell=0)}\lesssim 2$, see Fig.~\ref{fig:adjP}. This tension might result in a small allowed region.

\begin{table}[ht]
    \centering
    \caption{Approximate dimension (up to $1/N$ corrections) of operators in QED$_3$ and free fermion theory.
  The highlighted sectors distinguish the two theories  In particular, one can add proper gaps in the red (blue) sectors to exclude the free fermion theory (QED$_3$). We also write down the schematic form of operators. 
    \label{tab:QEDvsGFF}
    }
    \begin{tabular}{c|c|c|c} \hline\hline
        Operator sector & Dimension & QED$_3$ & Free fermion  \\ \hline $(+,S, \ell=0)$
        & $\Delta=4$ & $(\bar\psi \psi)^2$, $F_{\mu\nu}^2$ & $(\bar\psi \psi)^2$, $(\bar\psi \gamma_\mu \psi)^2$ \\  \hline 
       
       \multirow{1}{*}{$(-,S, \ell=0)$}  & $\Delta=2$ & $(\bar\psi \psi)$ & $(\bar\psi \psi)$ \\ \hline
       
       \multirow{2}{*}{$(+,S, \ell=2)$}  & $\Delta=3$ & $\bar\psi \gamma_\mu\partial_\nu \psi$ & $\bar\psi \gamma_\mu\partial_\nu \psi$ \\ 
       & $\Delta=4$ & $J^{top}_\mu J^{top}_\nu$ & $(\bar\psi \gamma_\mu \psi)( \bar\psi \gamma_\nu \psi)$ \\ \hline 
       
       \multirow{1}{*}{$(-,S, \ell=2)$} & $\Delta\leq4$ & None & None  \\ \hline
       
       \multirow{2}{*}{\color{red} $(+,Adj, \ell=0)$} & $\Delta=4$ & $(\bar\psi \psi) (\bar \psi^i \psi_j)$ & $(\bar\psi \psi) (\bar \psi^i \psi_j)$, $(\bar\psi \gamma_\mu \psi) (\bar \psi^i \gamma_\mu \psi_j)$ \\ 
      & $\Delta=5$ & $(\partial_\nu F_{\mu\nu}) (\bar\psi^i \gamma_\mu \psi_j)$ & None \\ \hline
      
      \multirow{2}{*}{\color{blue}$(-,Adj, \ell=0)$} & $\Delta=2$ & $\bar\psi^i \psi_j$ & $\bar\psi^i \psi_j$ \\
      & $\Delta=4$ & $(\bar\psi^i \gamma_\mu \psi_j) J_\mu^{top}$ & None  \\ \hline
      
      \multirow{2}{*}{\color{blue}$(+,Adj, \ell=1)$} & $\Delta=2$ & $\bar\psi^i \gamma_\mu \psi_j$ &  $\bar\psi^i \gamma_\mu \psi_j$ \\
      & $\Delta=4$ & $(\bar\psi^i \psi_j)  J_\mu^{top}$, $(\bar\psi^i \gamma_\mu \psi_j) F_{\mu\nu}$ & None \\ \hline
      
      \multirow{1}{*}{$(-,Adj, \ell=1)$} & $\Delta\leq 3$ & None & None \\ \hline 
      
      \multirow{2}{*}{\color{red}$(+,Adj, \ell=2)$} & $\Delta=3$ & $\bar\psi^i \gamma_\mu\partial_\nu \psi_j$ & $\bar\psi^i \gamma_\mu\partial_\nu \psi_j$ \\ 
      & $\Delta=4$ & None & $(\bar\psi \gamma_\mu \psi) (\bar \psi^i \gamma_\nu \psi_j)$ \\ 
\hline 
      
       \multirow{1}{*}{\color{blue}$(-,Adj, \ell=2)$} & $\Delta=4$ & $(\bar\psi^i \gamma_\mu \psi_j) J_\nu^{top}$ & None \\
  \hline 
        \multirow{2}{*}{$(+, A\bar A, \ell =0)$} & $\Delta=4$ & $\bar\psi^2 \psi^2$ & $\bar\psi^2 \psi^2$ \\ 
      & $\Delta=6$ & $\bar\psi^2 \partial^2 \psi^2$ &  $\bar\psi^2 \partial^2 \psi^2$ \\ 
  \hline \hline
       
     \end{tabular}
\end{table}

\subsection{An island for QED\texorpdfstring{$_3$}{}}

In this section we show how it is possible to create an island that is consistent with QED$_3$ but excludes the free fermion theory. In order to do this we need to impose aggressive assumptions on the spectrum. Let us justify them.

In the large-$N$ limit, the spectrum of QED$_3$ and free fermions are quite similar, with very few distinct features. Both theories have such $SU(N)$ flavor current. However, QED$_3$ also has a parity odd topological U(1) current

\begin{equation}
    J_\text{top}^\mu \sim \varepsilon^{\mu\nu\rho}F_{\nu\rho}\,,
\end{equation}
 while the free fermion theory has a parity even U(1) current $\bar\psi\gamma^\mu\psi$. In the QED$_3$ theory one can still write the operator $\bar\psi\gamma^\mu\psi$: it is, however, a descendent of the topological current due to the equation of motion,
\begin{equation}
    e^2 \bar\psi\gamma^\nu\psi \sim \partial_\mu F^{\mu \nu}\,.
\end{equation}
Thus, when classifying operators of QED$_3$ in the large-$N$ limit, one should use the above identification.

As shown in Table \ref{tab:QEDvsGFF}, there are various gap features to distinguish QED$_3$ from free fermion theory. We tried to explore whether free fermion theory can be excluded by exploiting those differences. The simplest setup corresponds to assume the existence of an isolated operator in the $(+, Adj, \ell=2)$ sector, with dimension $\Delta_{(+, Adj, \ell=2)}$, input a gap $\Delta\gtrsim 5$ for the next operator and then scan the value of $\Delta_{(+, Adj, \ell=2)}$ together with other quantities. Unfortunately this setup is not sufficient to exclude the free fermion solution, at least not at $\Lambda\leq 23$.

\begin{figure}[h]
    \centering
    \includegraphics[width=0.49\textwidth]{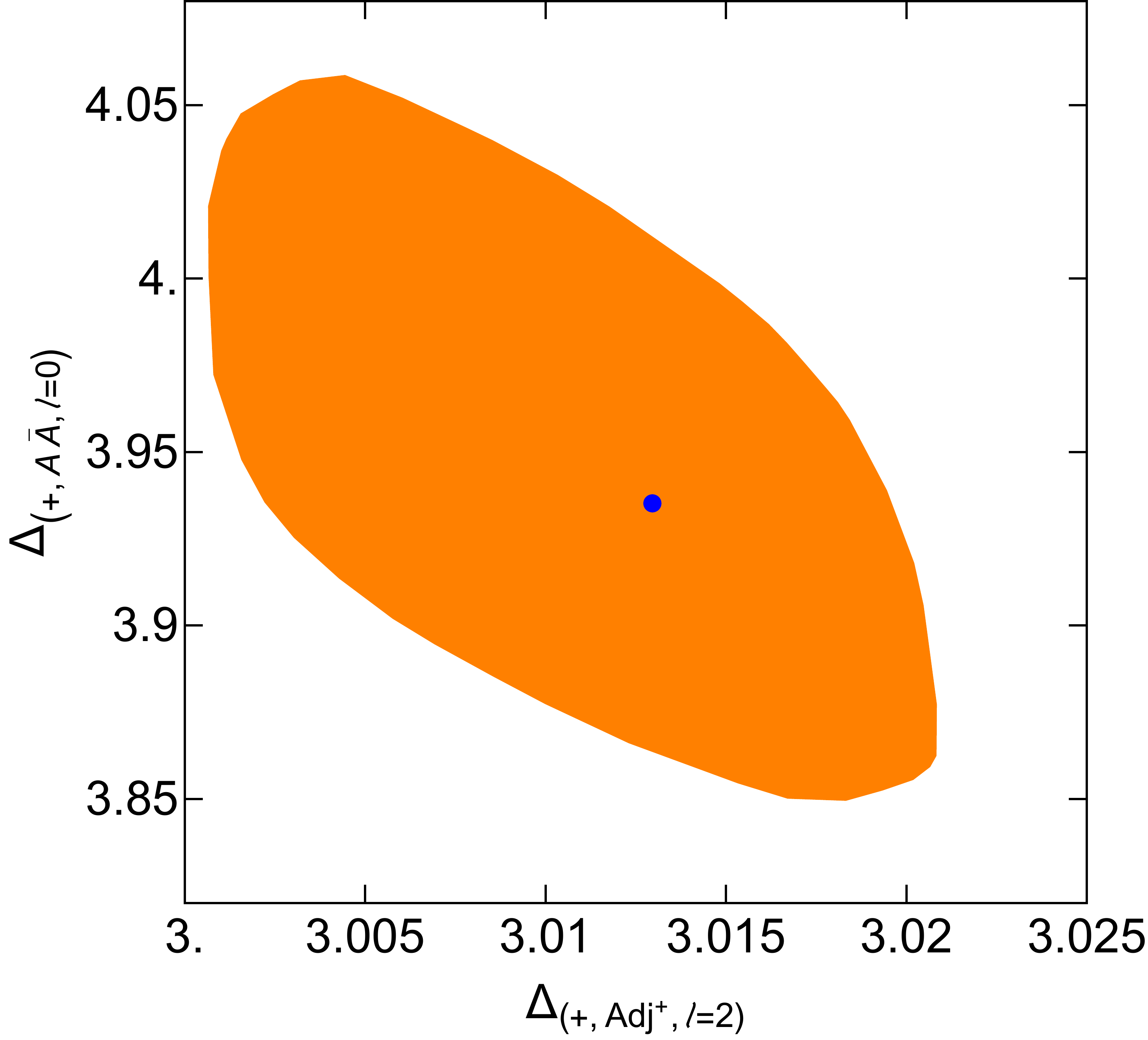}
    \caption{ An island of $SU(100)$ CFT in the  ($\Delta_{(+, A\bar{A}, \ell=0)}$, $\Delta_{(+, Adj, \ell=0)})$ space, which excludes the free fermion theory and is consistent with QED$_3$. The blue dot corresponds to the large-$N$ value~\cite{Zhou2022HigherSpin,Chester:2016ref} of $N=100$ QED$_3$. $\Lambda=23$ and the specific condition for the setup is in Appendix \ref{app:param}.}
    \label{fig:intervalpositivity}
\end{figure}

In order to restrict the space of allowed solution we impose a form of a spectrum close to the expected spectrum at large-$N$. The idea is to demand that operators should exist only within certain intervals centered around the large-$N$ prediction for QED$_3$. We take $N$ to be large and the size of the interval to be $10/N$. This assumption, together with the gap discussed in the previous paragraph turns out to be sufficient to exclude the free fermion solution, leaving an island in the plane  $(\Delta_{(+, A\bar A,\ell=0)},\Delta_{(+,Adj,\ell=2)})$ consistent only with QED$_3$. 

 The result at $\Lambda=23, N=100$ is shown in Figure \ref{fig:intervalpositivity}. If we relax the derivative order to $\Lambda=19$, the free fermion theory won't be excluded. It would be very interesting to see how the island converges at very large $\Lambda$. We leave it for the future.

\section{\texorpdfstring{$O(N)$}{} currents}
\label{sec:ON}

Let us now move to the case of $O(N)$ global symmetry currents.

\begin{table}[ht]
\setlength{\tabcolsep}{0.2cm}
\renewcommand{\arraystretch}{1.4}
    \centering
    \caption{Expected spectrum for theories with $O(N)$ symmetry up to 1/N correction. The first column indicates the parity, irrep and spin of the operator. \label{tab:ON} 
    }
    \begin{tabular}{c|ccccccc} \hline\hline
        Operator & Free boson & WF & free fermion  & GNY & GVFF & bQCD & fQCD  \\ \hline  
        $(+,S,0)$ & 1  & 2 & 4 & 2 & 4 & 2 & 4 \\
        $(+,T,0)$ & 1 & 1  & 4 & 3 & 4 & 1 & 4 \\ 
        $(+, Q1, 0)$ & 8
        & 8 & 7 & 7 & 4 & 4
        & 7\footnote{There could be differences between $O(N_c=3)$ QCD and $O(N_c>4)$ QCD.} \\
        $(-,Q1,0)$ & 5 
        & 5 & 5 & 5 & 5 & 5 & 5  \\
        $(+, Q1, 2)$ & 6
        & 6 & 4 & 4 & 4 & 4
        & 4 \\ 
        $(+, Q2, 0)$ & 4 & 4 & 4 & 4 & 4 & 4 & 4 \\
        $(+, Q2, 2)$ & 4 & 4 & 6 & 6 & 4 & 4 & 4 \\
        \hline \hline 
     \end{tabular}
\end{table}

\subsection{OPE and targets}

The tensor product of $O(N)$ adjoint $A$, i.e. rank-2 antisymmetric tensor, is
\begin{equation}
A \times A = S^+ + T^+ + Q1^+ + Q2^+ + A^- + Q3^-.
\end{equation}
$S$ refers to singlet, $T$ refers to rank-2 symmetric traceless tensor, $Q1, Q2, Q3$ are rank-4 tensors, and the superscript $\pm $ refers to symmetric or antisymmetric tensor product. 
The dimensions of the three rank-4 tensors are 
\begin{align}
\textrm{dim}[Q1] &= \frac{1}{24} (N-3) (N-2) (N-1) N, \\ \textrm{dim}[Q2]&=\frac{1}{12} (N-3) N (N+1 ) (N+2), \\ \textrm{dim}[Q3]&=\frac{1}{8} (N-3) (N-1) N (N+2).
\end{align}
It is worth noting that $Q1$ is a fully antisymmetric rank-4 tensor, so for $N=4$ it becomes the pseudo-singlet, namely it is $SO(4)$ singlet but is odd under the improper $Z_2$ of $O(4)$.

Theories with the $O(N)$ conserved current include free boson, free fermion, $O(N)$ Wilson-Fisher, $O(N)$ Gross-Neveu-Yukawa theory, generalized vector free field theory, and $SO(N_c)$ (or $O(N_c)$) gauge theories.~\footnote{Indeed theories without a gauge field can be thought of as a special type of gauge theory that has $N_c=1$.
On the other hand, $N_c=2$ corresponds to a $U(1)$ gauge theory, which has an enhanced flavor symmetry $O(N)\rightarrow SU(N)$.}
\begin{figure}[h]
    \centering
    \includegraphics[width=0.49\textwidth]{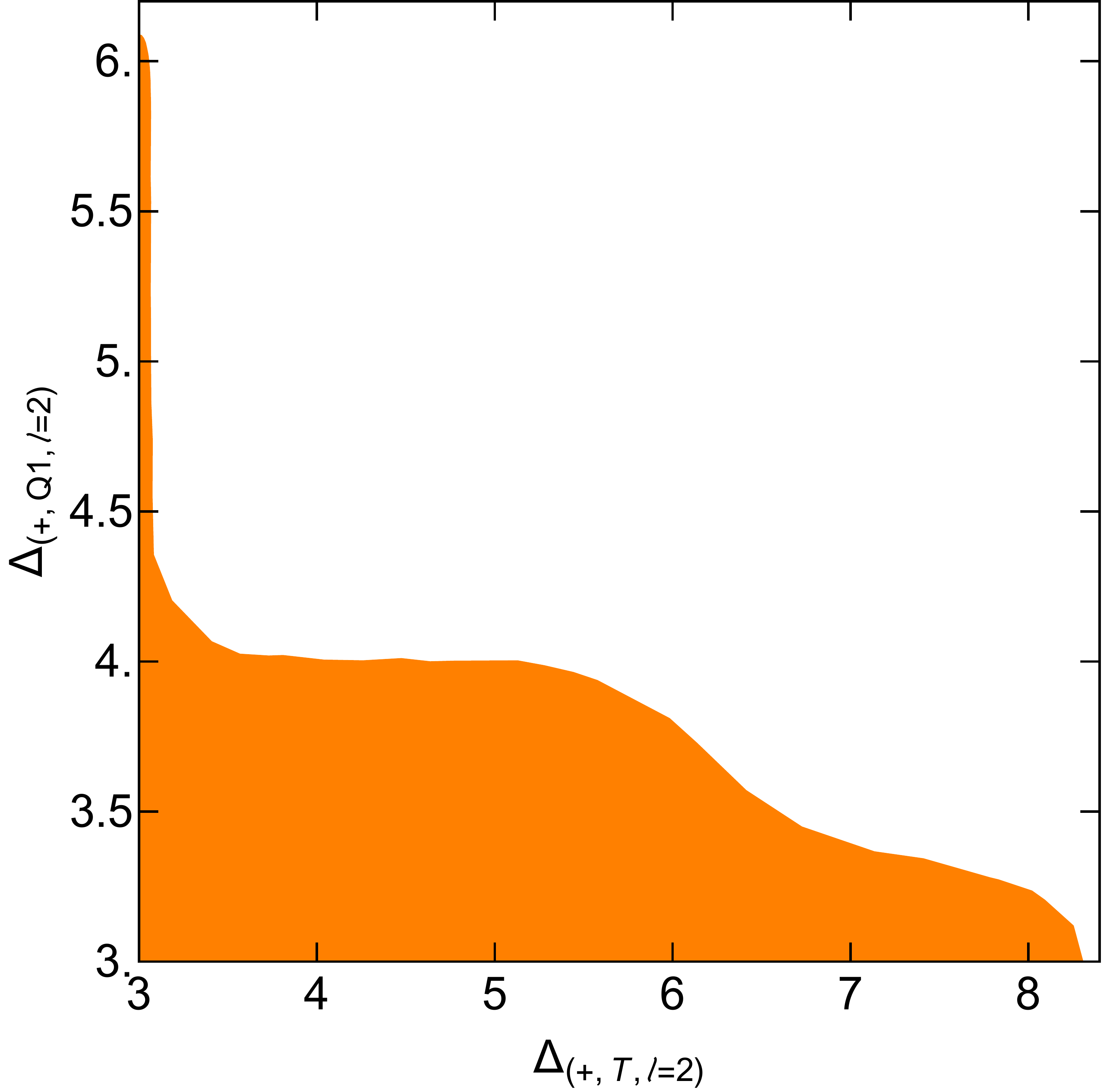}
    \includegraphics[width=0.49\textwidth]{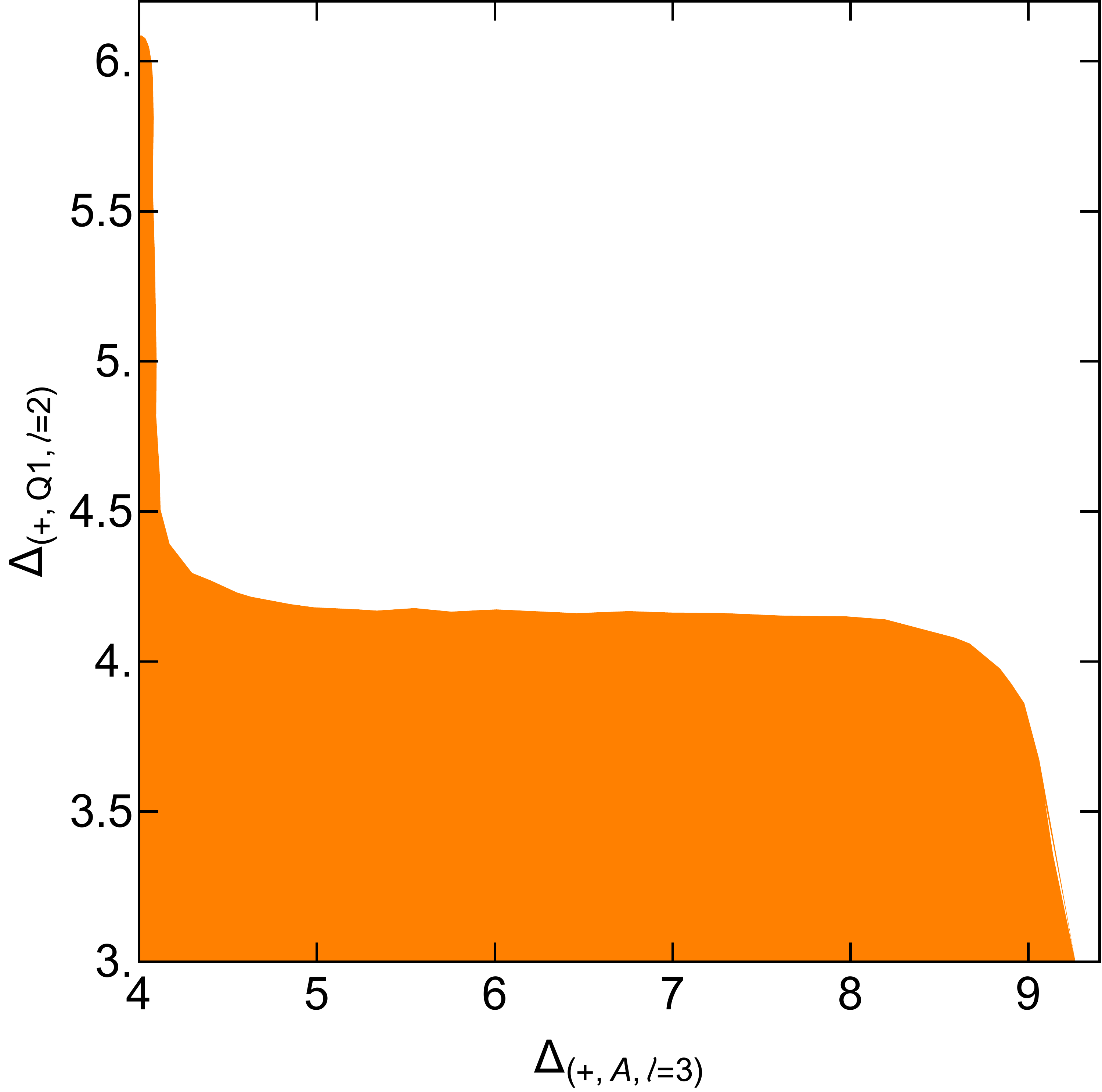}
    \caption{The lowest operator in the $(+, Q1, \ell=2)$ sector v.s. the sector $(+, T, \ell=2)$ and $(+, A, \ell=3)$: $O(100)$ CFT. $\Lambda=19$ and no gap is imposed.}
    \label{fig:Q1}
\end{figure}

\begin{figure}[h]
    \centering
    \includegraphics[width=0.49\textwidth]{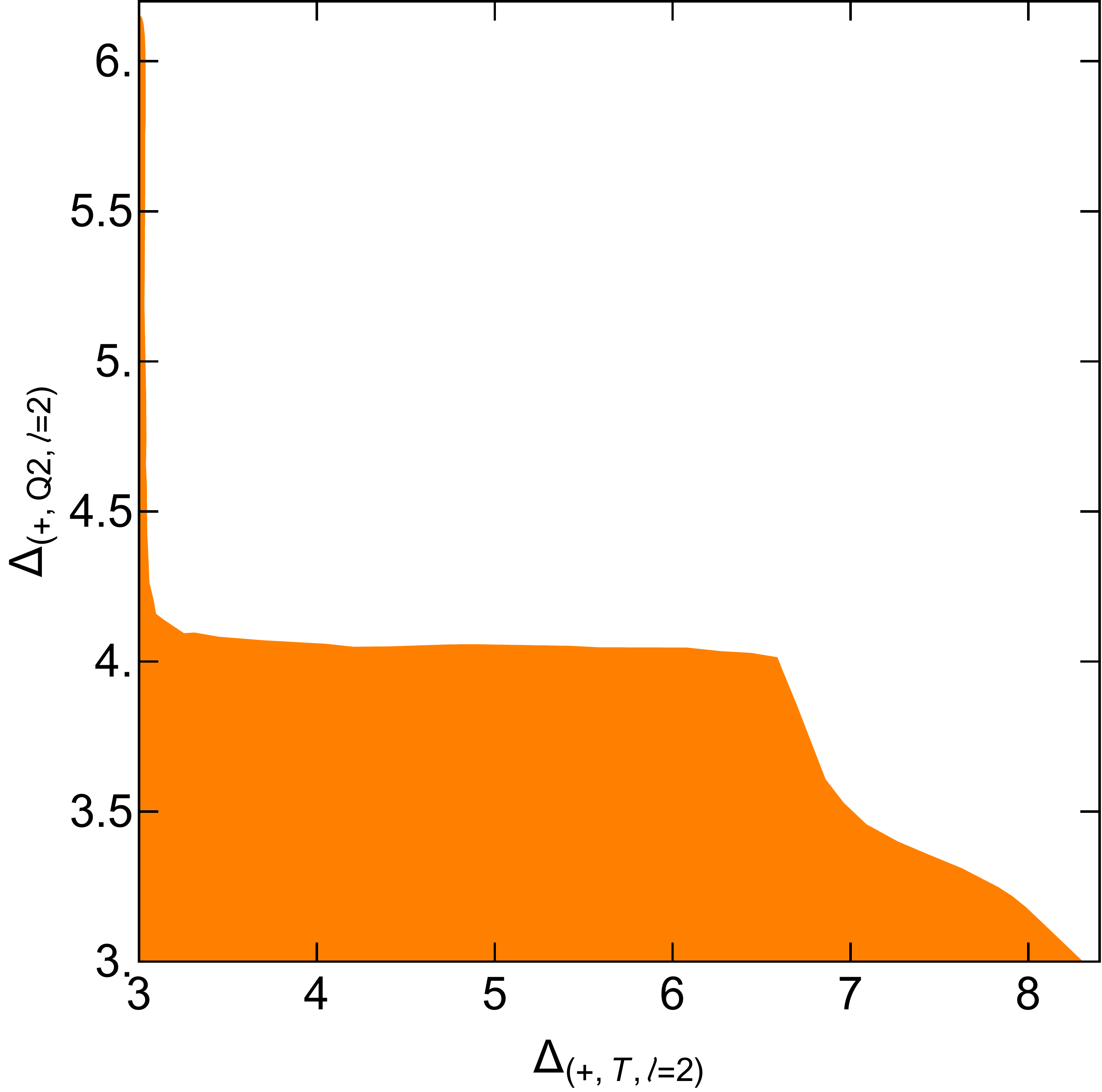}
    \includegraphics[width=0.49\textwidth]{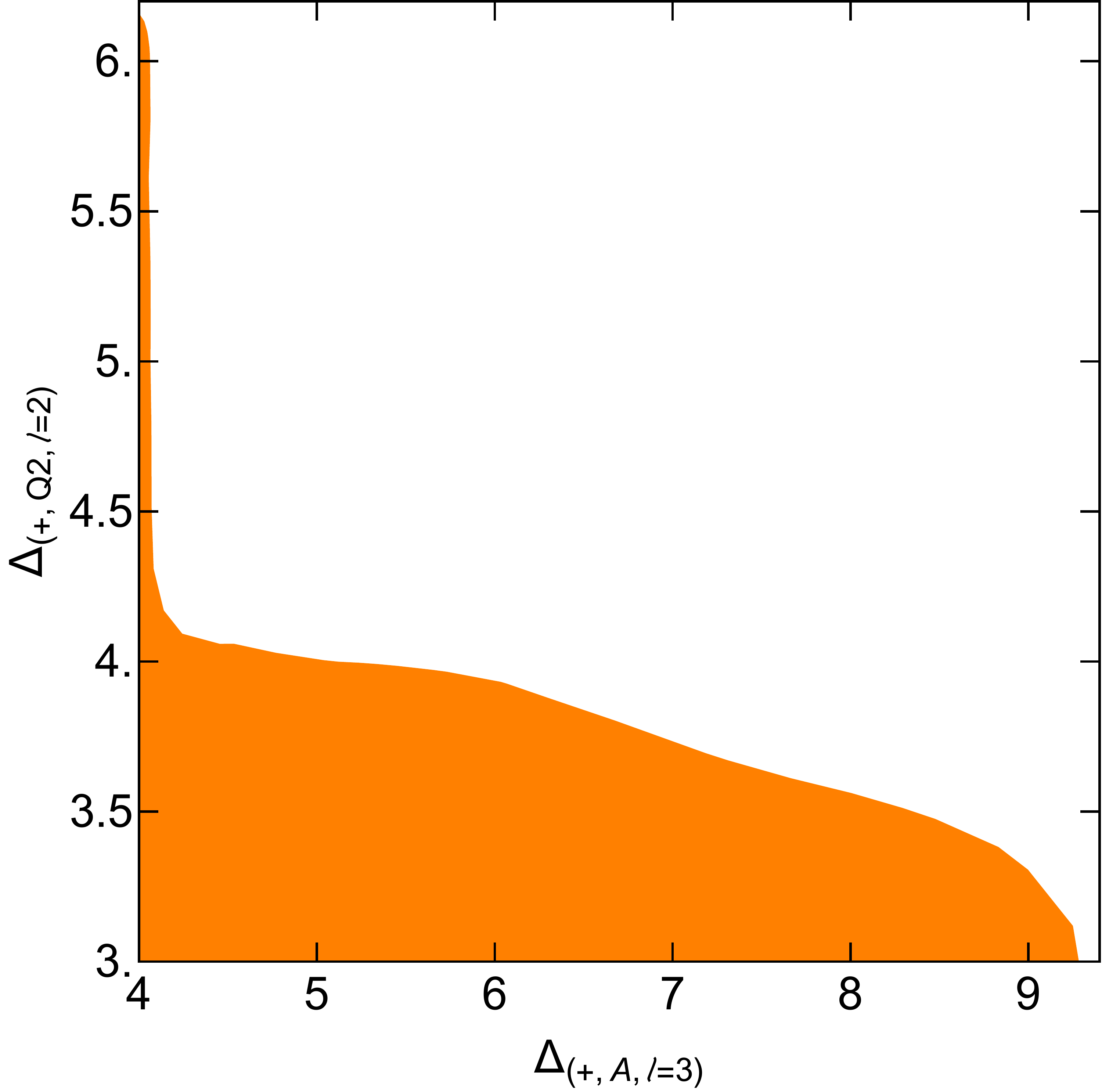}
    \caption{The lowest operator in the $(+, Q2, \ell=2)$ sector v.s. the sector $(+, T, \ell=2)$ and $(+, A, \ell=3)$: $O(100)$ CFT. $\Lambda=19$ and no gap is imposed.}
    \label{fig:Q2}
\end{figure}

\begin{figure}[h]
    \centering
    \includegraphics[width=0.49\textwidth]{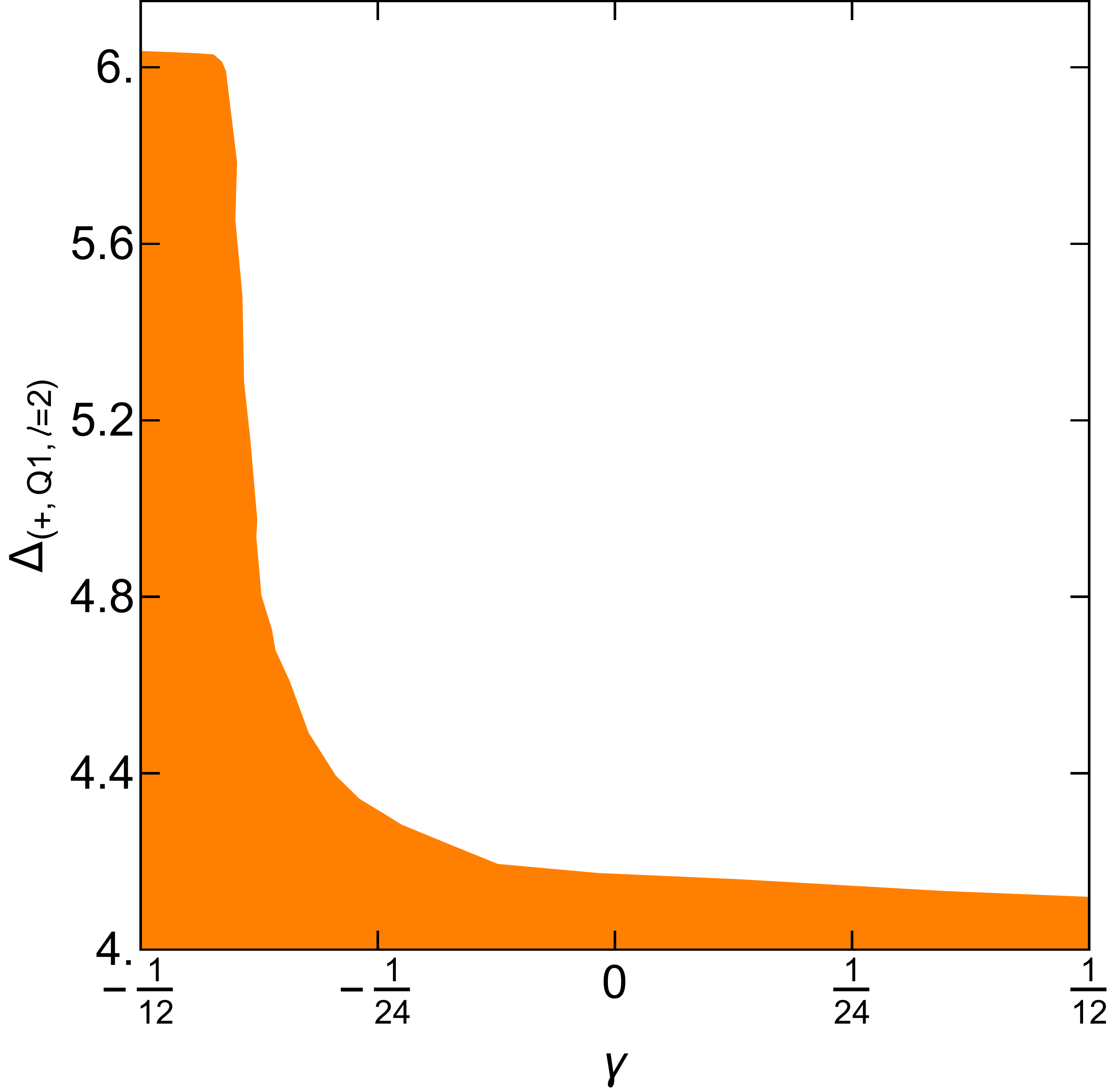}
    \includegraphics[width=0.49\textwidth]{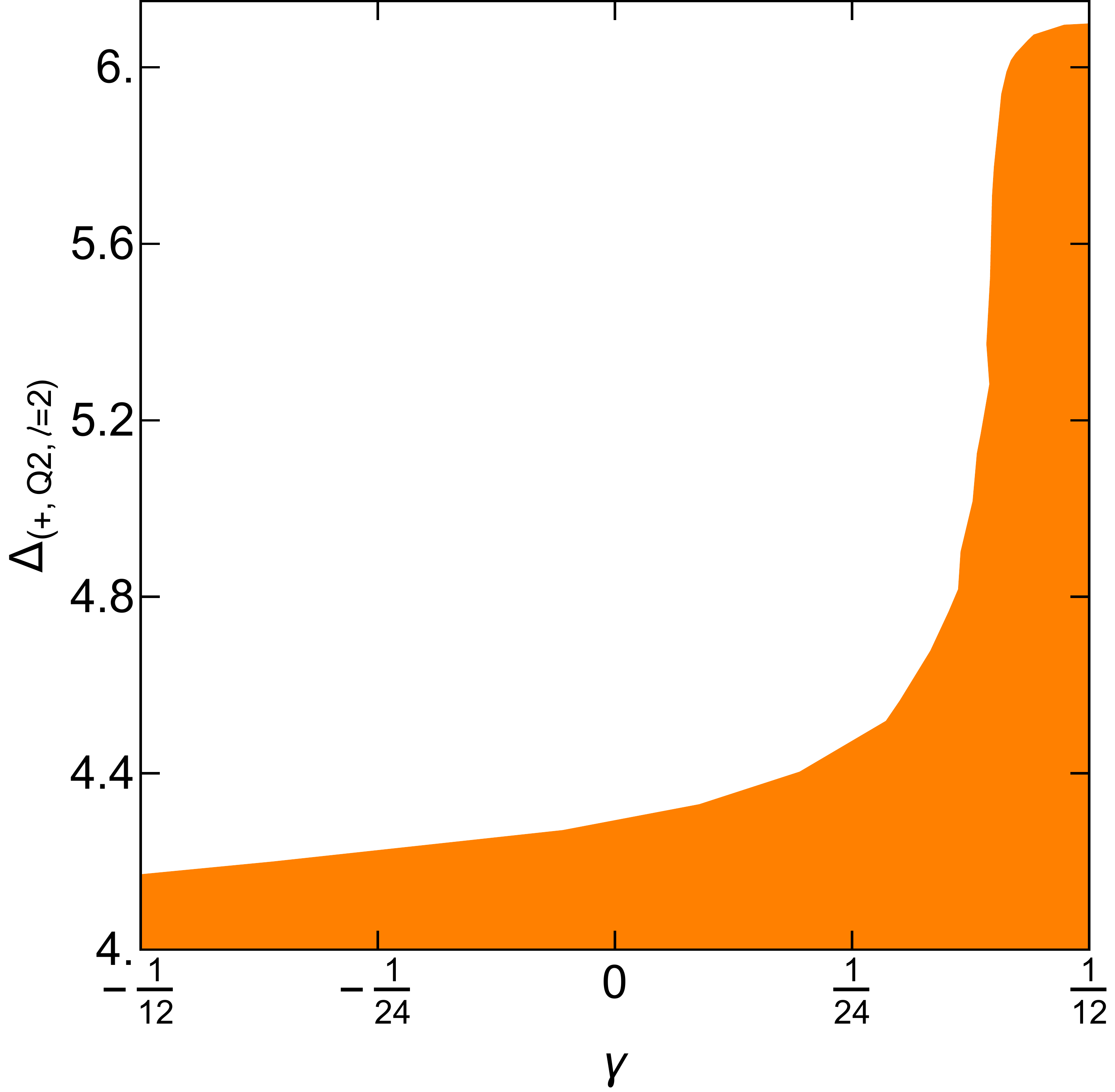}
    \caption{$\gamma$ v.s. the lowest operator in the $(+, Q1, \ell=2)$ or $(+, Q2, \ell=2)$ sector: $O(100)$ CFT. $\Lambda=19$ and mild gaps are imposed: $(+, S, \ell=0)\ge 1$, $(+, S, \ell=2)_{2nd}\ge 3.5$, $(+, Q_1, \ell=0)\ge 3$, $(+, Q_2, \ell=0)\ge 3$. These gaps are topological, namely the numerical bounds will not change if the gaps are either relaxed or tightened up. }
    \label{fig:gamma}
\end{figure}

\begin{figure}[h]
    \centering
    \includegraphics[width=0.49\textwidth]{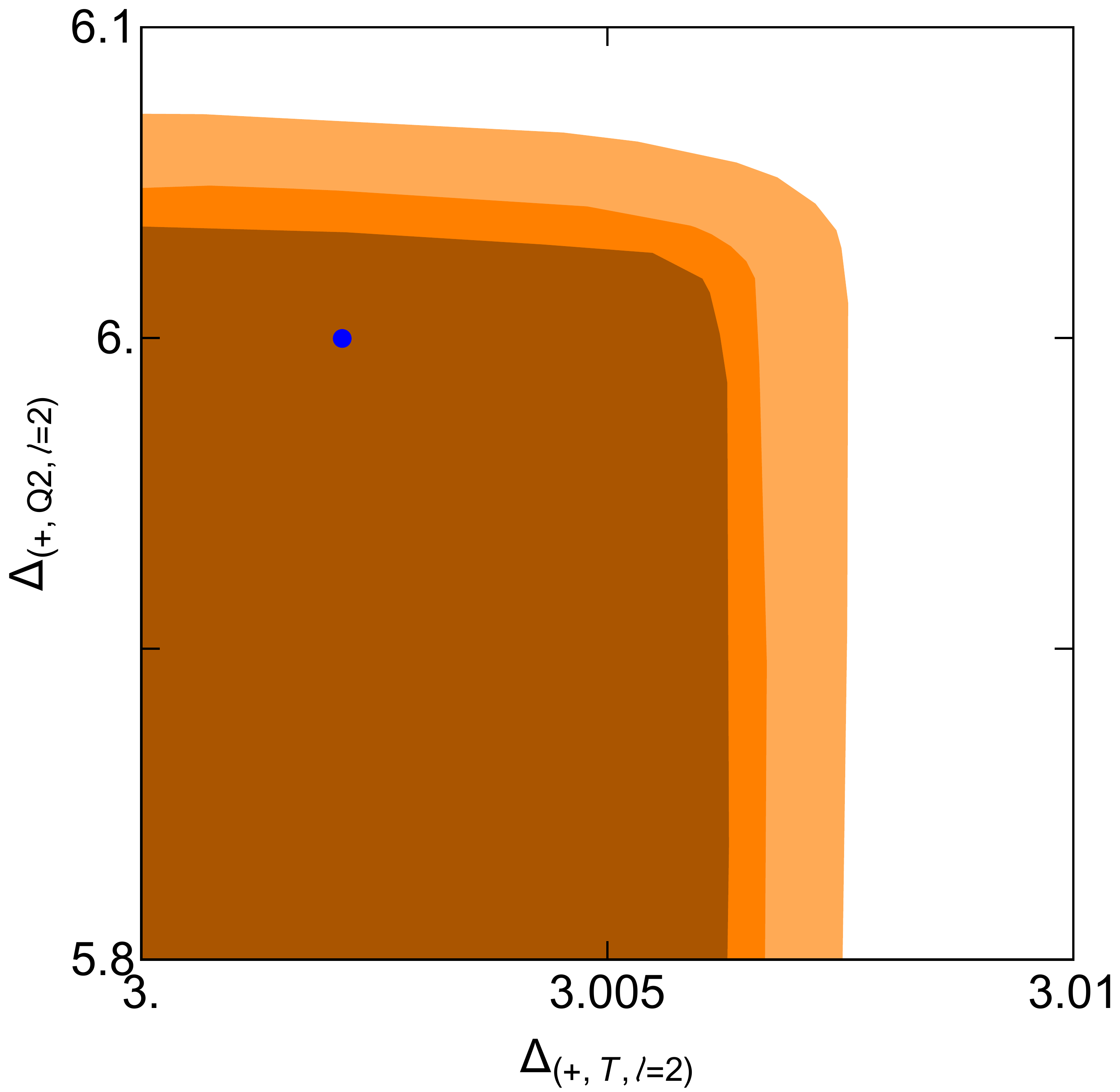}
    
    \caption{Haunting $O(100)$ GNY CFT: $\Delta_{(+,T,\ell=2)}$ v.s. $\Delta_{(+, Q2, \ell=2)}$. $\Lambda=19,23,27$ and  strong gaps in several sectors are imposed: $(+, S, \ell=0)\ge 1.5$, $(+, S, \ell=2)_{2nd}\ge 3.5$, $(+, Q1, \ell=0)\ge 6.5$, $(+, Q1, \ell=2)\ge 3.5$, $(+, T, \ell=0)\ge 3.5$, $(+, Q2, \ell=0)\ge 3.5$, $(+, A, \ell=1)_{2nd}\ge 2.5$.}
    \label{fig:Q2stronggap}
\end{figure}

Table~\ref{tab:ON} lists several low lying operators that appear in the $J\times J$ OPE. It is worth noting that the lowest operators in certain sectors have dimension that vary substantially depending on the theory. For instance, the sector $(+, Q1, 2)$ distinguishes free boson and Wilson-Fisher from all other theories such as fermionic theories, GVFF and QCD$_3$.
Similarly, the leading operator in the  $(+, Q2, 2)$ sector distinguishes free fermion and GNY from all other theories such as bosonic theories, GVFF and QCD$_3$. 

\subsection{Numerical results}

We begin by exploring the bounds on the lowest operators in sectors $(+, Q1, 2)$ and $(+, Q2, 2)$ discussed in the previous section. The peaks shown in Fig. \ref{fig:Q1} and Fig.~\ref{fig:Q2} correspond precisely to the large gaps expected in these sectors. In Fig. \ref{fig:gamma} and \ref{fig:Q2stronggap} further study these peaks by considering the upper bound as a function of the parameter $\gamma$ defined in \eqref{eq:gamma}, imposing additional gaps in various sectors: this verifies that the peaks correspond to either bosonic theories ($\gamma\sim -1/12$) or fermionic theories ($\gamma\sim 1/12$). 

Therefore, as in the case of $SU(N)$, imposing a large gap in the sectors $(+, Q1/Q2, 2)$ localizes the value of gamma toward the ends the allowed interval and it imposes the presence of slightly broken higher-spin currents.

Finally, in Fig.~\ref{fig:Q2stronggap} we compare the allowed region and the prediction from large-$N$ for two operators. We imposed a set of assumptions to remove the contamination from theories except for GNY. We observe that the peak is not far from the large-$N$ prediction~\cite{Manashov2017GNYHigherSpin}, but the convergence in $\Lambda$ is still slow.

\section{Discussion}
In this work, we considered the correlation functions of four conserved spin-1 currents associated with a non-abelian global symmetry. We systematically analyzed the consequences of conservation and permutations for three and four point functions and presented a complete set of linearly independent crossing equations. We applied the machinery of numerical conformal bootstrap, focusing on the case of $SU(N)$ or $O(N)$ symmetry. In both cases, we obtained  bounds on scaling dimensions of certain operators. We presented general bounds, which are valid for any unitary CFT in $3d$, but also considered ad-hoc assumptions, which allowed us to focus on specific cases of interest.  

One of the main motivations for the present work was to study the IR fixed point of QED$_3$. To achieve this, we analyzed the spectrum of various theories with $SU(N)$ symmetry at large-$N$, looking for distinct features that allow us to single out QED$_3$ from other solutions of crossing equations. We identified several such characteristics: a large gap in the parity even, spin-2 sector of the $SU(N)$ representation $S\bar{S}$, and a mild gap between the first and second operator in the parity-even, spin-2, adjoint representation.

We presented bounds on the scaling dimension of the first $S\bar S$ spin 2 operator as a functions of various quantities and we observed a sharp pike in the allowed region: it is tempting to identify this peak with the location of QED$_3$. We also observed that by demanding the $S\bar S$ spin 2 operator has dimension around 6, slightly broken higher spin currents must exist in the theory.

In order to completely isolate QED$_3$ we had to impose aggressive assumptions on the spectrum in the form of narrow intervals centered around the large-$N$ predictions. This assumption is somewhat justified at large-$N$ and allowed us to create a closed region in parameter space which excludes the free theory solution. 

In order to go beyond this result, especially for small values of $N$ where anomalous dimensions are expected to be $O(1)$, we will need an effective way to distinguish the QED$_3$ from free fermion. One way is to include in the bootstrap setup correlators of QED$_3$ operators that do not exist in the free fermion. Possible candidates are the topological current and monopole operators. 

Given the encouraging results of this work, it would be very interesting to obtain more large-$N$ predictions for more low lying operators, especially those transforming in more involved irreducible representations, as well as additional parameters, such as $\gamma$ and $\lambda_{JJJ}$.\footnote{In QED$_3$ there will be two distinct values of $\gamma$, one for the flavor currents and one for the topological current.}

\begin{acknowledgments}
We would like to thank Marten Reehorst for participating in the early stages of this project. We also thank Francesco Russo, Francesco Bertucci, Emilio Trevisani for discussions and Stefanos Kousvos for discussions and comments on the manuscript. Research at Perimeter Institute is supported in part by the Government of Canada through the Department of Innovation, Science and Industry Canada and by the Province of Ontario through the Ministry of Colleges and Universities. 
This project has received funding from the European Research Council (ERC) under the European Union’s Horizon 2020 research and innovation programme (grant agreement no. 758903).  The computations in this paper were run on the Symmetry cluster of Perimeter institute.
\end{acknowledgments}

\appendix
\section{Operator spectrum of theories with non-Abelian current}\label{sec:spectrum}

Our strategy is the following. We first calculate the partition function for the theories we care about. 
The partition function is the plethystic exponential of single particle characters, which was used in \cite{Meneses:2018xpu} to calculate the partition function of free theory in four dimensions. 
The SO(3) character for the global symmetry can be written as 
\begin{equation}
    \chi_\ell(x)=\sum_{i=-2\ell}^{2\ell} x^i,
\end{equation}
then the conformal character for a generic (long) multiplet can be written as 
\begin{equation}
    \chi_{\Delta,\ell}(q,x)=\frac{q^{\Delta} \chi_\ell(x)}{(1-q)(1-q/x)(1-q x)}.
\end{equation}
The numerator corresponds to the conformal primary, while the denominator comes from the $\partial_{\mu}$'s acting on the conformal primary.
For free theory scalar saturating the unitarity bound, due to shortening condition, the character becomes
\begin{equation}
    \chi^{short}_{\Delta=1/2,\ell=0}=\chi_{1/2,0}(q,x)-\chi_{5/2,0}(q,x).
\end{equation}
The partition function for free theory can then be written as 
\begin{equation}
    Z(q,x,a)=\text{PE}[\chi^{short}_{\Delta=1/2,\ell=0}(q,x) \cdot\chi_V(a)].
\end{equation}
Here PE stands for ``plethystic exponential''. The character $\chi_V(a)$ is the character for the global symmetry, which indicates that the scalar field is in the vector representation of O(N). For details of the characters of O(N), see Appendix B of \cite{Henriksson:2022rnm}.
Expanding the above partition function in terms consisting of conformal characters times the character for the global symmetry (which depends on the global symmetry representation), we get degeneracies of operators at each level. For Wilson-Fisher fixed point at large-$N$, one need to perform the Hubbard-Stratonovich transformation,
\begin{equation}
    Z(q,x,a)=\text{PE}[\chi^{short}_{\Delta=1/2,\ell=0}(q,x)\cdot \chi_V(a)-\chi_{1,0}(q,x)+\chi_{2,0}(q,x)].
\end{equation}
The partition function for other theories (at their perturbative limits) can be worked out in a similar manner.
Since the partition function calculation doesn't tell us about parity symmetry,
we still need to enumerate operators to understand their parity symmetry. 

 \newpage

\section{Review: abelian case}\label{sec:abeliancase}

\subsection{3pt functions in embedding formalism}
\label{sec3pt-:abeliancase}

In this section we construct the most generic 3pt function of two spin-1 conserved currents and a generic parity even/odd traceless symmetric operator in a CFT in $d=3$. We do this using the embedding formalism developed in \cite{Costa:2011mg,Costa:2011dw}.

Let us begin by introducing our basis of tensor structures in embedding space:
\begin{subequations}
\begin{align}
&    H_{ij} = \frac{(Z_i\cdot Z_j)(P_i\cdot P_j)-(Z_i\cdot P_j)(P_i\cdot Z_j)}{(P_i\cdot P_j)}\,\\
&    V_{i,jk} = \frac{(Z_i\cdot P_j)(P_i\cdot P_k)-(Z_i\cdot P_k)(P_i\cdot P_j)}{\sqrt{{-2 (P_i\cdot P_j)(P_i\cdot P_k)(P_j\cdot P_j)}}}\,\\
&    V_1 = V_{1,23}\,\qquad V_2 = V_{2,31}\, \qquad V_3 = V_{3,12}\,\\
&    \epsilon_{ij} = \varepsilon_{ABCDE} P^A_i P^B_j Z_1^C Z_2^D Z_3^E\,.  
\end{align}
\end{subequations}
where $P_i^A, Z_i^A$ are $d+2$ dimensional vectors representing the position and polarization vectors in embedding space. We can choose them such that $P_i^2 = Z_i^2 =P_i\cdot P_i =0 $.

The 3pt function of two currents and a third operator is
\begin{equation}\label{eq:general3pt}
    \langle J_1(P_1,Z_1)J_2(P_2,Z_2) \mathcal{O}_{\Delta,\ell} (P_3,Z_3) \rangle = \frac{\mathcal{T}_{\ell}(P_i,Z_i)}{ (-2P_1\cdot P_2)^{2-\Delta/2} (-2P_1\cdot P_3)^{\Delta/2}(-2P_2\cdot P_3)^{\Delta/2}}
\end{equation}
where the tensor structure for parity even operators reads
\begin{align}
&  \mathcal{T}_{0}(P_i,Z_i) =
        \lambda_1 V_1 V_2  +\lambda_2 H_{12}\\
&\mathcal{T}_{1}(P_i,Z_i) =
        \lambda_1 V_1 V_2 V_3 +\lambda_2 V_1 H_{23}  +\lambda_3 H_{13}V_2   +\lambda_4 H_{12} V_3\\
& \mathcal{T}_{\ell}(P_i,Z_i) =
        \lambda_1 V_1 V_2 V_3^\ell +\lambda_2 V_1 H_{23} V_3^{\ell-1} +\lambda_3 H_{13}V_2 V_3^{\ell-1} + \lambda_4 H_{13} H_{23} V_3^{\ell-2} +\lambda_5 H_{12} V_3^\ell
\end{align}
and we have suppressed the explicit dependence of the OPE coefficients $\lambda_i$ on the 3pt functions operators.

For non-abelian currents we will be interested in the case in which the 3pt function is both symmetric or antisymmetric under exchange of the currents at point 1 and 2. Hence we will analyze the two cases separately. Imposing the correct behavior under permutation and conservation at point 1 and 2, we get the relations among the OPE coefficients summarized in Table~\ref{tab:3pt-even-symm} and Table~\ref{tab:3pt-even-antisymm}.

\begin{table}[t!]
    \centering
    \begin{tabular}{c|ccc|c}    spin & \multicolumn{2}{c}{Parity-even: Symmetric \& cons} && $\#$ ind\\
    \hline
    $\ell=0$ & $\lambda _2 = -\frac{(2-\Delta ) \lambda _1}{\Delta } $ & & &1 \\
    $\ell=1$ &   & $\lambda_i=0$ & & 0\\
    $\ell>0$ even & $\lambda _1 =  -\frac{\lambda _2 (-\Delta +\ell +4)}{\Delta -2}-\frac{\lambda _5
   (-\Delta -\ell )}{\Delta -2}$;  & $\lambda _3 = \lambda _2 $; & $\lambda _4 = \frac{\lambda _5
   \ell }{-\Delta +\ell +2}-\frac{(\Delta -1) \lambda _2}{-\Delta +\ell +2}$ & 2 \\
    $\ell>1$ odd & &$\lambda_i=0$ &  & 0
    \end{tabular}
    \caption{OPE relations for conserved currents, symmetric under exchange $1\leftrightarrow 2$, and a third parity-even operator with spin-$\ell$}
    \label{tab:3pt-even-symm}
\end{table}

\begin{table}[t!]
    \centering
    \begin{tabular}{c|ccc|c}
    spin & \multicolumn{2}{c}{Parity-even: Anti-sym \& cons} && $\#$ ind\\
    \hline
    $\ell=0$ &  & $\lambda=0$ & & 0 \\
    $\ell=1$ & $\lambda _1 = (\Delta +3) \lambda _2$; & $\lambda _3 = \lambda _2$;&
    $\lambda _4 =
   (\Delta -1) \lambda _2  $ & 1 \\
    $J$ && $\lambda_2=\lambda_3 =\lambda_4$ &  & 2\\
    $\ell>0$ even & &$\lambda_i=0$ &  & 0 \\
    $\ell>1$ odd & $\lambda _1= -\frac{\lambda _2 (-\Delta +\ell +4)}{\Delta -2}-\frac{\lambda _5
   (-\Delta -\ell )}{\Delta -2}$ ;& $\lambda _3= \lambda _2$ ; & $\lambda _4 = \frac{\lambda _5   \ell }{-\Delta +\ell +2}-\frac{(\Delta -1) \lambda _2}{-\Delta +\ell +2}$ & 2
    \end{tabular}
    \caption{OPE relations for conserved currents, antisymmetric under exchange $1\leftrightarrow 2$, and a third parity-even operator with spin-$\ell$. Here $J$ represents a conserved spin-1 current at point 3. Notice that there is one extra OPE coefficient compared to a generic spin-1 non-conserved operator. This is because the rank of the linear relations among OPE coefficients is smaller for $\Delta=2, \ell=1$ compared to the generic case. The same phenomenon appears in $\langle TTT \rangle$.}
    \label{tab:3pt-even-antisymm}
\end{table}

The tensor structure for parity odd operators instead has the following form
\begin{align}
&        \mathcal{T}_{\ell}(P_i,Z_i) =
        \lambda_1 \varepsilon_{13} V_2 V_3^{\ell-1}+\lambda_2 \varepsilon_{23} V_1    V_3^{\ell-1}+
        \lambda_3 \varepsilon_{13} H_{23} V_3^{\ell-2}+\lambda_4 \varepsilon_{23} H_{13} V_3^{\ell-2}\\
& \mathcal{T}_{0}(P_i,Z_i) =
        \lambda_1 \varepsilon_{12} \\
& \mathcal{T}_{1}(P_i,Z_i) =
        \lambda_1 \varepsilon_{13} V_2 + \lambda_2 \varepsilon_{23} V_1 +\lambda_3 \varepsilon_{12} V_3  
\end{align}
Notice that for $\ell>1$ we removed one of the $\varepsilon_{ij}$ in terms of the other two using the liner relation 
\begin{equation}
    \varepsilon_{12} V_3^2 (P_1\cdot P_2) - \varepsilon_{12} (H_{13}+V_1 V_3) (P_2\cdot P_3) + \varepsilon_{13} (H_{23}+V_2 V_3) (P_1\cdot P_3) = 0
\end{equation}
which follows from the fact that in $d=5$ dimensions, any six vectors are linearly dependent \cite{Costa:2011mg}.

Imposing the correct behavior under permutation and conservation at point 1 and 2, we get the relations among the OPE coefficients summarized in Table~\ref{tab:3pt-odd-symm} and Table~\ref{tab:3pt-odd-antisymm}.

\begin{table}[h!]
    \centering
    \begin{tabular}{c|ccc|c}    spin & \multicolumn{2}{c}{Parity-odd: Symmetric \& cons} && $\#$ ind\\
    \hline
    $\ell=0$ & & / & &1 \\
    $\ell=1$ & & $\lambda_i=0$ &  & 0\\
    $J$ & &$ \lambda_i=0$ & & 0\\
    $\ell>0$ even & $\lambda _2= -\lambda _1$,&$\lambda _3= -\frac{(3-\Delta ) \lambda _1}{\Delta   -\ell -3}$;&$\lambda _4= \frac{(3-\Delta ) \lambda _1}{\Delta -\ell -3}$ & 1 \\
    $\ell>1$ odd &  $\lambda _2= \lambda _1$,&$\lambda _3= \frac{(\Delta-1) \lambda _1}{\Delta   -\ell -3}$;&$\lambda _4= \frac{(\Delta-1) \lambda _1}{\Delta -\ell -3}$ & 1 
    \end{tabular}
    \caption{OPE relations for conserved currents, symmetric under exchange $1\leftrightarrow 2$, and a third parity-odd operator with spin-$\ell$}
    \label{tab:3pt-odd-symm}
\end{table}

\begin{table}[h!]
    \centering
    \begin{tabular}{c|ccc|c}
    spin & \multicolumn{2}{c}{Parity-odd: Anti-sym \& cons} && $\#$ ind\\
    \hline
    $\ell=0$ &  & $\lambda_1=0$ & & 0 \\
    $\ell=1$ & $\lambda _2 =-\lambda_1 $; & $\lambda _3 =(\Delta-3) \lambda _1$;&  & 1 \\
    $J$ && $\lambda_2=\lambda_3 = -\lambda_1$ & & 1\\
    $\ell>1$ even &  $\lambda _2= \lambda _1$,&$\lambda _3= \frac{(\Delta-1) \lambda _1}{\Delta   -\ell -3}$;&$\lambda _4= \frac{(\Delta-1) \lambda _1}{\Delta -\ell -3}$ & 1 \\
    $\ell>0$ odd & $\lambda _2= -\lambda _1$ ,& $\lambda _3= -\frac{(3-\Delta ) \lambda _1}{\Delta   -\ell -3}$ ; & $\lambda _4= \frac{(3-\Delta ) \lambda _1}{\Delta -\ell -3}$ & 1 
    \end{tabular}
    \caption{OPE relations for conserved currents, antisymmetric under exchange $1\leftrightarrow 2$, and a third parity-odd operator with spin-$\ell$}
    \label{tab:3pt-odd-antisymm}
\end{table}

\subsubsection{From embedding to \texorpdfstring{$SO(3)$}{}-basis}
\label{sec:embedding-to-so3}

In this appendix we work out the change of basis between the embedding frame used in the previous section and the $SO(3)$ basis used by \texttt{block\_3d} \cite{Erramilli:2020rlr}.

Let us begin by introducing the projection from the embedding frame to the conformal frame. For three points we can choose:
\begin{align}\label{eq:conf-frame}
&x_1=(0,0,0)\nonumber\\
&x_2 = (0,0,1)\nonumber\\
&x_3 = (0,0,\infty)\nonumber
\end{align}
In addition, we introduce polarization spinors 
\begin{align}
s_i = \begin{pmatrix} \xi_i \\ \eta_i \end{pmatrix} 
\end{align}
The projection to this frame can be achieved by the choice of embedding vectors (see \cite{Erramilli:2020rlr}, appendix F)

\begin{align}\label{eq:conf-frame-embedding}
&P^A_1=(0,0,0,\frac12,\frac12)\nonumber\\
&P^A_2=(0,0,1,0,1)\nonumber\\
&P^A_3=(0,0,0,-\frac12,\frac12)\\
&Z^A_1=\left(\frac{\xi_1^2+\eta_1^2}{\sqrt2},\frac{\xi_1^2-\eta_1^2}{\sqrt2},-\sqrt{2} \xi_1\eta_1,0,0\right)\nonumber\\
&Z^A_2=\left(\frac{\xi_2^2+\eta_2^2}{\sqrt2},\frac{\xi_2^2-\eta_2^2}{\sqrt2},-\sqrt{2} \xi_2\eta_2,\sqrt{2} \xi_2\eta_2,-\sqrt{2} \xi_2\eta_2\right)\nonumber\\
&Z^A_3=\left(\frac{\xi_1^2+\eta_1^2}{\sqrt2},\frac{\xi_1^2-\eta_1^2}{\sqrt2},\sqrt{2} \xi_1\eta_1,0,0\right)
\end{align}
Plugging in the above expression in $H_{i,j}$ and $V_i$ one can express any 3pt function in conformal frame.\footnote{Our convention for the metric in embedding space is $\eta_{AB}=(-,+,+,+,-)$}

Let us now introduce the $SO(3)$-basis. First, let us define the Clebsch-Gordan $C(j_1,m_1; j_2, m_2 | j,m)$ to merge two $SO(3)$ representations $(j_1,m_1),(j_2,m_2)$ in a third representation $(j,m)$:

\begin{align}
   &   C(j_1,m_1; j_2, m_2 | j,m) =   \sqrt{\frac{(2 j+1) \left(j+j_1-j_2\right)! \left(j-j_1+j_2\right)!
   \left(-j+j_1+j_2\right)!}{\left(j+j_1+j_2+1\right)!}} \times \nonumber \\
 & \times \sqrt{(j-m)! (j+m)!
   \left(j_1-m_1\right)! \left(j_1+m_1\right)! \left(j_2-m_2\right)!
   \left(j_2+m_2\right)!} \times\\
  & \times \sum_k \frac{(-1)^k}{k! \left(-j+j_1+j_2-k\right)! \left(j_1-k-m_1\right)!
   \left(j-j_2+k+m_1\right)! \left(j-j_1+k-m_2\right)! \left(j_2-k+m_2\right)!}
\end{align}
where the sum over $k$ runs over all values such that:
\begin{align}
&    (j_1 + j_2 - j - k) \geq 0 & \quad\text{or}\quad& (j_1 - m_1 - k) \geq 0 &\nonumber \\
& (j_2 + m_2 - k) \geq   0 &  \quad\text{or}\quad&  (j - j_2 + m_1 + k) \geq 0 \\
& & \quad\text{or}\quad & (j - j_1 - m_2 + k) \geq 0\nonumber
\end{align}

With this definition, we can define the following state, which is an eigenstate of rotations at the three points $x_i$ (namely we diagonalize $J^2_i, J^z_i$)

\begin{align}
 \left| j_1,m_1,j_2,m_2,j_3,m_3 \right> = (-1)^{j_1-j_3+m_2} \sqrt{\binom{2 j_1}{j_1+m_1} \binom{2 j_2}{j_2+m_2} \binom{2 j_3}{j_3+m_3}} \times\nonumber\\
 \times\,\xi_1^{j_1+m_1}\eta_1^{j_1-m_1} \xi_2^{j_2+m_2}\eta_2^{j_2-m_2} \xi_3^{j_3+m_3}\eta_3^{j_3-m_3} \nonumber\\
 \qquad \text{for} \quad (m_1+m_2+m_3 = 0)
\end{align}

Next, we introduce a new basis to diagonalize $J_{12}^2$ and $J_{12}^z$, where $\vec J_{12} =\vec J_1 \vec J_2$. To do this, we trade the $m_1,m_2$ quantum numbers for $j_{12}, m_{12}$.
The new basis can be written in terms of the old one by:

\begin{align}
 \left| j_1,j_2 | j_{12},m_{12},j,m \right> =\sum_{m_1 = -j_1}^{j_1}\sum_{m_2 = -j_2}^{j_2} C(j_1,m_1,j_2,m_2| j_{12},m_{12}) \left| j_1,m_1,j_2,m_2,j,m \right>\nonumber\\ \qquad \text{for} \quad (m_1+m_2+m_3 = 0),\quad  |j_1-j_2|\leq j_{12} \leq j_1+j_2, \quad |m_{12}|\leq j_{12}
\end{align}

Finally, we make another change of variable and consider the total angular momentum of the three points $\vec J_{123} = J_{12}+J_3$ and consider eigenvalues of $J^2_i, J_{12}^2$ and $J_{123}^2$. The new basis reads:
\begin{align}
 \left| j_1,j_2,j_3 | j_{12},j_{123} \right> =\sum_{m = -j_3}^{j_3} C(j_{12},-m,j_3,m| j_{123},0) \left| j_1,j_2| j_{12},-m,j_3,m \right>\nonumber
\end{align}
The basis of 3pt functions tensor structures for $\langle O_{j_1}O_{j_2}O_{j_3}\rangle$ is finally given by the list 
\begin{align}
    \left\{\left| j_1,j_2,j_3 | j_{12},j_{123} \right>, \, \text{ with } |j_1-j_2|\leq j_{12} \leq j_1+j_2 , |j_3-j_{12}|\leq j_{123} \leq j_{12}+j_3\right\}
\end{align}

\begin{table}[h!]
    \centering
    \begin{tabular}{c|c|c|}
        $j_{12}$ & $j_{123}$ & parity \\
        \hline
        0 & $\ell$ & even \\ 
        1 & $\ell$ & even \\
        1 & $\ell-1, \ell+1$ & odd\\
        2 &  $\ell-2,\ell,\ell+2$ & even \\
        2 &  $\ell-1,\ell+1$ & odd
    \end{tabular}
    \caption{Values of the entries $j_{12}$ and $j_{123}$ for the case of currents. Last column is the space-time parity. When $\ell\leq1$ certain values are not present.}
    \label{tab:j-example}
\end{table}
A final comment concerns the space-time parity of the tensor structure, which is given by 
$(-1)^{j_1 - j_2 + j_3-j_{123}}$.

This is called $SO(3)$ basis and is the basis used by \texttt{block\_3d} \cite{Erramilli:2020rlr}.  In Table~\ref{tab:j-example} we explicitly list the elements of the basis for the case of $\langle O_{1}O_{1}O_{\ell}\rangle$.

Going through the chain of basis changes, we can express the embedding basis tensor structures in terms of the $SO(3)$ ones:
\begin{align}
\label{eq:relation-TEF-TOSO3}
    M(\ell)^{a}_{\,\,b} T_{SO(3)}^b=T^a_{EF}
\end{align}
where the matrices $M(\ell)$ are $5\times 5$ for parity even structures and $4\times 4$ for parity odd ones.

%\begin{align}
% M_+(\ell)^{a}_{\,\,b} =(-1)^\ell \sqrt{\frac{\sqrt{\pi } 2^{-\ell} \Gamma (\ell+1)}{3 \Gamma \left(\ell+\frac{1}%{2}\right)}}\begin{pmatrix}
% 1 & 0 &  \sqrt{ \frac{3(\ell -1) \ell }{(2 \ell -1) (2 \ell +1)}} & 
%   \sqrt{\frac{2\ell  (\ell +1)}{(2 \ell -1) (2 \ell +3)}} &  \sqrt{\frac{3(\ell
%   +1) (\ell +2)}{(2 \ell +1) (2 \ell +3)}} \\
% 1 & \sqrt{\frac{3}{2}} \sqrt{\frac{\ell +1}{\ell }} &  \sqrt{\frac{3(\ell -1) (2
%   \ell +1)}{\ell  (2 \ell -1)}} & \frac{\sqrt{\frac{(\ell +1) (2 \ell +3)}{\ell  (2 \ell
%   -1)}}}{\sqrt{2}} & 0 \\
% 1 & \sqrt{\frac{3}{2}} \sqrt{\frac{\ell +1}{\ell }} &  \sqrt{\frac{3(\ell -1)}{\ell
%    (2 \ell -1) (2 \ell +1)}} & \frac{\sqrt{\frac{(\ell +1) (2 \ell -3)^2}{\ell  (2 \ell
%   -1) (2 \ell +3)}}}{\sqrt{2}} &   \sqrt{\frac{12(\ell +1) (\ell +2)}{(2 \ell +1)
%   (2 \ell +3)}} \\
% 2 &  \sqrt{\frac{6(\ell +1)}{\ell }} &  \sqrt{\frac{3(2 \ell +1)}{(\ell -1)
%   \ell  (2 \ell -1)}} &  \sqrt{\frac{2(\ell +1) (2 \ell +3)}{\ell  (2 \ell -1)}}
%   & 0 \\
% 3 & 0 & 0 & 0 & 0 \\
% 9 & 0 & 0 & 0 & 0 \\
%    \end{pmatrix}
%\end{align}

\begin{align}
 M_+(\ell)^{a}_{\,\,b} =(-1)^\ell \sqrt{\frac{\sqrt{\pi } 2^{-\ell} \Gamma (\ell+1)}{3 \Gamma \left(\ell+\frac{1}{2}\right)}}\begin{pmatrix}
 1 & 0 &  -\sqrt{ \frac{3(\ell -1) \ell }{(2 \ell -1) (2 \ell +1)}} & 
   \sqrt{\frac{2\ell  (\ell +1)}{(2 \ell -1) (2 \ell +3)}} &  -\sqrt{\frac{3(\ell
   +1) (\ell +2)}{(2 \ell +1) (2 \ell +3)}} \\
 -1 & \sqrt{\frac{3}{2}} \sqrt{\frac{\ell +1}{\ell }} &  \sqrt{\frac{3(\ell -1) (2
   \ell +1)}{\ell  (2 \ell -1)}} & -\frac{\sqrt{\frac{(\ell +1) (2 \ell +3)}{\ell  (2 \ell
   -1)}}}{\sqrt{2}} & 0 \\
 -1 & \sqrt{\frac{3}{2}} \sqrt{\frac{\ell +1}{\ell }} &  -\sqrt{\frac{3(\ell -1)}{\ell
    (2 \ell -1) (2 \ell +1)}} & -\frac{\sqrt{\frac{(\ell +1) (2 \ell -3)^2}{\ell  (2 \ell
   -1) (2 \ell +3)}}}{\sqrt{2}} &   \sqrt{\frac{12(\ell +1) (\ell +2)}{(2 \ell +1)
   (2 \ell +3)}} \\
 2 &  -\sqrt{\frac{6(\ell +1)}{\ell }} &  \sqrt{\frac{3(2 \ell +1)}{(\ell -1)
   \ell  (2 \ell -1)}} &  \sqrt{\frac{2(\ell +1) (2 \ell +3)}{\ell  (2 \ell -1)}}
   & 0 \\
 -3 & 0 & 0 & 0 & 0 \\
    \end{pmatrix}
\end{align}

%$\begin{align}
% M_-(\ell)^{a}_{\,\,b} = (-1)^\ell
%\sqrt{ \frac{\sqrt{\pi } 2^{-\ell-4} (\ell+1) \Gamma (\ell+2)}{\ell \Gamma \left(\ell+\frac{3}{2}\right)}}
% \begin{pmatrix}
% 1 & \sqrt{\frac{\ell}{\ell+1}} & \sqrt{\frac{\ell-1}{\ell+1}} & \sqrt{\frac{\ell+2}{\ell+1}} \\
% 1 & \sqrt{\frac{\ell}{\ell+1}} & \sqrt{\frac{\ell-1}{\ell+1}} & \sqrt{\frac{\ell+2}{\ell+1}} \\
% \sqrt{\frac{(2 \ell+1)^2}{(\ell+1)^2}} & 0 & \sqrt{\frac{(2 \ell+1)^2}{\ell^2-1}} & 0 \\
% \sqrt{\frac{1}{(\ell+1)^2}} & 2 \sqrt{\frac{\ell}{\ell+1}} & 3 \sqrt{\frac{1}{\ell^2-1}} & 2
%   \sqrt{\frac{\ell+2}{\ell+1}} 
%    \end{pmatrix}
%\end{align}

\begin{align}
M_-(\ell)^{a}_{\,\,b} = (-1)^\ell
\sqrt{ \frac{\sqrt{\pi } 2^{-\ell-4} (\ell+1) \Gamma (\ell+2)}{\ell \Gamma \left(\ell+\frac{3}{2}\right)}}
 \begin{pmatrix}
-1 & -\sqrt{\frac{\ell}{1 + \ell}} & \sqrt{\frac{-1 + \ell}{1 + \ell}} & -\sqrt{\frac{2 + \ell}{1 + \ell}} \\
-1 & -\sqrt{\frac{\ell}{1 + \ell}} & -\sqrt{\frac{-1 + \ell}{1 + \ell}} & \sqrt{\frac{2 + \ell}{1 + \ell}} \\
\sqrt{\frac{(1 + 2 \ell)^2}{(1 + \ell)^2}} & 0 & -\sqrt{\frac{(1 + 2 \ell)^2}{-1 + \ell^2}} & 0 \\
\sqrt{\frac{1}{(1 + \ell)^2}} & 2 \sqrt{\frac{\ell}{1 + \ell}} & -3 \sqrt{\frac{1}{-1 + \ell^2}} & -2 \sqrt{\frac{2 + \ell}{1 + \ell}}
    \end{pmatrix}
\end{align}

For $\ell=0$ the parity-even matrix of change of basis is $2\times2$, while is just a number for parity-odd:
\begin{align}
 M_+(0)^{a}_{\,\,b} = 
 \begin{pmatrix}
\frac{1}{\sqrt{3}} & -\sqrt{\frac{2}{3}} \\
-\sqrt{3} & 0 \\
    \end{pmatrix},
    \qquad
M_-(0)=\frac1{\sqrt2}
\end{align}
Finally for $\ell=1$ the parity-even matrix of change of basis is $4\times4$ and the parity-odd matrix is $3\times3$:

\begin{align}
 M_+(1)^{a}_{\,\,b} =
 \begin{pmatrix}
-\frac{1}{\sqrt{3}} & 0 & -\frac{2}{\sqrt{15}} & \sqrt{\frac{2}{5}} \\
\frac{1}{\sqrt{3}} & -1 & \sqrt{\frac{5}{3}} & 0 \\
\frac{1}{\sqrt{3}} & -1 & -\frac{1}{\sqrt{15}} & -2 \sqrt{\frac{2}{5}} \\
\sqrt{3} & 0 & 0 & 0 \\
    \end{pmatrix}
\qquad 
 M_-(1)^{a}_{\,\,b} =
 \begin{pmatrix}
\frac{1}{\sqrt{6}} & \frac{1}{2\sqrt{3}} & \frac{1}{2} \\
\frac{1}{\sqrt{6}} & \frac{1}{2\sqrt{3}} & -\frac{1}{2} \\
\frac{1}{\sqrt{6}} & -\frac{1}{\sqrt{3}} & 0 \\
    \end{pmatrix}
\end{align}

\subsection{4pt function}

In this section we review the analysis of the 4pt functions of four identical conserved currents, thus translating the analysis of \cite{Dymarsky:2017xzb} to the conformal frame introduced in \cite{Kravchuk:2016qvl,Dymarsky:2017yzx}.

Our convention for the conformal frame of four points is
\begin{align}\label{eq:conf-frame-4tp}
&x_1=(0,0,0)\nonumber\\
&x_2 = (\frac{\bar{z}-z}2,\frac{z+\bar{z}}{2},0)\\
&x_3 = (0,1,0)\nonumber\\
&x_4 = (0,\infty,0)\nonumber
\end{align}
In addition, we introduce polarization spinors 
\begin{align}
s_i = \begin{pmatrix} \xi_i \\ \eta_i \end{pmatrix} 
\end{align}

In the following we will also use an alternative set of polarization variables, in terms of three dimensional vectors $Z^\mu_j$. They are related to the above spin variables by the simple relation
\begin{align}
Z^\mu_j = s_j \sigma^\mu \epsilon s_j = \left(\frac{w_j +\bar{w}_j}2, \frac{w_j -\bar{w}_j}{2i} , w^0_j \right)
\end{align}
where $\sigma^\mu$ are the Pauli matrices. In fact, we will use the variables  $w,\bar{w}$ and $w_0$.

A 4pt tensor structure will be indicated as
\begin{align}
\mathcal T^{(\ell_1,\ell_2,\ell_3,\ell_4)}_{[q_1,q_2,q_3,q_4]} =  \prod_{i_1}^4 \xi_i^{\ell_i+q_i} \eta_i^{\ell_i-q_i}
\end{align}

This is called the $q-$ basis. Parity invariance of the tensor structures restrict the values $q_i$ to those such that $\displaystyle \sum_{i=1}^4 q_i $ is an even quantity. 

All in all, we parametrize the four point function of four spin-1 conserved currents as
\begin{align}
\langle J_1(s_1,x_1)\ldots J_4(s_4,x_4)\rangle = \sum_{\substack{q_i =-1,0,1 , \\   q_1+q_2+q_3+q_4 \text{ even}} } \mathcal T^{(1,1,1,1)}_{[q_1,q_2,q_3,q_4]} f_{[q_1,q_2,q_3,q_4]} (z,\bar z) 
\end{align}
The sum in the r.h.s. produces 41 independent tensor structures. Alternatively, we can use the following parametrization

\begin{align}
\mathbb T^{(\ell_1,\ell_2,\ell_3,\ell_4)}_{[h_1,h_2,h_3,h_4]} =  \prod_{i_1}^4 [h_i]
\end{align}
where we define monomials of charge $h_i$ under rotation along the 3rd axis: 
\begin{align}
[-1] = \bar{w}\, \qquad [+1] = w\, \qquad [0] = w_0\,.
\end{align}

Finally, we can express the  the four point function of four spin-1 conserved currents in the $w-$basis as

\begin{align}
\langle J_1(w_1,x_1)\ldots J_4(w_4,x_4)\rangle = \sum_{\substack{h_i =-1,0,1 , \\   h_1+h_2+h_3+h_4 \text{ even}} }\mathbb T^{(1,1,1,1)}_{[h_1,h_2,h_3,h_4]}  g_{[h_1,h_2,h_3,h_4]} (z,\bar z) 
\end{align}

Again, there are 41 independent tensor structures. There is a simple one-to-one correspondence between the functions $f_{[q_1,q_2,q_3,q_4]} (z,\bar z) $ and $g_{[h_1,h_2,h_3,h_4]} (z,\bar z) $ which simply amounts to a numerical factor. (check!)

\subsubsection{Permutation symmetry}
\label{app:permutations}

Whenever the currents appearing in the correlation functions are identical, or belonging to the same irreducible representation of the global symmetry, the correlation function must obey certain permutation symmetries. This in turn translates into relations between the g-functions. 

In this section we only consider the case of identical currents. In this case the correlator is invariant under any exchange of labels $i\leftrightarrow j$. Among all possible combination, there is a subgroup of pairwise permutations that leaves the conformal invariants $z,\bar z$ un-altered:

 \begin{align}\label{eq:permutation}
1234 \leftrightarrow 2143  \leftrightarrow 4321  \leftrightarrow 3412
\end{align}

Invariance under this set of permutations will then imply an algebraic relation between the $g-$(and $f-$)functions. 

One has to be careful however when dealing with permutation-symmetry in the conformal frame: because the conformal frame is not permutation invariant, the exchange of any two points will bring us in a different frame. One then needs to combine the permutation with a space-time rotation to go back to the original conformal frame. Because the $s_i$ and $w_i$ are not scalars, they pick up a phase under this additional rotation. This analysis has been done in detail in \cite{Kravchuk:2016qvl} for the $q-$basis and in \cite{Dymarsky:2017yzx} for the $w-$basis. Here we simply summarize the recipe.

\begin{table}[ht!]
\begin{center}
\begin{tabular}{r || c | c | c | c }
			& $r_1$ 		& $r_2$ 		& $r_3$ 		& $r_4$ 		\\
\hline\hline
id			& $1$ 		& $1$ 		& $1$		&	 $1$			\\\hline
$(12)(34)$	& $-(1-z)$	& $-(1-\bar z)$		& $-(1-\bar z)$ 	&	$-(1-z)$	\\\hline
$(13)(24)$	& $\bar z(1-z)$		& $\bar z(1-\bar z)$		& $z(1-\bar z)$ 		&	$z(1-z)$			\\\hline
$(14)(23)$	& $-\bar z$		& $-\bar z$ 		& $-z$		&	$-z$		\\\hline
\end{tabular}
\end{center}
\caption{Permutation phases for a 4-point function of identical operators.}
\label{tab:perm}
\end{table}

A permutation $\pi \in \Pi^\mathrm{kin}$ acts on a monomials $[h_i]$ as
\begin{align}
\pi: [h_i] \longrightarrow  n(r_i(\pi))^{h_i}[h_{\pi(i)}],
\end{align}
where $n(x) = \sqrt{x/\bar x}$ is a phase and the $r_i(\pi)$ are given in the table~\ref{tab:perm}.

Applying repetitively the permutations \eqref{eq:permutation} we can group the various tensor structure 17 orbits, with at most 4 elements each. Let us choose a representative for each orbit and construct a basis of permutation invariant tensor structures:

\begin{align}
\widetilde {\mathbb T}{[h_1,h_2,h_3,h_4]} =& \Bigg(\mathbb T^{(1,1,1,1)}_{[h_1 h_2 h_3 h_4]}+ n(1-z)^{-h_1+h_2+h_3-h_4} \mathbb T^{(1,1,1,1)}_{[h_2 h_1 h_4 h_3]}\nonumber\\
& + n(z)^{h_1+h_2-h_3-h_4} \mathbb T^{(1,1,1,1)}_{[h_4 h_3 h_2 h_1]}\nonumber\\
& + n(z)^{h_1+h_2-h_3-h_4} n(1-z)^{-h_1+h_2+h_3-h_4}\mathbb T^{(1,1,1,1)}_{[h_3 h_4 h_1 h_2]}\Bigg)
\end{align}
where we decided to take as orbit representatives the following set:
\begin{align}
\label{eq:list4pt-structure}
[h_1,h_2,h_3,h_4] \in \mathcal S = \{ & [-1, -1, -1, -1], [-1, -1, -1, 1], [-1, -1, 0, 0], [0, -1, -1,  0], \nonumber\\
&[-1, -1, 1, 1], [1, -1, -1, 1], [-1, 0, -1, 0], [-1, 0, 0, 1],\nonumber\\
& [0, 0, -1, 1], [0, 1, 0, -1], [-1, 1, -1, 1], [1, 1, 1, -1], [0,0, 0, 0], \nonumber\\
&[1, 1, 0, 0], [0, 1, 1, 0], [1, 0, 1, 0], [1, 1, 1, 1]\}
\end{align}

with this choice, the currents 4pt function can be written as

\begin{align}\label{eq:4pt-perm}
\langle J_1(w_1,x_1)\ldots J_4(w_4,x_4)\rangle = \sum_{[h_i] \in \mathcal S }\widetilde {\mathbb T}{[h_1, h_2, h_3, h_4]}  g_{[h_1,h_2,h_3,h_4]} (z,\bar z) 
\end{align}

\subsubsection{Conservation}
\label{app:conservation}

Up to this point we have reduced the size of the problem to 17 independent functions of two variables. Let us now review how the conservation of the currents can reduce the problem further. As shown in \cite{Dymarsky:2013wla,Dymarsky:2017xzb,Dymarsky:2017yzx}, conservation implies a system of coupled first order differential equations. We should stress that, once we impose the conformal block decomposition and we use the relation between OPE coefficients obtained in the previous section, the correlation function will automatically satisfy the differential equations. In fact, each conformal block will. Hence, the constraints presented in this section should not be imposed again. Rather, they represent a non trivial relation among the $g-$functions. In particular, we can regard the conservation equation as an evolution equation from a certain initial condition. The evolution kernel is crossing symmetric, therefore it will evolve crossing symmetric initial conditions to fully crossing symmetric functions. As a bottom line, it is sufficient to impose crossing symmetry on the initial conditions and not on the full set of $g-$functions.

Before discussing how to find linearly independent bootstrap equations, we would like to emphasize the necessity of this procedure for numerical applications. Naively, one might think that even if we don’t remove the linearly dependent equations, the equations will just be redundant but not wrong. But in practice, the linear dependence can actually lead to catastrophic errors in the numerics. The key reason is that we will not be able to exactly implement those equations. Instead, in numerical bootstrap, we use rational approximations, which may or may not keep the exact linear dependence among the equations. Assuming our approximation slightly breaks the linear dependence, which is the case for the bootstrap equations in this work, it is possible one may be able to show the system of equations is inconsistent regardless of the choice of parameters \footnote{This phenomenon has already been observed in \cite{Rong:2018okz} and \cite{Iliesiu:2015qra}.} . This means the error of the approximation could lead to completely uncontrolled errors in the allowed region of the parameter space. On the other hand, if we only keep the linearly dependent equations, although the boundary of the allowed region would still be affected by the error raised from the rational approximation, the error is small and controllable.

Let us proceed by steps. First let us introduce the conservation operator in conformal frame
\begin{align}\label{eq:conservationOp}
D_i = \left(\frac12 - w_j \partial_{w_j} \right)\partial_{\bar{z}} \partial_{\bar{w}_j} + \left(\frac12 - \bar{w}_j \partial_{\bar{w}_j} \right)\partial_z \partial_{w_j}  - \frac{3i}{\bar{z}-z} \mathcal D_j \mathcal L_{23}
\end{align}
where $\mathcal D_j$ is the Todorov operator while $\mathcal L_{23}$ is the generator of rotations around the first direction in the $w-$representation. Applying the differential operator \eqref{eq:conservationOp} to \eqref{eq:4pt-perm} produces a system of equations of the form
\begin{align}\label{eq:conservationOp-zzb}
\left(A_z \partial_z +B_{\bar{z}} \partial_{\bar{z}} + C \right) \cdot \vec{g}(z,\bar{z}) = 0\,,
\end{align}
where we have collected the $g-$functions in a 17-dimensional vector $\vec g (z,\bar z)$ and $A_z , B_{\bar{z}}, C$ are $17\times14$ matrices.\footnote{The number of equations equals the number of independent tensor structures $\mathbb T^{(0,1,1,1)}_{[h_1,h_2,h_3,h_4]}$, which is 14.}

Let us interpret \eqref{eq:conservationOp-zzb} as an evolution equation along a certain time direction, which for the time being we will identify with the $\bar{z}$ direction. It turns out that the matrix $B_{\bar{z}}$ has only rank 12: in fact one can immediately see that all functions not dependent on $\bar{w}_j$ will not appear in this term. Hence the functions 
\begin{align}\label{eq:bulk-functions}
g_{[0, 0, 0, 0]}\,, \quad g_{[1, 1, 0, 0]}, \quad g_{[0, 1, 1, 0]}, \quad g_{[1, 0, 1, 0]},\quad g_{[1, 1, 1, 1]}
\end{align}
cannot be time-evolved and instead must be known in the whole $(z,\bar{z})$ plane. We call the above set \emph{bulk functions}.
All other functions in \eqref{eq:4pt-perm} instead can be time-evolved starting from an initial condition: we call them \emph{boundary functions}.

Let us now rotate the time direction and choose a different one. We choose 
\begin{align}
z = \frac12 + x + t\, \qquad \bar{z} = \frac12 + x - t\,.
\end{align}
Now the initial time corresponds to the line $z=\bar z$, while the time direction corresponds to the orthogonal direction. When making this change of coordinates, the conservation equations also transform in a simple way:

\begin{align}\label{eq:conservationOp-xt}
\left(A'_t \partial_t +B'_{x} \partial_{x} + C' \right) \cdot \vec{g}(x,t) = 0\,,
\end{align}

 In particular, the set of functions that cannot be time-evolved is going to be a linear combination of \eqref{eq:bulk-functions} and the other 12. The same is true for those that can be evolved. Hence, we can maintain the same distinction between bulk and boundary functions. In fact we can maintain the same boundary functions as long as they are not completely contained in the kernel of the time evolution operator.\\
 This would not be true if, for instance, we choose $z$ as the new time direction. In that case some  of the functions that cannot be time-evolved would be exactly contained among the 12  boundary ones. 

Up to this point we have made use of 12 equations to reduce the boundary functions to 12 single variable functions on the $z=\bar{z}$ line. The system \eqref{eq:conservationOp-xt}, however, contains two additional equations we haven't exploited. To do this, we compute the left kernel of $A_t$,
$\vec v_1$ and $\vec v_2$ such that $\vec v_k \cdot A'_t = 0$. Hence the resulting equations only contain $x$-derivatives and can be regarded as evolution equations in the $x-$direction. We can use them to evolve at most two functions from an initial condition at the  crossing symmetric point $z=\bar{z}=1/2$. We will come back to the most convenient choice after the next subsection.

\subsubsection{Smoothness condition}
\label{app:smoothness}

The conformal frame \ref{eq:conf-frame-4tp} has a $\mathbb Z_2$ stabilizer,  i.e. a set of conformal transformations that leaves it invariant. This symmetry is however enhanced in the special configuration when $z=\bar z$. In that case indeed the four points are aligned in the 2nd direction and one is free to rotate the system around it and the stabilizer symmetry is enhanced to $SO(2)$. When the points are not aligned the symmetry is explicitly broken. We can parametrize such breaking by the vector orthogonal to the $z=\bar z$ line: $\vec{n} = (z-\bar z) \hat e_1$, where $\hat e_1$ is a unit vector point in the direction $1$.

As a preliminary step, we can classify the 4pt functions tensor structures according to their irreducible representations under $SO(2)$. Clearly we expect to find singlets and non-singlets. The important observation is that only the former can survive in the limit $z=\bar{z}$. In addition, the rate at which non singlets vanish in this limit is linked to the $SO(2)$ irrep. Let us see this in detail. 

Let us define the action of the generator of rotations around direction 1 in the spinor $s_k$ 
\begin{align}
    \mathcal{L}^{(j)}_{23} s_k= i \delta_{jk}\sigma^1  s_k
\end{align}
where $\sigma^1$ is the first Pauli matrix.
Then the quadratic $SO(2)$ Casimir operator acting on a 4pt tensor structure is given by:
\begin{align}
    C^{SO(2)}_2 \mathcal T^{(1,1,1,1)}_{[q_1,q_2,q_3,q_4]}  =  \left(\sum_j \mathcal{L}^{(j)}_{23}\right)^2 T^{(1,1,1,1)}_{[q_1,q_2,q_3,q_4]} 
\end{align}

For any choice $[q_1,q_2,q_3,q_4]$ one can express the action of the Casimir in terms of elements of the basis $ T^{(1,1,1,1)}_{[q_1,q_2,q_3,q_4]}$. Diagonalizing the resulting $17\times 17$ matrix we can find the following $SO(2)$ representations

\begin{align}
    0^{7} \oplus 1^{4} \oplus 2^{4} \oplus 3 \oplus 4
\end{align}

and the corresponding eigenvectors. We conclude that there exist only 7 combinations of tensor structures that are allowed to be non zero on the line $z=\bar z$. All other combinations must vanish. \footnote{These are traceless symmetric representations of $SO(3)$ with spin $j=1,2,3,4$. Outside the line $z=\bar z$ one should be able to write them as $t_{\mu_1\ldots \mu_j} n^{\mu_1}\ldots n^{\mu_j}$ which is therefore $O(|z-\bar z|^j)$.}

\subsubsection{Final choice of crossing condition}

Let us combine the analysis of the previous two subsections into the final choice of  crossing condition. In Sec.~\ref{app:conservation} we saw that using conservation equations we can reconstruct the 12 of the 17 functions $g$ if we know the other 5 everywhere. We also need boundary conditions for these 12. Moreover, using the results of Sec.~\ref{app:smoothness} we know that at most 7 functions can be non zero on the line $z=\bar z$. 

One can explicitly check that the projection of the bulk functions \ref{eq:bulk-functions} into the space of $SO(2)$ singlets is full rank, which means that when restricted to $z=\bar z$ they still are independent. Hence we need to specify only two more boundary conditions. 
At the end of Sec.~\ref{app:conservation} we saw that we can use two differential equations restricted on $z=\bar z$ to evolve at most two functions from an initial condition, say at $z=\bar z=1/2$. Again one can inspect the kernel and find a pair of functions that do not belong to the kernel, have a non zero overlap with the singlet sector, are independent from the bulk functions and are mapped into each other under crossing. For this pair we only need to specify boundary conditions on a single point.
A convenient choice is $[-1, -1, 1, 1]$ and $[1, -1, -1, 1]$. We call them \emph{point functions}.

All other functions are determined entirely by the knowledge of the bulk functions and the point functions. For the specific case of abelian currents we don't have any line functions left.

In conclusion, we will impose crossing symmetry for generic $z, \bar z$ on bulk functions:
\begin{align}\label{eq:crossing-bulk}
& g_{[0, 0, 0, 0]}(z,\bar z) = g_{[0, 0, 0, 0]}(1-z,1-\bar z) \,,\nonumber\\
& g_{[1, 1, 0, 0]}(z,\bar z) = g_{[0, 1, 1, 0]}(1-z,1-\bar z) \,,\nonumber\\
& g_{[1, 0, 1, 0]}(z,\bar z) = g_{[1, 0, 1, 0]}(1-z,1-\bar z) \,,\nonumber\\
& g_{[1, 1, 1, 1]}(z,\bar z) = g_{[1, 1, 1, 1]}(1-z,1-\bar z)  \,,
\end{align}
and for $z= \bar z=1/2$ on point functions:
\begin{align}\label{eq:crossing-point}
& g_{[-1, -1, 1, 1]}(1/2,1/2) = g_{[1, -1, -1, 1]}(1/2,1/2) 
\end{align}

\section{Non-abelian case}\label{sec:nonabeliancase}

The generalization to non-abelian currents is quite straightforward. Let us outline the main steps. The 4pt function of four currents transforming in the adjoint (Adj) representation of a global symmetry $\mathcal G$ is

\begin{align}
\label{eq:non-abelian-4pt}
\langle J^a(w_1,x_1) J^b(w_2,x_2) J^c(w_3,x_3) J^d(w_4,x_4)\rangle \nonumber\\ =\sum_{r\in Adj\otimes Adj} T^{abcd}_r
\sum_{\substack{h_i =-1,0,1 , \\   h_1+h_2+h_3+h_4 \text{ even}} }  \mathbb T^{(1,1,1,1)}_{[h_1,h_2,h_3,h_4]}  g^{(r)}_{[h_1,h_2,h_3,h_4]} (z,\bar z) 
\end{align}
where $T^{abcd}_r$ is a tensor structure in flavor space and we introduced different functions $g^{(r)}_{[h_1,h_2,h_3,h_4]}$ for each irrep appearing in the tensor product $Adj\otimes Adj$.

Given the above parametrization we can repeat the analysis of permutation, conservation and smoothness conditions. Clearly the last two are not affected by the flavor indices, since the global symmetry commutes with space-time symmetries.

Permutation (and crossing) require some care. If we assume the symmetry properties

\begin{align}
    T^{abcd}_r = T^{badc}_r = T^{dcba}_r = T^{cdab}_r \,,\quad \forall \, r \in Adj\otimes Adj
\end{align}
then the analysis of Sec.\ref{app:permutations} applies without any change. Within each representation then we can reduce to bulk functions and point functions. 
Crossing equations however will mix different representations through a crossing matrix $\mathcal M$:
\begin{align}\label{eq:crossing-bulk-nonabelian}
& g^{(r)}_{[0, 0, 0, 0]}(z,\bar z) = \mathcal M^{r}_{\,\,s}\, g^{(s)}_{[0, 0, 0, 0]}(1-z,1-\bar z) \,,\nonumber\\
& g^{(r)}_{[1, 1, 0, 0]}(z,\bar z) = \mathcal M^{r}_{\,\,s} \,g^{(s)}_{[0, 1, 1, 0]}(1-z,1-\bar z) \,,\nonumber\\
& g^{(r)}_{[1, 0, 1, 0]}(z,\bar z) = \mathcal M^{r}_{\,\,s} \,g^{(s)}_{[1, 0, 1, 0]}(1-z,1-\bar z) \,,\nonumber\\
& g^{(r)}_{[1, 1, 1, 1]}(z,\bar z) = \mathcal M^{r}_{\,\,s} \,g^{(s)}_{[1, 1, 1, 1]}(1-z,1-\bar z)  \,,\nonumber\\
& g^{(r)}_{[-1, -1, 1, 1]}(1/2,1/2) = \mathcal M^{r}_{\,\,s} \,g^{(s)}_{[1, -1, -1, 1]}(1/2,1/2) 
\end{align}

The crossing matrices for $SU(N)$ and $O(N)$ are:
\begin{align} 
\mathcal M^{O(N)}=
 \resizebox{0.9\textwidth}{!}{$
\left(
\begin{array}{cccccc}
 \frac{2}{(N-1) N} & \frac{\sqrt{2} (N+2)}{N \sqrt{N^2+N-2}} & \frac{\sqrt{N \left(N^3-7 N-6\right)}}{\sqrt{3} (N-1) N} & \frac{(N-3) (N-2)}{\sqrt{6} \sqrt{(N-3)
   (N-2) (N-1) N}} & -\frac{\sqrt{2}}{\sqrt{(N-1) N}} & -\frac{(N-3) (N+2)}{\sqrt{2} \sqrt{(N-3) (N-1) N (N+2)}} \\
 \frac{\sqrt{2} \sqrt{N^2+N-2}}{(N-1) N} & \frac{2}{N}+\frac{1}{2-N}+\frac{1}{2} & \frac{(N-4) \sqrt{N \left(N^3-7 N-6\right)}}{\sqrt{6} (N-2) N \sqrt{N^2+N-2}} &
   -\frac{(N-3) \sqrt{N^2+N-2}}{\sqrt{3} \sqrt{(N-3) (N-2) (N-1) N}} & -\frac{(N-4) \sqrt{N^2+N-2}}{2 (N-2) \sqrt{(N-1) N}} & \frac{2 \sqrt{(N-3) (N-1) N
   (N+2)}}{(N-2) N \sqrt{N^2+N-2}} \\
 \frac{\sqrt{N \left(N^3-7 N-6\right)}}{\sqrt{3} (N-1) N} & \frac{(N-4) \sqrt{N \left(N^3-7 N-6\right)}}{\sqrt{6} (N-2) N \sqrt{N^2+N-2}} &
   -\frac{2}{N-1}+\frac{1}{N-2}+\frac{1}{3} & \frac{\sqrt{N \left(N^3-7 N-6\right)}}{3 \sqrt{2} \sqrt{(N-3) (N-2) (N-1) N}} & \frac{\sqrt{N \left(N^3-7
   N-6\right)}}{\sqrt{6} (N-2) \sqrt{(N-1) N}} & \frac{(N-4) \sqrt{N \left(N^3-7 N-6\right)}}{\sqrt{6} (N-2) \sqrt{(N-3) (N-1) N (N+2)}} \\
 \frac{(N-3) (N-2)}{\sqrt{6} \sqrt{(N-3) (N-2) (N-1) N}} & -\frac{(N-3) (N-1) (N+2)}{\sqrt{3} \sqrt{(N-3) (N-2) (N-1) N} \sqrt{N^2+N-2}} & \frac{(N-3) N (N+1)
   (N+2)}{3 \sqrt{2} \sqrt{(N-3) (N-2) (N-1) N} \sqrt{N \left(N^3-7 N-6\right)}} & \frac{1}{6} & -\frac{\sqrt{(N-3) (N-2) (N-1) N}}{\sqrt{3} (N-2) \sqrt{(N-1) N}} &
   \frac{\sqrt{(N-3) (N-2) (N-1) N} (N+2)}{2 \sqrt{3} (N-2) \sqrt{(N-3) (N-1) N (N+2)}} \\
 -\frac{\sqrt{2}}{\sqrt{(N-1) N}} & -\frac{(N-4) \sqrt{(N-1) N} (N+2)}{2 (N-2) N \sqrt{N^2+N-2}} & \frac{\sqrt{N \left(N^3-7 N-6\right)}}{\sqrt{6} (N-2) \sqrt{(N-1)
   N}} & -\frac{(N-3) \sqrt{(N-1) N}}{\sqrt{3} \sqrt{(N-3) (N-2) (N-1) N}} & \frac{1}{2} & 0 \\
 \frac{-N^2+N+6}{\sqrt{2} \sqrt{(N-3) (N-1) N (N+2)}} & \frac{2 \sqrt{(N-3) (N-1) N (N+2)}}{(N-2) N \sqrt{N^2+N-2}} & \frac{(N-4) (N+1) \sqrt{(N-3) (N-1) N
   (N+2)}}{\sqrt{6} (N-2) (N-1) \sqrt{N \left(N^3-7 N-6\right)}} & \frac{\sqrt{(N-3) (N-1) N (N+2)}}{2 \sqrt{3} \sqrt{(N-3) (N-2) (N-1) N}} & 0 & \frac{1}{2} \\
\end{array}
\right)$
}\nonumber\\
\end{align}

\begin{align}
\mathcal M^{SU(N)}=
 \resizebox{0.8\textwidth}{!}{%
$\left(
\begin{array}{cccccc}
 \frac{1}{N^2-1} & \frac{1}{\sqrt{N^2-1}} & \frac{\sqrt{(N-3) N^2 (N+1)}}{2 \left(N^2-1\right)} & \frac{\sqrt{(N-1) N^2 (N+3)}}{2 \left(N^2-1\right)} &
   -\frac{1}{\sqrt{N^2-1}} & -\frac{N^2-4}{\sqrt{2} \sqrt{N^4-5 N^2+4}} \\
 \frac{1}{\sqrt{N^2-1}} & \frac{1}{\frac{16}{N^2-12}+2} & -\frac{\sqrt{(N-3) N^2 (N+1)}}{2 (N-2) \sqrt{N^2-1}} & \frac{\sqrt{(N-1) N^2 (N+3)}}{2 (N+2) \sqrt{N^2-1}}
   & -\frac{1}{2} & \frac{\sqrt{2} \sqrt{N^2-1}}{\sqrt{N^4-5 N^2+4}} \\
 \frac{\sqrt{(N-3) N^2 (N+1)}}{2 \left(N^2-1\right)} & -\frac{\sqrt{(N-3) N^2 (N+1)} \sqrt{N^2-1}}{2 (N-2) (N-1) (N+1)} & \frac{1}{N-2}+\frac{1}{2-2 N}+\frac{1}{4} &
   \frac{(N-3) \sqrt{(N-1) N^2 (N+3)}}{4 (N-1) \sqrt{(N-3) N^2 (N+1)}} & -\frac{\sqrt{(N-3) N^2 (N+1)}}{2 N \sqrt{N^2-1}} & \frac{(N-3) N \sqrt{N^4-5 N^2+4}}{2
   \sqrt{2} \sqrt{(N-3) N^2 (N+1)} \left(N^2-3 N+2\right)} \\
 \frac{\sqrt{(N-1) N^2 (N+3)}}{2 \left(N^2-1\right)} & \frac{N^2 (N+3) \sqrt{N^2-1}}{2 (N+1) (N+2) \sqrt{(N-1) N^2 (N+3)}} & \frac{\sqrt{(N-3) N^2 (N+1)}
   (N+3)}{\sqrt{(N-1) N^2 (N+3)} (4 N+4)} & \frac{1}{\frac{8 N}{N^2+N+2}+4} & \frac{\sqrt{(N-1) N^2 (N+3)}}{2 N \sqrt{N^2-1}} & \frac{N (N+3) \sqrt{N^4-5 N^2+4}}{2
   \sqrt{2} (N+1) (N+2) \sqrt{(N-1) N^2 (N+3)}} \\
 -\frac{1}{\sqrt{N^2-1}} & -\frac{1}{2} & -\frac{\sqrt{(N-3) N^2 (N+1)}}{2 N \sqrt{N^2-1}} & \frac{\sqrt{(N-1) N^2 (N+3)}}{2 N \sqrt{N^2-1}} & \frac{1}{2} & 0 \\
 -\frac{N^2-4}{\sqrt{2} \sqrt{N^4-5 N^2+4}} & \frac{\sqrt{2} \sqrt{N^2-1}}{\sqrt{N^4-5 N^2+4}} & \frac{\left(\frac{1}{N}+\frac{1}{2}\right) \sqrt{(N-3) N^2
   (N+1)}}{\sqrt{2} \sqrt{N^4-5 N^2+4}} & \frac{\left(\frac{1}{2}-\frac{1}{N}\right) \sqrt{(N-1) N^2 (N+3)}}{\sqrt{2} \sqrt{N^4-5 N^2+4}} & 0 & \frac{1}{2} \\
\end{array}
\right)$}\nonumber\\
\end{align}
where, for $\mathcal M^{O(N)}$, the indices run over the representation $S^+, T^+, Q2^+, Q1^+, A^-, Q3^-$, and for $\mathcal M^{O(N)}$, the indices run over the representation $ S^+, Adj^+, A\bar{A}^+, S\bar{S}^+, +Adj^-, S\bar{A}^-$.

\section{Conformal blocks decomposition}
\label{sec:cb}

All functions appearing in \eqref{eq:non-abelian-4pt} admit a conformal blocks decomposition over parity even/odd operators $O^{\pm}_{
    \Delta,\ell}$ transforming in a given irreducible representation $r$ of $\mathcal G$. It reads as follows:
\begin{align}
    g^{(r)}_{\vec h}(z,\bar z) =& \sum_{\mathcal O^{+}_{
    \Delta,\ell} \in r }\sum_{a,b=1}^{5} \lambda_{SO(3)}^a \lambda_{SO(3)}^b g^{a,b}_{+,{\vec h},\Delta,\ell}(z,\bar{z})\,\nonumber \\ +
  &  \sum_{\mathcal O^{-}_{
    \Delta,\ell} \in r }\sum_{c,d=1}^{4} \tilde\lambda_{SO(3)}^c \tilde\lambda_{SO(3)}^c g^{a,b}_{-,{\vec h},\Delta,\ell}(z,\bar{z})
\end{align}
where $\lambda_{SO(3)}^a$ are OPE coefficients for a given tensor structure in the $SO(3)$ basis for $\langle JJ \mathcal O^{+}_{ \Delta,\ell}\rangle$ discussed in Sec.\ref{sec:embedding-to-so3} and $\tilde\lambda_{SO(3)}^c$ are OPE coefficients for a given tensor structure in the $SO(3)$ basis for $\langle JJ \mathcal O^{-}_{ \Delta,\ell}\rangle$.
They are related to the more familiar embedding basis by
\begin{align}
    \langle JJ \mathcal O^{+}_{
    \Delta,\ell}\rangle & =\sum_{a=1}^5 \lambda_{SO(3)}^{a} T_{SO(3)}^a = \sum_{a=1}^5 \lambda_{EF}^{a} T_{EF}^a \,\\
\langle JJ \mathcal O^{-}_{
    \Delta,\ell}\rangle & =\sum_{c=1}^4 \tilde\lambda_{SO(3)}^{c} \tilde{T}_{SO(3)}^a = \sum_{c=1}^4 \tilde\lambda_{EF}^{a} \tilde{T}_{EF}^a \,
\end{align}
where the two basis of 3pt-functions are related by \eqref{eq:relation-TEF-TOSO3}. 

All the ingredients discussed in this appendix were implemented in
\texttt{simpleboot}\footnote{The program is available at \hyperlink{https://gitlab.com/bootstrapcollaboration/simpleboot}{https://gitlab.com/bootstrapcollaboration/simpleboot}}, which also interfaces with the conformal block generator \texttt{block\_3d} \cite{Erramilli:2020rlr}. The semi-definite problems have been solved using \texttt{SDPB} program \cite{Simmons-Duffin:2015qma}.

\section{Parameters in numerics and gap assumptions}\label{app:param}

We used the following choices for the set of spins at each value of $\Lambda$:
\begin{align}\label{eq:spinsets}
S_{19} &= \{0,\dots,26\}\cup \{49, 52\}\,,\nonumber\\
S_{23} &= \{0,\dots,26\}\cup \{29,30,33,34,37,38,41,42,45,46,49,50\}\,,\nonumber\\
S_{27} &= \{0,\dots,26\}\cup \{29,30,33,34,37,38,41,42,45,46,49,50\}\,,\nonumber\\
\end{align}

The parameters used in the bootstrap softwares are summarized in Table \ref{tab:params}.

\begin{table}
\begin{center}
\begin{tabular}{@{}|c|c|c|c|c@{}}
	\hline
$\Lambda$ &  19 & 23 & 27 \\
{\small\texttt{keptPoleOrder}}& 14 & 18 & 20 \\
{\small\texttt{order}}& 56 & 72 & 80 \\
{\small\texttt{spins}} & $S_{19}$ & $S_{23}$ & $S_{27}$  \\
{\small\texttt{precision}} & 765 & 765 & 765 \\
{\small\texttt{dualityGapThreshold}} &  $10^{-20}$ & $10^{-20}$  & $10^{-20}$ \\
{\small\texttt{primalErrorThreshold}}&  $10^{-25}$ & $10^{-25}$ & $10^{-20}$ \\
{\small\texttt{dualErrorThreshold}} & $10^{-25}$ & $10^{-25}$ & $10^{-20}$\\ 
{\small\texttt{initialMatrixScalePrimal}} & $10^{20}$& $10^{20}$ & $10^{20}$\\
{\small\texttt{initialMatrixScaleDual}} &  $10^{20}$& $10^{20}$ & $10^{20}$\\
{\small\texttt{maxComplementarity}} &  $10^{100}$& $10^{100}$ & $10^{100}$\\
{\small\texttt{detectPrimalFeasibleJump}} & yes & yes & yes\\
{\small\texttt{detectDualFeasibleJump}} & yes & yes & yes\\
 \hline
\end{tabular}
\caption{\label{tab:params}Parameters used in \texttt{blocks\_3d} and \texttt{SDPB}. For feasibility computations, we specified ``--detectPrimalFeasibleJump --detectDualFeasibleJump" in \texttt{SDPB}. We used default parameters if unspecified above.  The sets $S_{\Lambda}$ are defined in (\ref{eq:spinsets}).}
\end{center}
\end{table}

For the computation of Figure \ref{fig:intervalpositivity}, we demand that many operators should exist within an interval $(a-0.1,a+0.1)$, where $a$ is the large-$N$ prediction, and put gaps after those individual operators. The individual operators and gaps are summarized in Table \ref{tab:gaps}. 

\begin{table}[H]
\begin{center}
\begin{tabular}{@{}|c|c|c|@{}}
	\hline
\text{Representation} & $\Delta$ & \text{Gap}  \\
$(+,S,\ell=0)$ & 4 & 4.5 \\
$(+,S,\ell=2)$ & 4 & 2 \\
$(+,Adj^+,\ell=0)$ & 4 & 4.5 \\
$(+,Adj^+,\ell=2)$ & 5 & 5.5 \\
$(+,A\bar{A},\ell=0)$ & 6 & 6.5 \\
$(+,S\bar{S},\ell=0)$ & 4, 6 & 6.5 \\
$(+,S\bar{S},\ell=2)$ & 6 & 6.5 \\
$(+,Adj^-,\ell=1)$ &  & 3.5 \\
$(+,S\bar{A},\ell=1)$ & 5 & 5.5 \\
$(-,S,\ell=0)$ & 2, 5 & 5.5 \\
$(-,Adj^+,\ell=0)$ & 2, 5 & 5.5 \\
$(-,Adj^+,\ell=2)$ &   & 3.5 \\
$(-,A\bar{A},\ell=0)$ & 5 & 5.5 \\
$(-,A\bar{A},\ell=2)$ &   & 4.5 \\
$(-,S\bar{S},\ell=0)$ & 6  & 6.5 \\
$(-,S\bar{S},\ell=2)$ & 5  & 5.5 \\
$(-,Adj^-,\ell=1)$ &    & 3.5 \\
$(-,Adj^-,\ell=2)$ &    & 4.5 \\
$(-,S\bar{A},\ell=1)$ &    & 3.5 \\
$(-,S\bar{A},\ell=2)$ &    & 4.5 \\
 \hline
\end{tabular}
\label{tab:gaps}
\caption{The bootstrap condition for the computation of Figure \ref{fig:intervalpositivity}.}
\end{center}
\end{table}

\addtocontents{toc}{\protect\enlargethispage{\baselineskip}}

\newpage
\bibliography{ref.bib}

\begin{thebibliography}{62}%
\makeatletter
\providecommand \@ifxundefined [1]{%
 \@ifx{#1\undefined}
}%
\providecommand \@ifnum [1]{%
 \ifnum #1\expandafter \@firstoftwo
 \else \expandafter \@secondoftwo
 \fi
}%
\providecommand \@ifx [1]{%
 \ifx #1\expandafter \@firstoftwo
 \else \expandafter \@secondoftwo
 \fi
}%
\providecommand \natexlab [1]{#1}%
\providecommand \enquote  [1]{``#1''}%
\providecommand \bibnamefont  [1]{#1}%
\providecommand \bibfnamefont [1]{#1}%
\providecommand \citenamefont [1]{#1}%
\providecommand \href@noop [0]{\@secondoftwo}%
\providecommand \href [0]{\begingroup \@sanitize@url \@href}%
\providecommand \@href[1]{\@@startlink{#1}\@@href}%
\providecommand \@@href[1]{\endgroup#1\@@endlink}%
\providecommand \@sanitize@url [0]{\catcode `\\12\catcode `\$12\catcode
  `\&12\catcode `\#12\catcode `\^12\catcode `\_12\catcode `\%12\relax}%
\providecommand \@@startlink[1]{}%
\providecommand \@@endlink[0]{}%
\providecommand \url  [0]{\begingroup\@sanitize@url \@url }%
\providecommand \@url [1]{\endgroup\@href {#1}{\urlprefix }}%
\providecommand \urlprefix  [0]{URL }%
\providecommand \Eprint [0]{\href }%
\providecommand \doibase [0]{http://dx.doi.org/}%
\providecommand \selectlanguage [0]{\@gobble}%
\providecommand \bibinfo  [0]{\@secondoftwo}%
\providecommand \bibfield  [0]{\@secondoftwo}%
\providecommand \translation [1]{[#1]}%
\providecommand \BibitemOpen [0]{}%
\providecommand \bibitemStop [0]{}%
\providecommand \bibitemNoStop [0]{.\EOS\space}%
\providecommand \EOS [0]{\spacefactor3000\relax}%
\providecommand \BibitemShut  [1]{\csname bibitem#1\endcsname}%
\let\auto@bib@innerbib\@empty
%</preamble>
\bibitem [{\citenamefont {Rattazzi}\ \emph {et~al.}(2008)\citenamefont
  {Rattazzi}, \citenamefont {Rychkov}, \citenamefont {Tonni},\ and\
  \citenamefont {Vichi}}]{Rattazzi:2008pe}%
  \BibitemOpen
  \bibfield  {author} {\bibinfo {author} {\bibfnamefont {Riccardo}\
  \bibnamefont {Rattazzi}}, \bibinfo {author} {\bibfnamefont {Vyacheslav~S.}\
  \bibnamefont {Rychkov}}, \bibinfo {author} {\bibfnamefont {Erik}\
  \bibnamefont {Tonni}}, \ and\ \bibinfo {author} {\bibfnamefont {Alessandro}\
  \bibnamefont {Vichi}},\ }\bibfield  {title} {\enquote {\bibinfo {title}
  {{Bounding scalar operator dimensions in 4D CFT}},}\ }\href {\doibase
  10.1088/1126-6708/2008/12/031} {\bibfield  {journal} {\bibinfo  {journal}
  {JHEP}\ }\textbf {\bibinfo {volume} {0812}},\ \bibinfo {pages} {031}
  (\bibinfo {year} {2008})},\ \Eprint {http://arxiv.org/abs/0807.0004}
  {arXiv:0807.0004 [hep-th]} \BibitemShut {NoStop}%
%%CITATION = ARXIV:0807.0004;%%
\bibitem [{\citenamefont {Rychkov}\ and\ \citenamefont
  {Vichi}(2009)}]{Rychkov:2009ij}%
  \BibitemOpen
  \bibfield  {author} {\bibinfo {author} {\bibfnamefont {Vyacheslav~S.}\
  \bibnamefont {Rychkov}}\ and\ \bibinfo {author} {\bibfnamefont {Alessandro}\
  \bibnamefont {Vichi}},\ }\bibfield  {title} {\enquote {\bibinfo {title}
  {{Universal Constraints on Conformal Operator Dimensions}},}\ }\href
  {\doibase 10.1103/PhysRevD.80.045006} {\bibfield  {journal} {\bibinfo
  {journal} {Phys.Rev.}\ }\textbf {\bibinfo {volume} {D80}},\ \bibinfo {pages}
  {045006} (\bibinfo {year} {2009})},\ \Eprint {http://arxiv.org/abs/0905.2211}
  {arXiv:0905.2211 [hep-th]} \BibitemShut {NoStop}%
%%CITATION = ARXIV:0905.2211;%%
\bibitem [{\citenamefont {Kos}\ \emph {et~al.}(2016)\citenamefont {Kos},
  \citenamefont {Poland}, \citenamefont {Simmons-Duffin},\ and\ \citenamefont
  {Vichi}}]{Kos:2016ysd}%
  \BibitemOpen
  \bibfield  {author} {\bibinfo {author} {\bibfnamefont {Filip}\ \bibnamefont
  {Kos}}, \bibinfo {author} {\bibfnamefont {David}\ \bibnamefont {Poland}},
  \bibinfo {author} {\bibfnamefont {David}\ \bibnamefont {Simmons-Duffin}}, \
  and\ \bibinfo {author} {\bibfnamefont {Alessandro}\ \bibnamefont {Vichi}},\
  }\bibfield  {title} {\enquote {\bibinfo {title} {{Precision Islands in the
  Ising and $O(N)$ Models}},}\ }\href {\doibase 10.1007/JHEP08(2016)036}
  {\bibfield  {journal} {\bibinfo  {journal} {JHEP}\ }\textbf {\bibinfo
  {volume} {08}},\ \bibinfo {pages} {036} (\bibinfo {year} {2016})},\ \Eprint
  {http://arxiv.org/abs/1603.04436} {arXiv:1603.04436 [hep-th]} \BibitemShut
  {NoStop}%
%%CITATION = ARXIV:1603.04436;%%
\bibitem [{\citenamefont {Atanasov}\ \emph {et~al.}(2022)\citenamefont
  {Atanasov}, \citenamefont {Hillman}, \citenamefont {Poland}, \citenamefont
  {Rong},\ and\ \citenamefont {Su}}]{Atanasov:2022bpi}%
  \BibitemOpen
  \bibfield  {author} {\bibinfo {author} {\bibfnamefont {Alexander}\
  \bibnamefont {Atanasov}}, \bibinfo {author} {\bibfnamefont {Aaron}\
  \bibnamefont {Hillman}}, \bibinfo {author} {\bibfnamefont {David}\
  \bibnamefont {Poland}}, \bibinfo {author} {\bibfnamefont {Junchen}\
  \bibnamefont {Rong}}, \ and\ \bibinfo {author} {\bibfnamefont {Ning}\
  \bibnamefont {Su}},\ }\bibfield  {title} {\enquote {\bibinfo {title}
  {{Precision bootstrap for the $ \mathcal{N} $ = 1 super-Ising model}},}\
  }\href {\doibase 10.1007/JHEP08(2022)136} {\bibfield  {journal} {\bibinfo
  {journal} {JHEP}\ }\textbf {\bibinfo {volume} {08}},\ \bibinfo {pages} {136}
  (\bibinfo {year} {2022})},\ \Eprint {http://arxiv.org/abs/2201.02206}
  {arXiv:2201.02206 [hep-th]} \BibitemShut {NoStop}%
\bibitem [{\citenamefont {Chester}\ \emph {et~al.}(2020)\citenamefont
  {Chester}, \citenamefont {Landry}, \citenamefont {Liu}, \citenamefont
  {Poland}, \citenamefont {Simmons-Duffin}, \citenamefont {Su},\ and\
  \citenamefont {Vichi}}]{Chester:2019ifh}%
  \BibitemOpen
  \bibfield  {author} {\bibinfo {author} {\bibfnamefont {Shai~M.}\ \bibnamefont
  {Chester}}, \bibinfo {author} {\bibfnamefont {Walter}\ \bibnamefont
  {Landry}}, \bibinfo {author} {\bibfnamefont {Junyu}\ \bibnamefont {Liu}},
  \bibinfo {author} {\bibfnamefont {David}\ \bibnamefont {Poland}}, \bibinfo
  {author} {\bibfnamefont {David}\ \bibnamefont {Simmons-Duffin}}, \bibinfo
  {author} {\bibfnamefont {Ning}\ \bibnamefont {Su}}, \ and\ \bibinfo {author}
  {\bibfnamefont {Alessandro}\ \bibnamefont {Vichi}},\ }\bibfield  {title}
  {\enquote {\bibinfo {title} {{Carving out OPE space and precise $O(2)$ model
  critical exponents}},}\ }\href {\doibase 10.1007/JHEP06(2020)142} {\bibfield
  {journal} {\bibinfo  {journal} {JHEP}\ }\textbf {\bibinfo {volume} {06}},\
  \bibinfo {pages} {142} (\bibinfo {year} {2020})},\ \Eprint
  {http://arxiv.org/abs/1912.03324} {arXiv:1912.03324 [hep-th]} \BibitemShut
  {NoStop}%
\bibitem [{\citenamefont {Chester}\ \emph {et~al.}(2021)\citenamefont
  {Chester}, \citenamefont {Landry}, \citenamefont {Liu}, \citenamefont
  {Poland}, \citenamefont {Simmons-Duffin}, \citenamefont {Su},\ and\
  \citenamefont {Vichi}}]{Chester:2020iyt}%
  \BibitemOpen
  \bibfield  {author} {\bibinfo {author} {\bibfnamefont {Shai~M.}\ \bibnamefont
  {Chester}}, \bibinfo {author} {\bibfnamefont {Walter}\ \bibnamefont
  {Landry}}, \bibinfo {author} {\bibfnamefont {Junyu}\ \bibnamefont {Liu}},
  \bibinfo {author} {\bibfnamefont {David}\ \bibnamefont {Poland}}, \bibinfo
  {author} {\bibfnamefont {David}\ \bibnamefont {Simmons-Duffin}}, \bibinfo
  {author} {\bibfnamefont {Ning}\ \bibnamefont {Su}}, \ and\ \bibinfo {author}
  {\bibfnamefont {Alessandro}\ \bibnamefont {Vichi}},\ }\bibfield  {title}
  {\enquote {\bibinfo {title} {{Bootstrapping Heisenberg magnets and their
  cubic instability}},}\ }\href {\doibase 10.1103/PhysRevD.104.105013}
  {\bibfield  {journal} {\bibinfo  {journal} {Phys. Rev. D}\ }\textbf {\bibinfo
  {volume} {104}},\ \bibinfo {pages} {105013} (\bibinfo {year} {2021})},\
  \Eprint {http://arxiv.org/abs/2011.14647} {arXiv:2011.14647 [hep-th]}
  \BibitemShut {NoStop}%
\bibitem [{\citenamefont {Erramilli}\ \emph {et~al.}(2022)\citenamefont
  {Erramilli}, \citenamefont {Iliesiu}, \citenamefont {Kravchuk}, \citenamefont
  {Liu}, \citenamefont {Poland},\ and\ \citenamefont
  {Simmons-Duffin}}]{Erramilli:2022kgp}%
  \BibitemOpen
  \bibfield  {author} {\bibinfo {author} {\bibfnamefont {Rajeev~S.}\
  \bibnamefont {Erramilli}}, \bibinfo {author} {\bibfnamefont {Luca~V.}\
  \bibnamefont {Iliesiu}}, \bibinfo {author} {\bibfnamefont {Petr}\
  \bibnamefont {Kravchuk}}, \bibinfo {author} {\bibfnamefont {Aike}\
  \bibnamefont {Liu}}, \bibinfo {author} {\bibfnamefont {David}\ \bibnamefont
  {Poland}}, \ and\ \bibinfo {author} {\bibfnamefont {David}\ \bibnamefont
  {Simmons-Duffin}},\ }\bibfield  {title} {\enquote {\bibinfo {title} {{The
  Gross-Neveu-Yukawa Archipelago}},}\ }\href@noop {} {\  (\bibinfo {year}
  {2022})},\ \Eprint {http://arxiv.org/abs/2210.02492} {arXiv:2210.02492
  [hep-th]} \BibitemShut {NoStop}%
\bibitem [{\citenamefont {Poland}\ \emph {et~al.}(2019)\citenamefont {Poland},
  \citenamefont {Rychkov},\ and\ \citenamefont {Vichi}}]{Poland:2018epd}%
  \BibitemOpen
  \bibfield  {author} {\bibinfo {author} {\bibfnamefont {David}\ \bibnamefont
  {Poland}}, \bibinfo {author} {\bibfnamefont {Slava}\ \bibnamefont {Rychkov}},
  \ and\ \bibinfo {author} {\bibfnamefont {Alessandro}\ \bibnamefont {Vichi}},\
  }\bibfield  {title} {\enquote {\bibinfo {title} {{The Conformal Bootstrap:
  Theory, Numerical Techniques, and Applications}},}\ }\href {\doibase
  10.1103/RevModPhys.91.015002} {\bibfield  {journal} {\bibinfo  {journal}
  {Rev. Mod. Phys.}\ }\textbf {\bibinfo {volume} {91}},\ \bibinfo {pages}
  {15002} (\bibinfo {year} {2019})},\ \Eprint {http://arxiv.org/abs/1805.04405}
  {arXiv:1805.04405 [hep-th]} \BibitemShut {NoStop}%
%%CITATION = ARXIV:1805.04405;%%
\bibitem [{\citenamefont {Chester}\ and\ \citenamefont
  {Pufu}(2016{\natexlab{a}})}]{Chester:2016wrc}%
  \BibitemOpen
  \bibfield  {author} {\bibinfo {author} {\bibfnamefont {Shai~M.}\ \bibnamefont
  {Chester}}\ and\ \bibinfo {author} {\bibfnamefont {Silviu~S.}\ \bibnamefont
  {Pufu}},\ }\bibfield  {title} {\enquote {\bibinfo {title} {{Towards
  bootstrapping QED$_{3}$}},}\ }\href {\doibase 10.1007/JHEP08(2016)019}
  {\bibfield  {journal} {\bibinfo  {journal} {JHEP}\ }\textbf {\bibinfo
  {volume} {08}},\ \bibinfo {pages} {019} (\bibinfo {year}
  {2016}{\natexlab{a}})},\ \Eprint {http://arxiv.org/abs/1601.03476}
  {arXiv:1601.03476 [hep-th]} \BibitemShut {NoStop}%
%%CITATION = ARXIV:1601.03476;%%
\bibitem [{\citenamefont {Li}(2018)}]{Li:2018lyb}%
  \BibitemOpen
  \bibfield  {author} {\bibinfo {author} {\bibfnamefont {Zhijin}\ \bibnamefont
  {Li}},\ }\bibfield  {title} {\enquote {\bibinfo {title} {{Solving QED$_3$
  with Conformal Bootstrap}},}\ }\href@noop {} {\  (\bibinfo {year} {2018})},\
  \Eprint {http://arxiv.org/abs/1812.09281} {arXiv:1812.09281 [hep-th]}
  \BibitemShut {NoStop}%
%%CITATION = ARXIV:1812.09281;%%
\bibitem [{\citenamefont {He}\ \emph {et~al.}(2021{\natexlab{a}})\citenamefont
  {He}, \citenamefont {Rong},\ and\ \citenamefont {Su}}]{He:2020azu}%
  \BibitemOpen
  \bibfield  {author} {\bibinfo {author} {\bibfnamefont {Yin-Chen}\
  \bibnamefont {He}}, \bibinfo {author} {\bibfnamefont {Junchen}\ \bibnamefont
  {Rong}}, \ and\ \bibinfo {author} {\bibfnamefont {Ning}\ \bibnamefont {Su}},\
  }\bibfield  {title} {\enquote {\bibinfo {title} {{Non-Wilson-Fisher kinks of
  $O(N)$ numerical bootstrap: from the deconfined phase transition to a
  putative new family of CFTs}},}\ }\href {\doibase
  10.21468/SciPostPhys.10.5.115} {\bibfield  {journal} {\bibinfo  {journal}
  {SciPost Phys.}\ }\textbf {\bibinfo {volume} {10}},\ \bibinfo {pages} {115}
  (\bibinfo {year} {2021}{\natexlab{a}})},\ \Eprint
  {http://arxiv.org/abs/2005.04250} {arXiv:2005.04250 [hep-th]} \BibitemShut
  {NoStop}%
\bibitem [{\citenamefont {Li}\ and\ \citenamefont {Poland}(2021)}]{Li:2020bnb}%
  \BibitemOpen
  \bibfield  {author} {\bibinfo {author} {\bibfnamefont {Zhijin}\ \bibnamefont
  {Li}}\ and\ \bibinfo {author} {\bibfnamefont {David}\ \bibnamefont
  {Poland}},\ }\bibfield  {title} {\enquote {\bibinfo {title} {{Searching for
  gauge theories with the conformal bootstrap}},}\ }\href {\doibase
  10.1007/JHEP03(2021)172} {\bibfield  {journal} {\bibinfo  {journal} {JHEP}\
  }\textbf {\bibinfo {volume} {03}},\ \bibinfo {pages} {172} (\bibinfo {year}
  {2021})},\ \Eprint {http://arxiv.org/abs/2005.01721} {arXiv:2005.01721
  [hep-th]} \BibitemShut {NoStop}%
\bibitem [{\citenamefont {He}\ \emph {et~al.}(2021{\natexlab{b}})\citenamefont
  {He}, \citenamefont {Rong},\ and\ \citenamefont {Su}}]{He:2021xvg}%
  \BibitemOpen
  \bibfield  {author} {\bibinfo {author} {\bibfnamefont {Yin-Chen}\
  \bibnamefont {He}}, \bibinfo {author} {\bibfnamefont {Junchen}\ \bibnamefont
  {Rong}}, \ and\ \bibinfo {author} {\bibfnamefont {Ning}\ \bibnamefont {Su}},\
  }\bibfield  {title} {\enquote {\bibinfo {title} {{A roadmap for bootstrapping
  critical gauge theories: decoupling operators of conformal field theories in
  $d>2$ dimensions}},}\ }\href {\doibase 10.21468/SciPostPhys.11.6.111}
  {\bibfield  {journal} {\bibinfo  {journal} {SciPost Phys.}\ }\textbf
  {\bibinfo {volume} {11}},\ \bibinfo {pages} {111} (\bibinfo {year}
  {2021}{\natexlab{b}})},\ \Eprint {http://arxiv.org/abs/2101.07262}
  {arXiv:2101.07262 [hep-th]} \BibitemShut {NoStop}%
\bibitem [{\citenamefont {Albayrak}\ \emph {et~al.}(2022)\citenamefont
  {Albayrak}, \citenamefont {Erramilli}, \citenamefont {Li}, \citenamefont
  {Poland},\ and\ \citenamefont {Xin}}]{Albayrak:2021xtd}%
  \BibitemOpen
  \bibfield  {author} {\bibinfo {author} {\bibfnamefont {Soner}\ \bibnamefont
  {Albayrak}}, \bibinfo {author} {\bibfnamefont {Rajeev~S.}\ \bibnamefont
  {Erramilli}}, \bibinfo {author} {\bibfnamefont {Zhijin}\ \bibnamefont {Li}},
  \bibinfo {author} {\bibfnamefont {David}\ \bibnamefont {Poland}}, \ and\
  \bibinfo {author} {\bibfnamefont {Yuan}\ \bibnamefont {Xin}},\ }\bibfield
  {title} {\enquote {\bibinfo {title} {{Bootstrapping $N_f$=4 conformal
  QED$_3$}},}\ }\href {\doibase 10.1103/PhysRevD.105.085008} {\bibfield
  {journal} {\bibinfo  {journal} {Phys. Rev. D}\ }\textbf {\bibinfo {volume}
  {105}},\ \bibinfo {pages} {085008} (\bibinfo {year} {2022})},\ \Eprint
  {http://arxiv.org/abs/2112.02106} {arXiv:2112.02106 [hep-th]} \BibitemShut
  {NoStop}%
\bibitem [{\citenamefont {He}\ \emph {et~al.}(2022)\citenamefont {He},
  \citenamefont {Rong},\ and\ \citenamefont {Su}}]{He:2021sto}%
  \BibitemOpen
  \bibfield  {author} {\bibinfo {author} {\bibfnamefont {Yin-Chen}\
  \bibnamefont {He}}, \bibinfo {author} {\bibfnamefont {Junchen}\ \bibnamefont
  {Rong}}, \ and\ \bibinfo {author} {\bibfnamefont {Ning}\ \bibnamefont {Su}},\
  }\bibfield  {title} {\enquote {\bibinfo {title} {{Conformal bootstrap bounds
  for the $U(1)$ Dirac spin liquid and $N=7$ Stiefel liquid}},}\ }\href
  {\doibase 10.21468/SciPostPhys.13.2.014} {\bibfield  {journal} {\bibinfo
  {journal} {SciPost Phys.}\ }\textbf {\bibinfo {volume} {13}},\ \bibinfo
  {pages} {014} (\bibinfo {year} {2022})},\ \Eprint
  {http://arxiv.org/abs/2107.14637} {arXiv:2107.14637 [cond-mat.str-el]}
  \BibitemShut {NoStop}%
\bibitem [{\citenamefont {{Li}}(2022)}]{Li:2021emd}%
  \BibitemOpen
  \bibfield  {author} {\bibinfo {author} {\bibfnamefont {Zhijin}\ \bibnamefont
  {{Li}}},\ }\bibfield  {title} {\enquote {\bibinfo {title} {{Conformality and
  self-duality of N$_{f}$ = 2 QED$_{3}$}},}\ }\href {\doibase
  10.1016/j.physletb.2022.137192} {\bibfield  {journal} {\bibinfo  {journal}
  {Physics Letters B}\ }\textbf {\bibinfo {volume} {831}},\ \bibinfo {eid}
  {137192} (\bibinfo {year} {2022})},\ \Eprint
  {http://arxiv.org/abs/2107.09020} {arXiv:2107.09020 [hep-th]} \BibitemShut
  {NoStop}%
\bibitem [{\citenamefont {Reehorst}\ \emph {et~al.}(2020)\citenamefont
  {Reehorst}, \citenamefont {Refinetti},\ and\ \citenamefont
  {Vichi}}]{Reehorst:2020phk}%
  \BibitemOpen
  \bibfield  {author} {\bibinfo {author} {\bibfnamefont {Marten}\ \bibnamefont
  {Reehorst}}, \bibinfo {author} {\bibfnamefont {Maria}\ \bibnamefont
  {Refinetti}}, \ and\ \bibinfo {author} {\bibfnamefont {Alessandro}\
  \bibnamefont {Vichi}},\ }\bibfield  {title} {\enquote {\bibinfo {title}
  {{Bootstrapping traceless symmetric $O(N)$ scalars}},}\ }\href@noop {} {\
  (\bibinfo {year} {2020})},\ \Eprint {http://arxiv.org/abs/2012.08533}
  {arXiv:2012.08533 [hep-th]} \BibitemShut {NoStop}%
\bibitem [{\citenamefont {Manenti}\ and\ \citenamefont
  {Vichi}(2021)}]{Manenti:2021elk}%
  \BibitemOpen
  \bibfield  {author} {\bibinfo {author} {\bibfnamefont {Andrea}\ \bibnamefont
  {Manenti}}\ and\ \bibinfo {author} {\bibfnamefont {Alessandro}\ \bibnamefont
  {Vichi}},\ }\bibfield  {title} {\enquote {\bibinfo {title} {{Exploring
  $SU(N)$ adjoint correlators in $3d$}},}\ }\href@noop {} {\  (\bibinfo {year}
  {2021})},\ \Eprint {http://arxiv.org/abs/2101.07318} {arXiv:2101.07318
  [hep-th]} \BibitemShut {NoStop}%
\bibitem [{\citenamefont {Affleck}\ and\ \citenamefont
  {Marston}(1987)}]{affleck1988large}%
  \BibitemOpen
  \bibfield  {author} {\bibinfo {author} {\bibfnamefont {Ian}\ \bibnamefont
  {Affleck}}\ and\ \bibinfo {author} {\bibfnamefont {J.~Brad}\ \bibnamefont
  {Marston}},\ }\bibfield  {title} {\enquote {\bibinfo {title} {{Large n limit
  of the Heisenberg-Hubbard model: Implications for high t(c)
  supconductors}},}\ }\href {\doibase 10.1103/PhysRevB.37.3774} {\bibfield
  {journal} {\bibinfo  {journal} {Phys. Rev. B}\ }\textbf {\bibinfo {volume}
  {37}},\ \bibinfo {pages} {3774--3777} (\bibinfo {year} {1987})}\BibitemShut
  {NoStop}%
\bibitem [{\citenamefont {Wen}\ and\ \citenamefont
  {Lee}(1996)}]{wen1996theory}%
  \BibitemOpen
  \bibfield  {author} {\bibinfo {author} {\bibfnamefont {Xiao-Gang}\
  \bibnamefont {Wen}}\ and\ \bibinfo {author} {\bibfnamefont {Patrick~A.}\
  \bibnamefont {Lee}},\ }\bibfield  {title} {\enquote {\bibinfo {title}
  {{Theory of Underdoped Cuprates}},}\ }\href {\doibase
  10.1103/PhysRevLett.76.503} {\bibfield  {journal} {\bibinfo  {journal} {Phys.
  Rev. Lett.}\ }\textbf {\bibinfo {volume} {76}},\ \bibinfo {pages} {503--506}
  (\bibinfo {year} {1996})},\ \Eprint {http://arxiv.org/abs/cond-mat/9506065}
  {arXiv:cond-mat/9506065} \BibitemShut {NoStop}%
\bibitem [{\citenamefont {Hastings}(2000)}]{hastings2000dirac}%
  \BibitemOpen
  \bibfield  {author} {\bibinfo {author} {\bibfnamefont {MB}~\bibnamefont
  {Hastings}},\ }\bibfield  {title} {\enquote {\bibinfo {title} {Dirac
  structure, rvb, and goldstone modes in the kagom{\'e} antiferromagnet},}\
  }\href@noop {} {\bibfield  {journal} {\bibinfo  {journal} {Physical Review
  B}\ }\textbf {\bibinfo {volume} {63}},\ \bibinfo {pages} {014413} (\bibinfo
  {year} {2000})}\BibitemShut {NoStop}%
\bibitem [{\citenamefont {Hermele}\ \emph {et~al.}(2005)\citenamefont
  {Hermele}, \citenamefont {Senthil},\ and\ \citenamefont
  {Fisher}}]{hermele2005algebraic}%
  \BibitemOpen
  \bibfield  {author} {\bibinfo {author} {\bibfnamefont {Michael}\ \bibnamefont
  {Hermele}}, \bibinfo {author} {\bibfnamefont {T}~\bibnamefont {Senthil}}, \
  and\ \bibinfo {author} {\bibfnamefont {Matthew~PA}\ \bibnamefont {Fisher}},\
  }\bibfield  {title} {\enquote {\bibinfo {title} {Algebraic spin liquid as the
  mother of many competing orders},}\ }\href@noop {} {\bibfield  {journal}
  {\bibinfo  {journal} {Physical Review B}\ }\textbf {\bibinfo {volume} {72}},\
  \bibinfo {pages} {104404} (\bibinfo {year} {2005})}\BibitemShut {NoStop}%
\bibitem [{\citenamefont {Hermele}\ \emph {et~al.}(2008)\citenamefont
  {Hermele}, \citenamefont {Ran}, \citenamefont {Lee},\ and\ \citenamefont
  {Wen}}]{hermele2008properties}%
  \BibitemOpen
  \bibfield  {author} {\bibinfo {author} {\bibfnamefont {Michael}\ \bibnamefont
  {Hermele}}, \bibinfo {author} {\bibfnamefont {Ying}\ \bibnamefont {Ran}},
  \bibinfo {author} {\bibfnamefont {Patrick~A}\ \bibnamefont {Lee}}, \ and\
  \bibinfo {author} {\bibfnamefont {Xiao-Gang}\ \bibnamefont {Wen}},\
  }\bibfield  {title} {\enquote {\bibinfo {title} {Properties of an algebraic
  spin liquid on the kagome lattice},}\ }\href@noop {} {\bibfield  {journal}
  {\bibinfo  {journal} {Physical Review B}\ }\textbf {\bibinfo {volume} {77}},\
  \bibinfo {pages} {224413} (\bibinfo {year} {2008})}\BibitemShut {NoStop}%
\bibitem [{\citenamefont {Song}\ \emph {et~al.}(2020)\citenamefont {Song},
  \citenamefont {He}, \citenamefont {Vishwanath},\ and\ \citenamefont
  {Wang}}]{song2020spinon}%
  \BibitemOpen
  \bibfield  {author} {\bibinfo {author} {\bibfnamefont {Xue-Yang}\
  \bibnamefont {Song}}, \bibinfo {author} {\bibfnamefont {Yin-Chen}\
  \bibnamefont {He}}, \bibinfo {author} {\bibfnamefont {Ashvin}\ \bibnamefont
  {Vishwanath}}, \ and\ \bibinfo {author} {\bibfnamefont {Chong}\ \bibnamefont
  {Wang}},\ }\bibfield  {title} {\enquote {\bibinfo {title} {From spinon band
  topology to the symmetry quantum numbers of monopoles in dirac spin
  liquids},}\ }\href@noop {} {\bibfield  {journal} {\bibinfo  {journal}
  {Physical Review X}\ }\textbf {\bibinfo {volume} {10}},\ \bibinfo {pages}
  {011033} (\bibinfo {year} {2020})}\BibitemShut {NoStop}%
\bibitem [{\citenamefont {Song}\ \emph {et~al.}(2019)\citenamefont {Song},
  \citenamefont {Wang}, \citenamefont {Vishwanath},\ and\ \citenamefont
  {He}}]{song2019unifying}%
  \BibitemOpen
  \bibfield  {author} {\bibinfo {author} {\bibfnamefont {Xue-Yang}\
  \bibnamefont {Song}}, \bibinfo {author} {\bibfnamefont {Chong}\ \bibnamefont
  {Wang}}, \bibinfo {author} {\bibfnamefont {Ashvin}\ \bibnamefont
  {Vishwanath}}, \ and\ \bibinfo {author} {\bibfnamefont {Yin-Chen}\
  \bibnamefont {He}},\ }\bibfield  {title} {\enquote {\bibinfo {title}
  {Unifying description of competing orders in two-dimensional quantum
  magnets},}\ }\href@noop {} {\bibfield  {journal} {\bibinfo  {journal} {Nature
  communications}\ }\textbf {\bibinfo {volume} {10}},\ \bibinfo {pages} {1--12}
  (\bibinfo {year} {2019})}\BibitemShut {NoStop}%
\bibitem [{\citenamefont {Kosterlitz}\ and\ \citenamefont
  {Thouless}(1973)}]{kosterlitz1973ordering}%
  \BibitemOpen
  \bibfield  {author} {\bibinfo {author} {\bibfnamefont {John~Michael}\
  \bibnamefont {Kosterlitz}}\ and\ \bibinfo {author} {\bibfnamefont
  {David~James}\ \bibnamefont {Thouless}},\ }\bibfield  {title} {\enquote
  {\bibinfo {title} {Ordering, metastability and phase transitions in
  two-dimensional systems},}\ }\href@noop {} {\bibfield  {journal} {\bibinfo
  {journal} {Journal of Physics C: Solid State Physics}\ }\textbf {\bibinfo
  {volume} {6}},\ \bibinfo {pages} {1181} (\bibinfo {year} {1973})}\BibitemShut
  {NoStop}%
\bibitem [{\citenamefont {Ran}\ \emph {et~al.}(2007)\citenamefont {Ran},
  \citenamefont {Hermele}, \citenamefont {Lee},\ and\ \citenamefont
  {Wen}}]{ran2007projected}%
  \BibitemOpen
  \bibfield  {author} {\bibinfo {author} {\bibfnamefont {Ying}\ \bibnamefont
  {Ran}}, \bibinfo {author} {\bibfnamefont {Michael}\ \bibnamefont {Hermele}},
  \bibinfo {author} {\bibfnamefont {Patrick~A}\ \bibnamefont {Lee}}, \ and\
  \bibinfo {author} {\bibfnamefont {Xiao-Gang}\ \bibnamefont {Wen}},\
  }\bibfield  {title} {\enquote {\bibinfo {title} {Projected-wave-function
  study of the spin-1/2 heisenberg model on the kagom{\'e} lattice},}\
  }\href@noop {} {\bibfield  {journal} {\bibinfo  {journal} {Physical review
  letters}\ }\textbf {\bibinfo {volume} {98}},\ \bibinfo {pages} {117205}
  (\bibinfo {year} {2007})}\BibitemShut {NoStop}%
\bibitem [{\citenamefont {Iqbal}\ \emph {et~al.}(2013)\citenamefont {Iqbal},
  \citenamefont {Becca}, \citenamefont {Sorella},\ and\ \citenamefont
  {Poilblanc}}]{iqbal2013gapless}%
  \BibitemOpen
  \bibfield  {author} {\bibinfo {author} {\bibfnamefont {Yasir}\ \bibnamefont
  {Iqbal}}, \bibinfo {author} {\bibfnamefont {Federico}\ \bibnamefont {Becca}},
  \bibinfo {author} {\bibfnamefont {Sandro}\ \bibnamefont {Sorella}}, \ and\
  \bibinfo {author} {\bibfnamefont {Didier}\ \bibnamefont {Poilblanc}},\
  }\bibfield  {title} {\enquote {\bibinfo {title} {Gapless spin-liquid phase in
  the kagome spin-1 2 heisenberg antiferromagnet},}\ }\href@noop {} {\bibfield
  {journal} {\bibinfo  {journal} {Physical Review B}\ }\textbf {\bibinfo
  {volume} {87}},\ \bibinfo {pages} {060405} (\bibinfo {year}
  {2013})}\BibitemShut {NoStop}%
\bibitem [{\citenamefont {Iqbal}\ \emph {et~al.}(2016)\citenamefont {Iqbal},
  \citenamefont {Hu}, \citenamefont {Thomale}, \citenamefont {Poilblanc},\ and\
  \citenamefont {Becca}}]{iqbal2016spin}%
  \BibitemOpen
  \bibfield  {author} {\bibinfo {author} {\bibfnamefont {Yasir}\ \bibnamefont
  {Iqbal}}, \bibinfo {author} {\bibfnamefont {Wen-Jun}\ \bibnamefont {Hu}},
  \bibinfo {author} {\bibfnamefont {Ronny}\ \bibnamefont {Thomale}}, \bibinfo
  {author} {\bibfnamefont {Didier}\ \bibnamefont {Poilblanc}}, \ and\ \bibinfo
  {author} {\bibfnamefont {Federico}\ \bibnamefont {Becca}},\ }\bibfield
  {title} {\enquote {\bibinfo {title} {Spin liquid nature in the heisenberg j
  1- j 2 triangular antiferromagnet},}\ }\href@noop {} {\bibfield  {journal}
  {\bibinfo  {journal} {Physical Review B}\ }\textbf {\bibinfo {volume} {93}},\
  \bibinfo {pages} {144411} (\bibinfo {year} {2016})}\BibitemShut {NoStop}%
\bibitem [{\citenamefont {He}\ \emph {et~al.}(2017)\citenamefont {He},
  \citenamefont {Zaletel}, \citenamefont {Oshikawa},\ and\ \citenamefont
  {Pollmann}}]{he2017signatures}%
  \BibitemOpen
  \bibfield  {author} {\bibinfo {author} {\bibfnamefont {Yin-Chen}\
  \bibnamefont {He}}, \bibinfo {author} {\bibfnamefont {Michael~P}\
  \bibnamefont {Zaletel}}, \bibinfo {author} {\bibfnamefont {Masaki}\
  \bibnamefont {Oshikawa}}, \ and\ \bibinfo {author} {\bibfnamefont {Frank}\
  \bibnamefont {Pollmann}},\ }\bibfield  {title} {\enquote {\bibinfo {title}
  {Signatures of dirac cones in a dmrg study of the kagome heisenberg model},}\
  }\href@noop {} {\bibfield  {journal} {\bibinfo  {journal} {Physical Review
  X}\ }\textbf {\bibinfo {volume} {7}},\ \bibinfo {pages} {031020} (\bibinfo
  {year} {2017})}\BibitemShut {NoStop}%
\bibitem [{\citenamefont {Hu}\ \emph {et~al.}(2019)\citenamefont {Hu},
  \citenamefont {Zhu}, \citenamefont {Eggert},\ and\ \citenamefont
  {He}}]{hu2019dirac}%
  \BibitemOpen
  \bibfield  {author} {\bibinfo {author} {\bibfnamefont {Shijie}\ \bibnamefont
  {Hu}}, \bibinfo {author} {\bibfnamefont {W}~\bibnamefont {Zhu}}, \bibinfo
  {author} {\bibfnamefont {Sebastian}\ \bibnamefont {Eggert}}, \ and\ \bibinfo
  {author} {\bibfnamefont {Yin-Chen}\ \bibnamefont {He}},\ }\bibfield  {title}
  {\enquote {\bibinfo {title} {Dirac spin liquid on the spin-1/2 triangular
  heisenberg antiferromagnet},}\ }\href@noop {} {\bibfield  {journal} {\bibinfo
   {journal} {Physical review letters}\ }\textbf {\bibinfo {volume} {123}},\
  \bibinfo {pages} {207203} (\bibinfo {year} {2019})}\BibitemShut {NoStop}%
\bibitem [{\citenamefont {Belavin}\ and\ \citenamefont
  {Migdal}(1974)}]{Belavin:1974gu}%
  \BibitemOpen
  \bibfield  {author} {\bibinfo {author} {\bibfnamefont {A.~A.}\ \bibnamefont
  {Belavin}}\ and\ \bibinfo {author} {\bibfnamefont {A.~A.}\ \bibnamefont
  {Migdal}},\ }\bibfield  {title} {\enquote {\bibinfo {title} {{Calculation of
  anomalous dimensions in non-abelian gauge field theories}},}\ }\href@noop {}
  {\bibfield  {journal} {\bibinfo  {journal} {Pisma Zh. Eksp. Teor. Fiz.}\
  }\textbf {\bibinfo {volume} {19}},\ \bibinfo {pages} {317--320} (\bibinfo
  {year} {1974})}\BibitemShut {NoStop}%
\bibitem [{\citenamefont {Caswell}(1974)}]{caswell1974asymptotic}%
  \BibitemOpen
  \bibfield  {author} {\bibinfo {author} {\bibfnamefont {William~E}\
  \bibnamefont {Caswell}},\ }\bibfield  {title} {\enquote {\bibinfo {title}
  {Asymptotic behavior of non-abelian gauge theories to two-loop order},}\
  }\href@noop {} {\bibfield  {journal} {\bibinfo  {journal} {Physical Review
  Letters}\ }\textbf {\bibinfo {volume} {33}},\ \bibinfo {pages} {244}
  (\bibinfo {year} {1974})}\BibitemShut {NoStop}%
\bibitem [{\citenamefont {Banks}\ and\ \citenamefont
  {Zaks}(1982)}]{banks1982phase}%
  \BibitemOpen
  \bibfield  {author} {\bibinfo {author} {\bibfnamefont {Tom}\ \bibnamefont
  {Banks}}\ and\ \bibinfo {author} {\bibfnamefont {A.}~\bibnamefont {Zaks}},\
  }\bibfield  {title} {\enquote {\bibinfo {title} {{On the Phase Structure of
  Vector-Like Gauge Theories with Massless Fermions}},}\ }\href {\doibase
  10.1016/0550-3213(82)90035-9} {\bibfield  {journal} {\bibinfo  {journal}
  {Nucl. Phys. B}\ }\textbf {\bibinfo {volume} {196}},\ \bibinfo {pages}
  {189--204} (\bibinfo {year} {1982})}\BibitemShut {NoStop}%
\bibitem [{\citenamefont {Dagotto}\ \emph {et~al.}(1990)\citenamefont
  {Dagotto}, \citenamefont {Koci{\'c}},\ and\ \citenamefont
  {Kogut}}]{dagotto1990chiral}%
  \BibitemOpen
  \bibfield  {author} {\bibinfo {author} {\bibfnamefont {Elbio}\ \bibnamefont
  {Dagotto}}, \bibinfo {author} {\bibfnamefont {Aleksandar}\ \bibnamefont
  {Koci{\'c}}}, \ and\ \bibinfo {author} {\bibfnamefont {JB}~\bibnamefont
  {Kogut}},\ }\bibfield  {title} {\enquote {\bibinfo {title} {Chiral symmetry
  breaking in three dimensional qed with nf flavors},}\ }\href@noop {}
  {\bibfield  {journal} {\bibinfo  {journal} {Nuclear Physics B}\ }\textbf
  {\bibinfo {volume} {334}},\ \bibinfo {pages} {279--301} (\bibinfo {year}
  {1990})}\BibitemShut {NoStop}%
\bibitem [{\citenamefont {{Kaveh}}\ and\ \citenamefont
  {{Herbut}}(2005)}]{Kaveh2005}%
  \BibitemOpen
  \bibfield  {author} {\bibinfo {author} {\bibfnamefont {Kamran}\ \bibnamefont
  {{Kaveh}}}\ and\ \bibinfo {author} {\bibfnamefont {Igor~F.}\ \bibnamefont
  {{Herbut}}},\ }\bibfield  {title} {\enquote {\bibinfo {title} {{Chiral
  symmetry breaking in three-dimensional quantum electrodynamics in the
  presence of irrelevant interactions: A renormalization group study}},}\
  }\href {\doibase 10.1103/PhysRevB.71.184519} {\bibfield  {journal} {\bibinfo
  {journal} {\prb}\ }\textbf {\bibinfo {volume} {71}},\ \bibinfo {eid} {184519}
  (\bibinfo {year} {2005})},\ \Eprint {http://arxiv.org/abs/cond-mat/0411594}
  {arXiv:cond-mat/0411594 [cond-mat.supr-con]} \BibitemShut {NoStop}%
\bibitem [{\citenamefont {{Grover}}(2012)}]{Grover2012Chiral}%
  \BibitemOpen
  \bibfield  {author} {\bibinfo {author} {\bibfnamefont {Tarun}\ \bibnamefont
  {{Grover}}},\ }\bibfield  {title} {\enquote {\bibinfo {title} {{Chiral
  Symmetry Breaking, Deconfinement and Entanglement Monotonicity}},}\ }\href
  {\doibase 10.48550/arXiv.1211.1392} {\bibfield  {journal} {\bibinfo
  {journal} {arXiv e-prints}\ ,\ \bibinfo {eid} {arXiv:1211.1392}} (\bibinfo
  {year} {2012})},\ \Eprint {http://arxiv.org/abs/1211.1392} {arXiv:1211.1392
  [hep-th]} \BibitemShut {NoStop}%
\bibitem [{\citenamefont {{Karthik}}\ and\ \citenamefont
  {{Narayanan}}(2016{\natexlab{a}})}]{Karthik2015}%
  \BibitemOpen
  \bibfield  {author} {\bibinfo {author} {\bibfnamefont {Nikhil}\ \bibnamefont
  {{Karthik}}}\ and\ \bibinfo {author} {\bibfnamefont {Rajamani}\ \bibnamefont
  {{Narayanan}}},\ }\bibfield  {title} {\enquote {\bibinfo {title} {{No
  evidence for bilinear condensate in parity-invariant three-dimensional QED
  with massless fermions}},}\ }\href {\doibase 10.1103/PhysRevD.93.045020}
  {\bibfield  {journal} {\bibinfo  {journal} {\prd}\ }\textbf {\bibinfo
  {volume} {93}},\ \bibinfo {eid} {045020} (\bibinfo {year}
  {2016}{\natexlab{a}})},\ \Eprint {http://arxiv.org/abs/1512.02993}
  {arXiv:1512.02993 [hep-lat]} \BibitemShut {NoStop}%
\bibitem [{\citenamefont {{Karthik}}\ and\ \citenamefont
  {{Narayanan}}(2016{\natexlab{b}})}]{Karthik2016}%
  \BibitemOpen
  \bibfield  {author} {\bibinfo {author} {\bibfnamefont {Nikhil}\ \bibnamefont
  {{Karthik}}}\ and\ \bibinfo {author} {\bibfnamefont {Rajamani}\ \bibnamefont
  {{Narayanan}}},\ }\bibfield  {title} {\enquote {\bibinfo {title} {{Scale
  invariance of parity-invariant three-dimensional QED}},}\ }\href {\doibase
  10.1103/PhysRevD.94.065026} {\bibfield  {journal} {\bibinfo  {journal}
  {\prd}\ }\textbf {\bibinfo {volume} {94}},\ \bibinfo {eid} {065026} (\bibinfo
  {year} {2016}{\natexlab{b}})},\ \Eprint {http://arxiv.org/abs/1606.04109}
  {arXiv:1606.04109 [hep-th]} \BibitemShut {NoStop}%
\bibitem [{\citenamefont {Giombi}\ \emph {et~al.}(2016)\citenamefont {Giombi},
  \citenamefont {Klebanov},\ and\ \citenamefont {Tarnopolsky}}]{Giombi2015}%
  \BibitemOpen
  \bibfield  {author} {\bibinfo {author} {\bibfnamefont {Simone}\ \bibnamefont
  {Giombi}}, \bibinfo {author} {\bibfnamefont {Igor~R.}\ \bibnamefont
  {Klebanov}}, \ and\ \bibinfo {author} {\bibfnamefont {Grigory}\ \bibnamefont
  {Tarnopolsky}},\ }\bibfield  {title} {\enquote {\bibinfo {title} {{Conformal
  QED$_d$, $F$-Theorem and the $\epsilon$ Expansion}},}\ }\href {\doibase
  10.1088/1751-8113/49/13/135403} {\bibfield  {journal} {\bibinfo  {journal}
  {J. Phys. A}\ }\textbf {\bibinfo {volume} {49}},\ \bibinfo {pages} {135403}
  (\bibinfo {year} {2016})},\ \Eprint {http://arxiv.org/abs/1508.06354}
  {arXiv:1508.06354 [hep-th]} \BibitemShut {NoStop}%
\bibitem [{\citenamefont {{Di Pietro}}\ \emph {et~al.}(2016)\citenamefont {{Di
  Pietro}}, \citenamefont {{Komargodski}}, \citenamefont {{Shamir}},\ and\
  \citenamefont {{Stamou}}}]{Lorenzo2016}%
  \BibitemOpen
  \bibfield  {author} {\bibinfo {author} {\bibfnamefont {Lorenzo}\ \bibnamefont
  {{Di Pietro}}}, \bibinfo {author} {\bibfnamefont {Zohar}\ \bibnamefont
  {{Komargodski}}}, \bibinfo {author} {\bibfnamefont {Itamar}\ \bibnamefont
  {{Shamir}}}, \ and\ \bibinfo {author} {\bibfnamefont {Emmanuel}\ \bibnamefont
  {{Stamou}}},\ }\bibfield  {title} {\enquote {\bibinfo {title} {{Quantum
  Electrodynamics in d =3 from the $\epsilon$ Expansion}},}\ }\href {\doibase
  10.1103/PhysRevLett.116.131601} {\bibfield  {journal} {\bibinfo  {journal}
  {\prl}\ }\textbf {\bibinfo {volume} {116}},\ \bibinfo {eid} {131601}
  (\bibinfo {year} {2016})},\ \Eprint {http://arxiv.org/abs/1508.06278}
  {arXiv:1508.06278 [hep-th]} \BibitemShut {NoStop}%
\bibitem [{\citenamefont {Reehorst}\ \emph {et~al.}(2021)\citenamefont
  {Reehorst}, \citenamefont {Rychkov}, \citenamefont {Simmons-Duffin},
  \citenamefont {Sirois}, \citenamefont {Su},\ and\ \citenamefont {van
  Rees}}]{reehorst2021navigator}%
  \BibitemOpen
  \bibfield  {author} {\bibinfo {author} {\bibfnamefont {Marten}\ \bibnamefont
  {Reehorst}}, \bibinfo {author} {\bibfnamefont {Slava}\ \bibnamefont
  {Rychkov}}, \bibinfo {author} {\bibfnamefont {David}\ \bibnamefont
  {Simmons-Duffin}}, \bibinfo {author} {\bibfnamefont {Benoit}\ \bibnamefont
  {Sirois}}, \bibinfo {author} {\bibfnamefont {Ning}\ \bibnamefont {Su}}, \
  and\ \bibinfo {author} {\bibfnamefont {Balt}\ \bibnamefont {van Rees}},\
  }\bibfield  {title} {\enquote {\bibinfo {title} {Navigator function for the
  conformal bootstrap},}\ }\href@noop {} {\bibfield  {journal} {\bibinfo
  {journal} {SciPost Physics}\ }\textbf {\bibinfo {volume} {11}},\ \bibinfo
  {pages} {072} (\bibinfo {year} {2021})}\BibitemShut {NoStop}%
\bibitem [{\citenamefont {Benoit}\ \emph {et~al.}(2023)\citenamefont {Benoit},
  \citenamefont {Reehorst}, \citenamefont {Rychkov},\ and\ \citenamefont
  {Van~Rees}}]{SlavaUnpublished2}%
  \BibitemOpen
  \bibfield  {author} {\bibinfo {author} {\bibfnamefont {S.}~\bibnamefont
  {Benoit}}, \bibinfo {author} {\bibfnamefont {M.}~\bibnamefont {Reehorst}},
  \bibinfo {author} {\bibfnamefont {S.}~\bibnamefont {Rychkov}}, \ and\
  \bibinfo {author} {\bibfnamefont {B.}~\bibnamefont {Van~Rees}},\ }\href@noop
  {} {\  (\bibinfo {year} {2023})},\ \bibinfo {note} {{\it to
  appear}}\BibitemShut {NoStop}%
\bibitem [{\citenamefont {Dymarsky}\ \emph {et~al.}(2019)\citenamefont
  {Dymarsky}, \citenamefont {Penedones}, \citenamefont {Trevisani},\ and\
  \citenamefont {Vichi}}]{Dymarsky:2017xzb}%
  \BibitemOpen
  \bibfield  {author} {\bibinfo {author} {\bibfnamefont {Anatoly}\ \bibnamefont
  {Dymarsky}}, \bibinfo {author} {\bibfnamefont {Joao}\ \bibnamefont
  {Penedones}}, \bibinfo {author} {\bibfnamefont {Emilio}\ \bibnamefont
  {Trevisani}}, \ and\ \bibinfo {author} {\bibfnamefont {Alessandro}\
  \bibnamefont {Vichi}},\ }\bibfield  {title} {\enquote {\bibinfo {title}
  {{Charting the space of 3D CFTs with a continuous global symmetry}},}\ }\href
  {\doibase 10.1007/JHEP05(2019)098} {\bibfield  {journal} {\bibinfo  {journal}
  {JHEP}\ }\textbf {\bibinfo {volume} {05}},\ \bibinfo {pages} {098} (\bibinfo
  {year} {2019})},\ \Eprint {http://arxiv.org/abs/1705.04278} {arXiv:1705.04278
  [hep-th]} \BibitemShut {NoStop}%
%%CITATION = ARXIV:1705.04278;%%
\bibitem [{\citenamefont {Reehorst}\ \emph {et~al.}(2019)\citenamefont
  {Reehorst}, \citenamefont {Trevisani},\ and\ \citenamefont
  {Vichi}}]{Reehorst:2019pzi}%
  \BibitemOpen
  \bibfield  {author} {\bibinfo {author} {\bibfnamefont {Marten}\ \bibnamefont
  {Reehorst}}, \bibinfo {author} {\bibfnamefont {Emilio}\ \bibnamefont
  {Trevisani}}, \ and\ \bibinfo {author} {\bibfnamefont {Alessandro}\
  \bibnamefont {Vichi}},\ }\bibfield  {title} {\enquote {\bibinfo {title}
  {{Mixed Scalar-Current bootstrap in three dimensions}},}\ }\href@noop {} {\
  (\bibinfo {year} {2019})},\ \Eprint {http://arxiv.org/abs/1911.05747}
  {arXiv:1911.05747 [hep-th]} \BibitemShut {NoStop}%
%%CITATION = ARXIV:1911.05747;%%
\bibitem [{\citenamefont {Dymarsky}\ \emph {et~al.}(2018)\citenamefont
  {Dymarsky}, \citenamefont {Kos}, \citenamefont {Kravchuk}, \citenamefont
  {Poland},\ and\ \citenamefont {Simmons-Duffin}}]{Dymarsky:2017yzx}%
  \BibitemOpen
  \bibfield  {author} {\bibinfo {author} {\bibfnamefont {Anatoly}\ \bibnamefont
  {Dymarsky}}, \bibinfo {author} {\bibfnamefont {Filip}\ \bibnamefont {Kos}},
  \bibinfo {author} {\bibfnamefont {Petr}\ \bibnamefont {Kravchuk}}, \bibinfo
  {author} {\bibfnamefont {David}\ \bibnamefont {Poland}}, \ and\ \bibinfo
  {author} {\bibfnamefont {David}\ \bibnamefont {Simmons-Duffin}},\ }\bibfield
  {title} {\enquote {\bibinfo {title} {{The 3d Stress-Tensor Bootstrap}},}\
  }\href {\doibase 10.1007/JHEP02(2018)164} {\bibfield  {journal} {\bibinfo
  {journal} {JHEP}\ }\textbf {\bibinfo {volume} {02}},\ \bibinfo {pages} {164}
  (\bibinfo {year} {2018})},\ \Eprint {http://arxiv.org/abs/1708.05718}
  {arXiv:1708.05718 [hep-th]} \BibitemShut {NoStop}%
%%CITATION = ARXIV:1708.05718;%%
\bibitem [{\citenamefont {Li}\ \emph {et~al.}(2016)\citenamefont {Li},
  \citenamefont {Meltzer},\ and\ \citenamefont {Poland}}]{Li:2015itl}%
  \BibitemOpen
  \bibfield  {author} {\bibinfo {author} {\bibfnamefont {Daliang}\ \bibnamefont
  {Li}}, \bibinfo {author} {\bibfnamefont {David}\ \bibnamefont {Meltzer}}, \
  and\ \bibinfo {author} {\bibfnamefont {David}\ \bibnamefont {Poland}},\
  }\bibfield  {title} {\enquote {\bibinfo {title} {{Conformal Collider Physics
  from the Lightcone Bootstrap}},}\ }\href {\doibase 10.1007/JHEP02(2016)143}
  {\bibfield  {journal} {\bibinfo  {journal} {JHEP}\ }\textbf {\bibinfo
  {volume} {02}},\ \bibinfo {pages} {143} (\bibinfo {year} {2016})},\ \Eprint
  {http://arxiv.org/abs/1511.08025} {arXiv:1511.08025 [hep-th]} \BibitemShut
  {NoStop}%
\bibitem [{\citenamefont {Costa}\ \emph
  {et~al.}(2011{\natexlab{a}})\citenamefont {Costa}, \citenamefont {Penedones},
  \citenamefont {Poland},\ and\ \citenamefont {Rychkov}}]{Costa:2011mg}%
  \BibitemOpen
  \bibfield  {author} {\bibinfo {author} {\bibfnamefont {Miguel~S.}\
  \bibnamefont {Costa}}, \bibinfo {author} {\bibfnamefont {Joao}\ \bibnamefont
  {Penedones}}, \bibinfo {author} {\bibfnamefont {David}\ \bibnamefont
  {Poland}}, \ and\ \bibinfo {author} {\bibfnamefont {Slava}\ \bibnamefont
  {Rychkov}},\ }\bibfield  {title} {\enquote {\bibinfo {title} {{Spinning
  Conformal Correlators}},}\ }\href {\doibase 10.1007/JHEP11(2011)071}
  {\bibfield  {journal} {\bibinfo  {journal} {JHEP}\ }\textbf {\bibinfo
  {volume} {11}},\ \bibinfo {pages} {071} (\bibinfo {year}
  {2011}{\natexlab{a}})},\ \Eprint {http://arxiv.org/abs/1107.3554}
  {arXiv:1107.3554 [hep-th]} \BibitemShut {NoStop}%
%%CITATION = ARXIV:1107.3554;%%
\bibitem [{\citenamefont {{Hofman}}\ and\ \citenamefont
  {{Maldacena}}(2008)}]{HofmanConformalCollider2008}%
  \BibitemOpen
  \bibfield  {author} {\bibinfo {author} {\bibfnamefont {Diego~M.}\
  \bibnamefont {{Hofman}}}\ and\ \bibinfo {author} {\bibfnamefont {Juan}\
  \bibnamefont {{Maldacena}}},\ }\bibfield  {title} {\enquote {\bibinfo {title}
  {{Conformal collider physics: energy and charge correlations}},}\ }\href
  {\doibase 10.1088/1126-6708/2008/05/012} {\bibfield  {journal} {\bibinfo
  {journal} {Journal of High Energy Physics}\ }\textbf {\bibinfo {volume}
  {2008}},\ \bibinfo {eid} {012} (\bibinfo {year} {2008})},\ \Eprint
  {http://arxiv.org/abs/0803.1467} {arXiv:0803.1467 [hep-th]} \BibitemShut
  {NoStop}%
\bibitem [{\citenamefont {{Zhou}}\ and\ \citenamefont
  {{He}}(2022)}]{Zhou2022HigherSpin}%
  \BibitemOpen
  \bibfield  {author} {\bibinfo {author} {\bibfnamefont {Zheng}\ \bibnamefont
  {{Zhou}}}\ and\ \bibinfo {author} {\bibfnamefont {Yin-Chen}\ \bibnamefont
  {{He}}},\ }\bibfield  {title} {\enquote {\bibinfo {title} {{Slightly broken
  higher-spin current in bosonic and fermionic QED in the large-$N$ limit}},}\
  }\href {\doibase 10.48550/arXiv.2205.07897} {\bibfield  {journal} {\bibinfo
  {journal} {arXiv e-prints}\ ,\ \bibinfo {eid} {arXiv:2205.07897}} (\bibinfo
  {year} {2022})},\ \Eprint {http://arxiv.org/abs/2205.07897} {arXiv:2205.07897
  [hep-th]} \BibitemShut {NoStop}%
\bibitem [{\citenamefont {Osborn}\ and\ \citenamefont
  {Petkou}(1994)}]{Osborn:1993cr}%
  \BibitemOpen
  \bibfield  {author} {\bibinfo {author} {\bibfnamefont {H.}~\bibnamefont
  {Osborn}}\ and\ \bibinfo {author} {\bibfnamefont {A.C.}\ \bibnamefont
  {Petkou}},\ }\bibfield  {title} {\enquote {\bibinfo {title} {{Implications of
  conformal invariance in field theories for general dimensions}},}\ }\href
  {\doibase 10.1006/aphy.1994.1045} {\bibfield  {journal} {\bibinfo  {journal}
  {Annals Phys.}\ }\textbf {\bibinfo {volume} {231}},\ \bibinfo {pages}
  {311--362} (\bibinfo {year} {1994})},\ \Eprint
  {http://arxiv.org/abs/hep-th/9307010} {arXiv:hep-th/9307010 [hep-th]}
  \BibitemShut {NoStop}%
%%CITATION = HEP-TH/9307010;%%
\bibitem [{\citenamefont {Chester}\ and\ \citenamefont
  {Pufu}(2016{\natexlab{b}})}]{Chester:2016ref}%
  \BibitemOpen
  \bibfield  {author} {\bibinfo {author} {\bibfnamefont {Shai~M.}\ \bibnamefont
  {Chester}}\ and\ \bibinfo {author} {\bibfnamefont {Silviu~S.}\ \bibnamefont
  {Pufu}},\ }\bibfield  {title} {\enquote {\bibinfo {title} {{Anomalous
  dimensions of scalar operators in QED$_{3}$}},}\ }\href {\doibase
  10.1007/JHEP08(2016)069} {\bibfield  {journal} {\bibinfo  {journal} {JHEP}\
  }\textbf {\bibinfo {volume} {08}},\ \bibinfo {pages} {069} (\bibinfo {year}
  {2016}{\natexlab{b}})},\ \Eprint {http://arxiv.org/abs/1603.05582}
  {arXiv:1603.05582 [hep-th]} \BibitemShut {NoStop}%
%%CITATION = ARXIV:1603.05582;%%
\bibitem [{\citenamefont {{Manashov}}\ and\ \citenamefont
  {{Skvortsov}}(2017)}]{Manashov2017GNYHigherSpin}%
  \BibitemOpen
  \bibfield  {author} {\bibinfo {author} {\bibfnamefont {A.~N.}\ \bibnamefont
  {{Manashov}}}\ and\ \bibinfo {author} {\bibfnamefont {E.~D.}\ \bibnamefont
  {{Skvortsov}}},\ }\bibfield  {title} {\enquote {\bibinfo {title}
  {{Higher-spin currents in the Gross-Neveu model at 1/n$^{2}$}},}\ }\href
  {\doibase 10.1007/JHEP01(2017)132} {\bibfield  {journal} {\bibinfo  {journal}
  {Journal of High Energy Physics}\ }\textbf {\bibinfo {volume} {2017}},\
  \bibinfo {eid} {132} (\bibinfo {year} {2017})},\ \Eprint
  {http://arxiv.org/abs/1610.06938} {arXiv:1610.06938 [hep-th]} \BibitemShut
  {NoStop}%
\bibitem [{\citenamefont {Meneses}\ \emph {et~al.}(2019)\citenamefont
  {Meneses}, \citenamefont {Penedones}, \citenamefont {Rychkov}, \citenamefont
  {Viana Parente~Lopes},\ and\ \citenamefont {Yvernay}}]{Meneses:2018xpu}%
  \BibitemOpen
  \bibfield  {author} {\bibinfo {author} {\bibfnamefont {Sim\~ao}\ \bibnamefont
  {Meneses}}, \bibinfo {author} {\bibfnamefont {Jo\~ao}\ \bibnamefont
  {Penedones}}, \bibinfo {author} {\bibfnamefont {Slava}\ \bibnamefont
  {Rychkov}}, \bibinfo {author} {\bibfnamefont {J.~M.}\ \bibnamefont {Viana
  Parente~Lopes}}, \ and\ \bibinfo {author} {\bibfnamefont {Pierre}\
  \bibnamefont {Yvernay}},\ }\bibfield  {title} {\enquote {\bibinfo {title} {{A
  structural test for the conformal invariance of the critical 3d Ising
  model}},}\ }\href {\doibase 10.1007/JHEP04(2019)115} {\bibfield  {journal}
  {\bibinfo  {journal} {JHEP}\ }\textbf {\bibinfo {volume} {04}},\ \bibinfo
  {pages} {115} (\bibinfo {year} {2019})},\ \Eprint
  {http://arxiv.org/abs/1802.02319} {arXiv:1802.02319 [hep-th]} \BibitemShut
  {NoStop}%
\bibitem [{\citenamefont {Henriksson}(2023)}]{Henriksson:2022rnm}%
  \BibitemOpen
  \bibfield  {author} {\bibinfo {author} {\bibfnamefont {Johan}\ \bibnamefont
  {Henriksson}},\ }\bibfield  {title} {\enquote {\bibinfo {title} {{The
  critical O(N) CFT: Methods and conformal data}},}\ }\href {\doibase
  10.1016/j.physrep.2022.12.002} {\bibfield  {journal} {\bibinfo  {journal}
  {Phys. Rept.}\ }\textbf {\bibinfo {volume} {1002}},\ \bibinfo {pages} {1--72}
  (\bibinfo {year} {2023})},\ \Eprint {http://arxiv.org/abs/2201.09520}
  {arXiv:2201.09520 [hep-th]} \BibitemShut {NoStop}%
\bibitem [{\citenamefont {Costa}\ \emph
  {et~al.}(2011{\natexlab{b}})\citenamefont {Costa}, \citenamefont {Penedones},
  \citenamefont {Poland},\ and\ \citenamefont {Rychkov}}]{Costa:2011dw}%
  \BibitemOpen
  \bibfield  {author} {\bibinfo {author} {\bibfnamefont {Miguel~S.}\
  \bibnamefont {Costa}}, \bibinfo {author} {\bibfnamefont {Joao}\ \bibnamefont
  {Penedones}}, \bibinfo {author} {\bibfnamefont {David}\ \bibnamefont
  {Poland}}, \ and\ \bibinfo {author} {\bibfnamefont {Slava}\ \bibnamefont
  {Rychkov}},\ }\bibfield  {title} {\enquote {\bibinfo {title} {{Spinning
  Conformal Blocks}},}\ }\href {\doibase 10.1007/JHEP11(2011)154} {\bibfield
  {journal} {\bibinfo  {journal} {JHEP}\ }\textbf {\bibinfo {volume} {1111}},\
  \bibinfo {pages} {154} (\bibinfo {year} {2011}{\natexlab{b}})},\ \Eprint
  {http://arxiv.org/abs/1109.6321} {arXiv:1109.6321 [hep-th]} \BibitemShut
  {NoStop}%
%%CITATION = ARXIV:1109.6321;%%
\bibitem [{\citenamefont {Erramilli}\ \emph {et~al.}(2021)\citenamefont
  {Erramilli}, \citenamefont {Iliesiu}, \citenamefont {Kravchuk}, \citenamefont
  {Landry}, \citenamefont {Poland},\ and\ \citenamefont
  {Simmons-Duffin}}]{Erramilli:2020rlr}%
  \BibitemOpen
  \bibfield  {author} {\bibinfo {author} {\bibfnamefont {Rajeev~S.}\
  \bibnamefont {Erramilli}}, \bibinfo {author} {\bibfnamefont {Luca~V.}\
  \bibnamefont {Iliesiu}}, \bibinfo {author} {\bibfnamefont {Petr}\
  \bibnamefont {Kravchuk}}, \bibinfo {author} {\bibfnamefont {Walter}\
  \bibnamefont {Landry}}, \bibinfo {author} {\bibfnamefont {David}\
  \bibnamefont {Poland}}, \ and\ \bibinfo {author} {\bibfnamefont {David}\
  \bibnamefont {Simmons-Duffin}},\ }\bibfield  {title} {\enquote {\bibinfo
  {title} {{blocks\_3d: software for general 3d conformal blocks}},}\ }\href
  {\doibase 10.1007/JHEP11(2021)006} {\bibfield  {journal} {\bibinfo  {journal}
  {JHEP}\ }\textbf {\bibinfo {volume} {11}},\ \bibinfo {pages} {006} (\bibinfo
  {year} {2021})},\ \Eprint {http://arxiv.org/abs/2011.01959} {arXiv:2011.01959
  [hep-th]} \BibitemShut {NoStop}%
\bibitem [{\citenamefont {Kravchuk}\ and\ \citenamefont
  {Simmons-Duffin}(2018)}]{Kravchuk:2016qvl}%
  \BibitemOpen
  \bibfield  {author} {\bibinfo {author} {\bibfnamefont {Petr}\ \bibnamefont
  {Kravchuk}}\ and\ \bibinfo {author} {\bibfnamefont {David}\ \bibnamefont
  {Simmons-Duffin}},\ }\bibfield  {title} {\enquote {\bibinfo {title}
  {{Counting Conformal Correlators}},}\ }\href {\doibase
  10.1007/JHEP02(2018)096} {\bibfield  {journal} {\bibinfo  {journal} {JHEP}\
  }\textbf {\bibinfo {volume} {02}},\ \bibinfo {pages} {096} (\bibinfo {year}
  {2018})},\ \Eprint {http://arxiv.org/abs/1612.08987} {arXiv:1612.08987
  [hep-th]} \BibitemShut {NoStop}%
\bibitem [{\citenamefont {Dymarsky}(2015)}]{Dymarsky:2013wla}%
  \BibitemOpen
  \bibfield  {author} {\bibinfo {author} {\bibfnamefont {Anatoly}\ \bibnamefont
  {Dymarsky}},\ }\bibfield  {title} {\enquote {\bibinfo {title} {{On the
  four-point function of the stress-energy tensors in a CFT}},}\ }\href
  {\doibase 10.1007/JHEP10(2015)075} {\bibfield  {journal} {\bibinfo  {journal}
  {JHEP}\ }\textbf {\bibinfo {volume} {10}},\ \bibinfo {pages} {075} (\bibinfo
  {year} {2015})},\ \Eprint {http://arxiv.org/abs/1311.4546} {arXiv:1311.4546
  [hep-th]} \BibitemShut {NoStop}%
\bibitem [{\citenamefont {Rong}\ and\ \citenamefont {Su}(2021)}]{Rong:2018okz}%
  \BibitemOpen
  \bibfield  {author} {\bibinfo {author} {\bibfnamefont {Junchen}\ \bibnamefont
  {Rong}}\ and\ \bibinfo {author} {\bibfnamefont {Ning}\ \bibnamefont {Su}},\
  }\bibfield  {title} {\enquote {\bibinfo {title} {{Bootstrapping the minimal $
  \mathcal{N} $ = 1 superconformal field theory in three dimensions}},}\ }\href
  {\doibase 10.1007/JHEP06(2021)154} {\bibfield  {journal} {\bibinfo  {journal}
  {JHEP}\ }\textbf {\bibinfo {volume} {06}},\ \bibinfo {pages} {154} (\bibinfo
  {year} {2021})},\ \Eprint {http://arxiv.org/abs/1807.04434} {arXiv:1807.04434
  [hep-th]} \BibitemShut {NoStop}%
\bibitem [{\citenamefont {Iliesiu}\ \emph {et~al.}(2016)\citenamefont
  {Iliesiu}, \citenamefont {Kos}, \citenamefont {Poland}, \citenamefont {Pufu},
  \citenamefont {Simmons-Duffin},\ and\ \citenamefont
  {Yacoby}}]{Iliesiu:2015qra}%
  \BibitemOpen
  \bibfield  {author} {\bibinfo {author} {\bibfnamefont {Luca}\ \bibnamefont
  {Iliesiu}}, \bibinfo {author} {\bibfnamefont {Filip}\ \bibnamefont {Kos}},
  \bibinfo {author} {\bibfnamefont {David}\ \bibnamefont {Poland}}, \bibinfo
  {author} {\bibfnamefont {Silviu~S.}\ \bibnamefont {Pufu}}, \bibinfo {author}
  {\bibfnamefont {David}\ \bibnamefont {Simmons-Duffin}}, \ and\ \bibinfo
  {author} {\bibfnamefont {Ran}\ \bibnamefont {Yacoby}},\ }\bibfield  {title}
  {\enquote {\bibinfo {title} {{Bootstrapping 3D Fermions}},}\ }\href {\doibase
  10.1007/JHEP03(2016)120} {\bibfield  {journal} {\bibinfo  {journal} {JHEP}\
  }\textbf {\bibinfo {volume} {03}},\ \bibinfo {pages} {120} (\bibinfo {year}
  {2016})},\ \Eprint {http://arxiv.org/abs/1508.00012} {arXiv:1508.00012
  [hep-th]} \BibitemShut {NoStop}%
%%CITATION = ARXIV:1508.00012;%%
\bibitem [{\citenamefont {Simmons-Duffin}(2015)}]{Simmons-Duffin:2015qma}%
  \BibitemOpen
  \bibfield  {author} {\bibinfo {author} {\bibfnamefont {David}\ \bibnamefont
  {Simmons-Duffin}},\ }\bibfield  {title} {\enquote {\bibinfo {title} {{A
  Semidefinite Program Solver for the Conformal Bootstrap}},}\ }\href {\doibase
  10.1007/JHEP06(2015)174} {\bibfield  {journal} {\bibinfo  {journal} {JHEP}\
  }\textbf {\bibinfo {volume} {06}},\ \bibinfo {pages} {174} (\bibinfo {year}
  {2015})},\ \Eprint {http://arxiv.org/abs/1502.02033} {arXiv:1502.02033
  [hep-th]} \BibitemShut {NoStop}%
%%CITATION = ARXIV:1502.02033;%%
\end{thebibliography}%
\end{document}